\documentclass[12pt]{article}

\usepackage{epsfig}
\usepackage[height=10in, width=7in]{geometry}
\usepackage{graphics}
\usepackage{amsfonts}
\usepackage{amssymb}
\usepackage{amsmath}
\def\ri{\mathrm i}
\def\e{\mathrm e}
\def\C{\mathbb C}
\def\R{\mathbb R}
\newcommand{\be}{\begin{equation}}
\newcommand{\ee}{\end{equation}}
\newcommand{\bea}{\begin{eqnarray}}
\newcommand{\eea}{\end{eqnarray}}
\newcommand{\ba}{\begin{array}}
\newcommand{\ea}{\end{array}}
\newcommand{\rf}[1] {(\ref{#1})}
\newcommand{\half}{\mbox{$\frac{1}{2}$}}
\newcommand{\eps}{\epsilon}
\newcommand{\cn}{\mbox{cn}}
\newcommand{\sn}{\mbox{sn}}
\newcommand{\dn}{\mbox{dn}}
\newcommand{\sech}{\mbox{sech}}
\newcommand{\nn}{\nonumber}

\begin{document}

  \title{Nonlinear damped spatially periodic breathers and the emergence of soliton-like rogue waves}
  
  \author{C.M. Schober and A. Islas\\
Department of Mathematics\\
University of Central FLorida\\
Orlando, FL 32816, USA}

  \date{December 31, 2021}
  \maketitle

\begin{abstract}
  The spatially periodic breather solutions (SPBs)  of the nonlinear Schrödinger  equation, prominent in modeling rogue
  waves, are unstable.
 In this paper we numerically investigate the routes to stability of the SPBs and related rogue wave activity in the 
 framework of the nonlinear damped higher order nonlinear Schrödinger  (NLD-HONLS) equation.
 The NLD-HONLS solutions are initialized  with single-mode and  two-mode 
 SPB data at different stages of their development.
 The  Floquet spectral theory of the NLS equation is applied to interpret and provide a characterization of the perturbed dynamics in terms of nearby solutions of the NLS equation. A broad categorization of the routes to stability of the SPBs is determined. Novel behavior related to the effects of nonlinear damping is obtained: tiny bands of complex spectrum develop  in the Floquet decomposition of the NLD-HONLS data, indicating the  
breakup of the SPB  into either a one or two ``soliton-like'' structure.
For solutions initialized in the early to middle stage of the development of the modulational instability, we find that all the rogue waves  in the NLD-HONLS flow occur when  the spectrum is in a one or two soliton-like configuration.
When the solutions are initialized as the modulational instability is saturating,  rogue waves may occur when the spectrum is not in a soliton-like state.
Another distinctive feature of the nonlinear damped dynamics  is that 
the growth of instabilities can be delayed and  expressed at higher order due to permanent frequency downshifting 
\end{abstract}

\section{Introduction}
The  nonlinear Schrödinger  (NLS)  equation arises in many fields  when
modeling phenomena in which modulational instability (MI) and  nonlinear focusing are important. In particular, the NLS equation has proven very useful for initiating studies of rogue waves in deep water.
The heteroclinic orbits of modulationally unstable Stokes waves, also termed
spatially periodic breathers (SPBs), are large amplitude solutions
of the NLS equation which  capture  the  salient features of rogue waves and
have come to be regarded as prototypes of deep water rogue waves \cite{oos00,cs02,asa09,aek87}.
More recently, heteroclinic orbits of unstable $N$-phase solutions
have been proposed as models of rogue waves over non-uniform backgrounds \cite{cs17,cp18,cpw19}.

The details of the MI for
spatially $L$-periodic solutions of the NLS equation depend on, amongst other things,  the amplitude of the background and
the size $L$ of the domain. For the Stokes wave of fixed amplitude $a$
and $L  > L_0$ (the threshold for instability),  as $L$ increases the number of unstable modes (UMs) increases.
In the case of the Stokes wave with $N$ UMs, the associated SPBs can have $M \le N$ modes active and are referred to as  $M$-mode SPBs
(the single mode SPB is the Akhmediev breather).
Given their prominence in modeling rogue wave phenomena, the stability of the SPBs is important.
The squared eigenfunction connection between the nonlinear spectrum of the NLS equation and the linear stability problem was successfully used to establish
the linear instability of the $M$-mode SPBs ($M < N$) \cite{cs14}.
The  exact nature of the instability of the special
``maximal'' SPBs  (when $M = N$) is currently under investigation \cite{gs21}.

An important phenomenon related to MI is frequency downshifting which  occurs when the energy transfer from the carrier wave to a lower sideband becomes permanent.
For sufficiently steep waves, frequency downshifting has been  observed in both laboratory experiments and 
in ocean waves and it was noted that dissipation is most likely essential for permanent downshift of a realistic wave spectrum \cite{hame91, dtks03}.
More realistic explorations of deep water wave dynamics can be achieved with the Dysthe equation which accurately predicts laboratory data for a wider range of wave parameters than the NLS equation \cite{dysthe79,chb2019,trul97}. 
Even so, frequency downshifting is not captured by either the linear damped NLS or Dysthe equations  as the momentum and energy decay at the same rate, thus preserving the spectral center.
Permanent downshifting is captured by a nonlinear damped higher order nonlinear Schrödinger (NLD-HONLS) equation:
\be
iu_t + u_{xx} + 2|u|^2 u 
+ i \eps\left(\half u_{xxx}
- 8|u|^2 u_x - 2ui(1 + i\beta)\left[\mathcal{ H}\left(|u|^2\right)\right]_x\right) = 0,
\label{dhonls}
\ee
where $0 < \epsilon, \beta << 1$, $u(x,t)$ is the complex envelope of the wave train and is
$L$-periodic in space.  Here $\mathcal{ H}\{f(x)\} = \frac{1}{\pi}\int_{-\infty}^\infty \frac{f(\xi)}{x-\xi}d\xi$ is the Hilbert transform of $f$.
In a comparative study,
it was shown to be one of the most accurate models for predicting the permanent downshift observed in a set of laboratory experiments
\cite{ss15, chb2019}.

In this article we investigate the routes to stability  of single-mode and multi-mode SPBs
over an unstable  Stokes wave background and related rogue wave activity
in the framework of the NLD-HONLS equation.
The initial data used in the numerical experiments is generated by evaluating  exact one or two mode SPB solutions of the integrable NLS
equation at time $T_0$,  i.e. $U^{(j)}(x, T_0)$ or $U^{(i,j)}(x, T_0)$ (see Equations~\rf{SPB1}-\rf{SPB2}).
We
systematically vary  $T_0 \in [-5,0]$ to allow the NLD-HONLS solutions to be initialized with SPB data at different stages of their development, i.e. from
the early  stage of  the MI when there is strong growth ($T_0 = -5$)
through the nonlinear stage  
when the MI is saturating ($T_0 = 0$).

Viewing the NLD-HONLS  dynamics
as near integrable, the perturbed flow is analyzed by appealing to the Floquet spectral theory of the NLS equation.
This type of normal mode analysis, where a characterization of the perturbed dynamics is obtained by projecting onto integrable nonlinear modes,
has been successfully applied in earlier studies of chaos and chaotic transport in perturbations of the sine-Gordon equation and  NLS
equations \cite{Forest1993,cems96,mo95,Ablowitz_Herbst_Schober_1996b}.

To characterize the routes to stability of the SPBs we address the following questions:
i) What are the distinguishing features of the NLD-HONLS flow as determined by
the Floquet spectrum and which
NLS states are relevant and can be used to model the perturbed flow? 
ii) How  does the NLD-HONLS evolution, its  Floquet spectral decomposition, and rogue wave activity 
depend on $T_0$?

The Floquet spectral  decomposition of the data determines which nonlinear modes are active and whether they are unstable.
The spectrum of an SPB contains degenerate complex elements of the periodic/antiperiodic spectrum. It is well known that exponential instabilities are
associated with complex ``double points''  and  play a role in generating complex behavior in perturbations of the NLS equation \cite{efm90,mo95}.
Recently, in the unrestricted solution space, complex critical points arising from transverse intersections of
bands of complex spectrum were numerically shown to be
associated with a weaker instability \cite{si21}.
Section 2  provides background on the NLS spectral theory and it's
use in distinguishing instabilities and the relevant nearby NLS states which are used to characterize the perturbed flow.

The results of a broad set of
numerical experiments using the NLD-HONLS equation are presented in Section 3.
Given that the Floquet spectral decomposition of the data evolves in time, the spectrum is
numerically computed for $t \in[0,100]$.
The Floquet spectral analysis is complemented by an  examination of the growth of small perturbations in the SPB initial data under the NLD-HONLS flow.
In section 2.3 we obtain a first order approximation for the NLD-HONLS solution and via a perturbation analysis present results which confirm the spectral evolutions observed in the numerical experiments for short time.

Significantly, in the NLD-HONLS evolution we observe novel behavior:
very tiny bands of complex spectrum pinch off, reflecting the  
breakup of the SPB  into a waveform that is close
to either a one or two ``soliton-like'' structure (see sections 2.4, 3.2.1 and 3.2.3).
The emergence of the one or two soliton-like structure in the NLD-HONLS dynamics, its relation to the occurence  of
rogue waves, and its dependence on $T_0$ and $\beta$
is examined in Subsection 3.2.3.
For wide ranges of $T_0$, 
i.e. for solutions initialized in the early to middle stage of the development of the MI, all rogue waves are observed to occur when  the spectrum is close to a one or two soliton-like state.
  When the solutions are initialized as the MI is saturating,  rogue waves also occur  after the spectrum has left a soliton-like state (for $T_0 =0$ considerably later)  and develops as
a result of superposition of the nonlinear modes.
Finally, for small $\beta$  long lived rogue waves can occur.

The  instabilities of the NLD-HONLS  are correlated with complex critical points and
complex double points in the spectrum. To complement the Floquet analysis we examine the growth of small perturbations in the SPB initial data under the NLD-HONLS equation.  We find the growth saturates and the solution stabilizes once all the complex critical and double points vanish in the spectral decompositions.
Not only are the routes to stability for the nonlinear damped SPBs determined
by correlating variations in the spectrum with certain NLS solutions, but this study provides further support for the relevance of complex critical points in identifying  instabilities.

Other interesting  features arise  which are specific to nonlinear damping.
Nonlinear damping is essentially effective only near the crest of the envelope,
 steeper waves result in significantly stronger damping (see Figure~\ref{Figure1}A), and 
is responsible for permanent downshift (defined in Subsection 3.2.2) in the NLD-HONLS. 
As a result of downshifting there are two observed timescales in the evolution of the spectrum.
In some experiments we find that certain unstable modes do not resonate with the NLD-HONLS perturbation and the corresponding instability (and complex double point) persist. Due to downshifting the  growth of the instability is  delayed and expressed at higher order.

\section{Analytical framework}
\subsection{Characterization of instabilities on periodic domains}

The nonlinear Schrödinger  equation (when $\epsilon = 0$ in Equation~\rf{dhonls}) is a completely integrable equation that arises as the compatibility condition of the Zakharov-Shabat (Z-S) system \cite{zs72}:

\bea
\label{Lax1}
\mathcal{L}^{(x)} \mathbf v& =\begin{pmatrix} \partial/\partial x + i\lambda & -u \\ u^* & \partial/\partial x -i\lambda \end{pmatrix} \mathbf v = 0,
 \\
\label{Lax2}
  {\cal L}^{(t)} \mathbf v & = \begin{pmatrix} \partial/\partial t -\ri\left(|u|^2 - 2 \lambda^2\right) &  -\ri u_x - 2\lambda u \\ -\ri u^*_x +2 \lambda u^* & \partial/\partial t +\ri\left(|u|^2 -  2\lambda^2\right) \end{pmatrix} \mathbf v = 0,
\eea
where $\lambda$ is the complex spectral parameter, $\mathbf v$ is a complex vector valued eigenfunction, and $u(x,t)$ is a solution of the NLS equation itself.
A general  $L-$periodic solution of the NLS equation can be represented in terms of a set of
nonlinear modes whose spatial structure and stability are
determined by its Floquet spectrum
\be
\sigma(u) 
:= \left\{\lambda \in \C \, | \mathcal{L}^{(x)} \mathbf v = 0, |\mathbf v|
\mbox{ bounded } \forall x\right\}.				
\ee

The principle object in Floquet theory is the  discriminant,
$
\Delta(u,\lambda) = \mbox{Trace}\left(\Psi(x + L; \lambda)\Psi^{-1}(x;\lambda)\right)
$,
which determines the growth of the eigenfunctions $\mathbf v$ as $x$ is shifted across one period $L$.
The Floquet spectrum has the following characterization in terms of the
 discriminant:
\be
\sigma(u) := \left\{ \lambda \in \C \, | \, \Delta(u,\lambda)\in\R,
 -2 \leq \Delta(u,\lambda) \leq 2 \right\}.
\ee 
 The discriminant $\Delta(\lambda)$  is invariant under the NLS evolution and encodes the infinite family of NLS constants of motion.  Thus, $\sigma(u)$ is constant in time and can be calculated by solving the eigenvalue problem \rf{Lax1} at $u(x,t_0)$, for a given $t_0$.

The spectrum for  an NLS solution consists of the entire real axis and curves or ``bands of spectrum'' in the complex $\lambda$ plane
($\mathcal{L}^{(x)}$ is not self-adjoint).
The periodic/antiperiodic points (abbreviated here as  periodic points) of the Floquet spectrum are
those at which $\Delta = \pm 2$. 
The  simple points of the periodic spectrum 
  $\sigma^s(u) = \{\lambda_j^s\,|\, \Delta(\lambda_j^s) = \pm 2, \partial \Delta/\partial \lambda \neq 0\}$ occur in complex conjugatge pairs off the real axis and determine the endpoints of the bands of spectrum.
Located within the bands of spectrum  are two  important spectral elements:
\begin{enumerate}
\item Critical points of spectrum, $\lambda_j^c$, determined by the condition
$\frac{\partial \Delta}{\partial \lambda} |_{\lambda_j^c} = 0$. 
\item Double points of periodic spectrum
$\sigma^d(u) = \{\lambda_j^d \,|\,
\Delta(\lambda_j^d) = \pm 2, \partial \Delta/\partial \lambda = 0,\;
\partial^2 \Delta/\partial \lambda^2 \neq 0\}$.
\label{critical}
\end{enumerate}
Although double points are among the critical points of $\Delta$,
 in this paper we only call the degenerate periodic spectrum
where $\Delta = \pm 2$ ``double points''. The term
``critical points'' is reserved for degenerate elements of the spectrum where
$\Delta \neq \pm 2$.

For generic initial data there are an infinite number of simple periodic points
$\lambda_j^s$. Each pair of simple periodic points, $(\lambda_{2j}^s, \lambda_{2j+1}^s)$,
corresponds to a stable active degree of freedom. 
NLS dynamics are often
well approximated by  $N$-phase quasiperiodic solutions  of the form
$u(x,t) = u_N(\theta_1, ...,\theta_N)$  where $N$ is finite. The $N$-phase solutions $u_N$ can be explicitly constructed in terms of Riemann theta functions associated with the Riemann surface $\cal R$ of
$\sqrt{\Delta^2(u,\lambda) - 4}$, with branch points at the $2N$ simple periodic points $\lambda_j^s  \in \sigma^s(u)$. The phase evolution, determined by
${\cal L}^{(t)}$, is given by 
$\theta_j = \kappa_j x + \omega_j t + \theta_j^{(0)}$ where the wave numbers $\kappa_j$ and frequencies $\omega_j$ are determined  by the $\lambda_j \in  \sigma^s$.

Critical points and double points play an important role in determining the stability of an $N-$phase solution. Real double points,
$\lambda_j^d \in \mathbb R$,  correspond to stable inactive modes.
Exponential instabilities are generally associated with complex double points  $\lambda_j^d \in \mathbb \C$  \cite{efm90}.
Recently,  
complex critical points arising from transverse intersections of bands of
 spectrum were shown to be associated with weak instabilities whose exact nature is under further investigation \cite{si21}.

\begin{figure}[ht!]
  \centerline{
    \includegraphics[width=.33\textwidth]{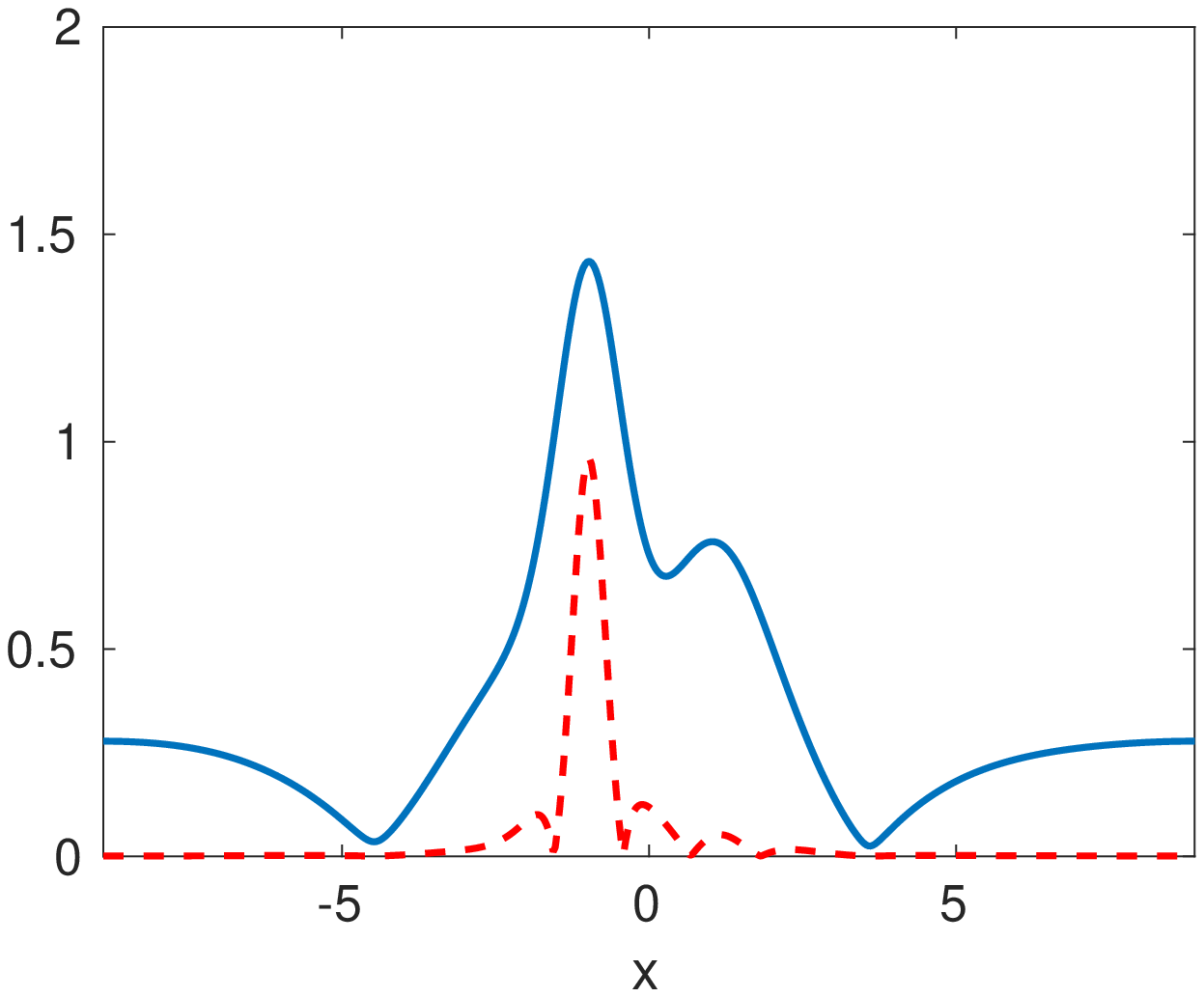}
    \includegraphics[width=.33\textwidth]{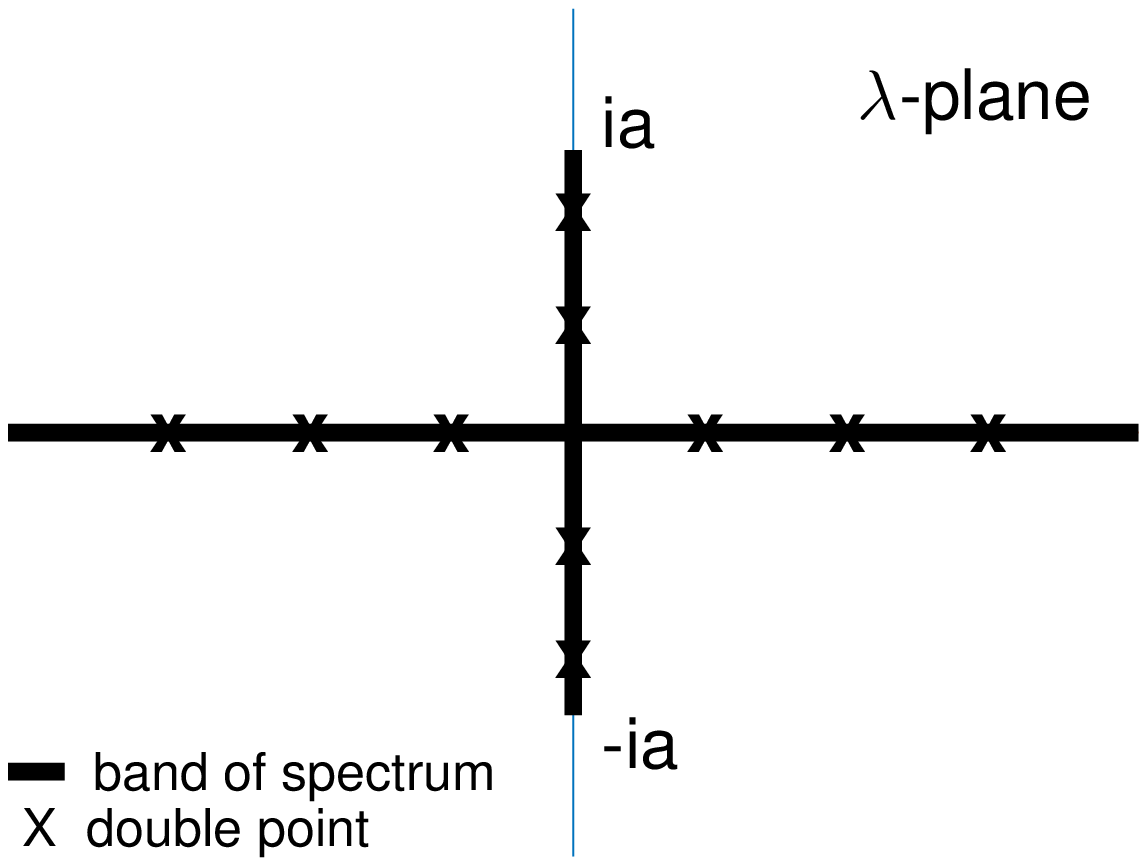}
    \includegraphics[width=.33\textwidth]{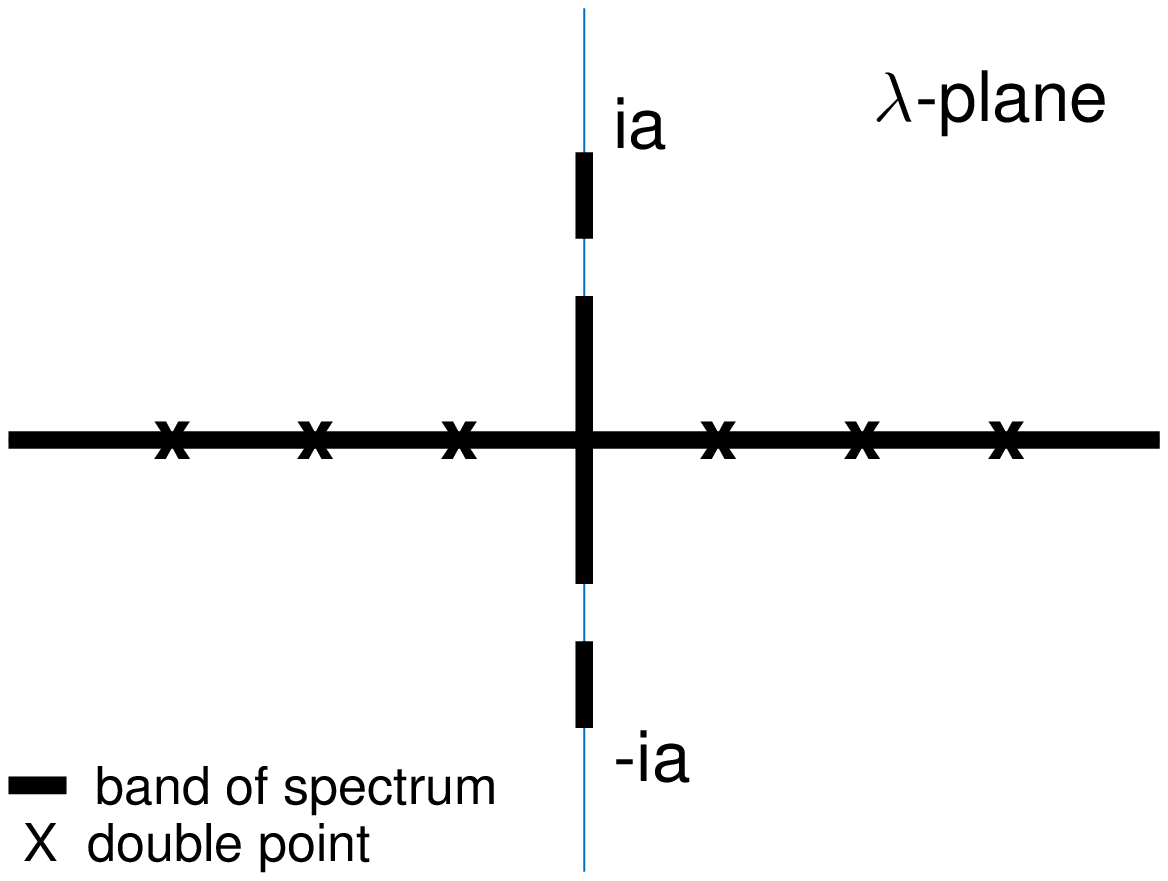}
  }
  \centerline{\bf A\hspace{2.25in} B\hspace{2.25in} C}
  \caption{(A) Sample time slice showing the amplitude of $u$ (solid line) versus magnitude of $\beta u\left[{\cal H}\left(|u|^2\right)\right]_x$ (dash line).  The Floquet spectrum of: (B)  the Stokes wave with $\displaystyle \lfloor {aL}/{\pi} \rfloor=2$ and  (C) the even three phase solution \rf{3phase}.
Due to the Schwarz symmetry of the Z-S problem, we subsequently 
only display the spectrum  in the upper half $\lambda$-plane.
}
\label{Figure1}
\end{figure}

The simplest solution  illustrating the correspondence between complex double
points in the spectrum and linear instabilities is the Stokes wave 
$u_a(t)=a \e^{\ri (2 a^2 t+\phi)}$.
The Floquet discriminant for the Stokes wave is  $\Delta = 2 \cos(\sqrt{a^2 + \lambda^2} L)$.
The Floquet spectrum (shown in Figure~\ref{Figure1}B)
consists of continuous bands $\R \bigcup$ $ [-\ri a,\ri a]$
and the discrete spectrum containing $\lambda_0^s =  \pm \ri |a|$ 
and infinitely many  double points 
\be
(\lambda_j^d)^2 = \left(\frac{j\pi}{L}\right)^2 - a^2,\quad j\in\mathbb{Z}, \quad j \ne 0.
\label{dps}
\ee
Complex double points $\lambda^d_j$ are obtained if  
\be
0 < (j\pi/L)^2 < |a|^2.
\label{imdps}
\ee
The number of complex double points is the largest integer M such that
$0 < M < |a|L/\pi$. The remaining
$ \lambda_j^d$ for $|j| >M$ are real double points. 
The condition for  $\lambda_j^d$ to be  complex is
precisely the condition  for small perturbations of the Stokes wave 
$u_{\eps}(x,t) = u_a(t)(1 + \eps \e^{\ri\mu_j x + \sigma_j t})$, $\mu_j = 2\pi j/L$,
to be unstable. 
According to linear stability analysis all modes with  $\mu_j$ satisfying
(\ref{imdps})  will initially grow exponentially
with 
$\sigma_j^2 = \mu_j^2\left(4|a|^2 - \mu_j^2 \right)$ before
saturating due to the nonlinear terms.

\subsection{Spatially periodic breather solutions of the NLS equation}
\begin{figure}[ht!]
  \centerline{
\includegraphics[width=.33\textwidth]{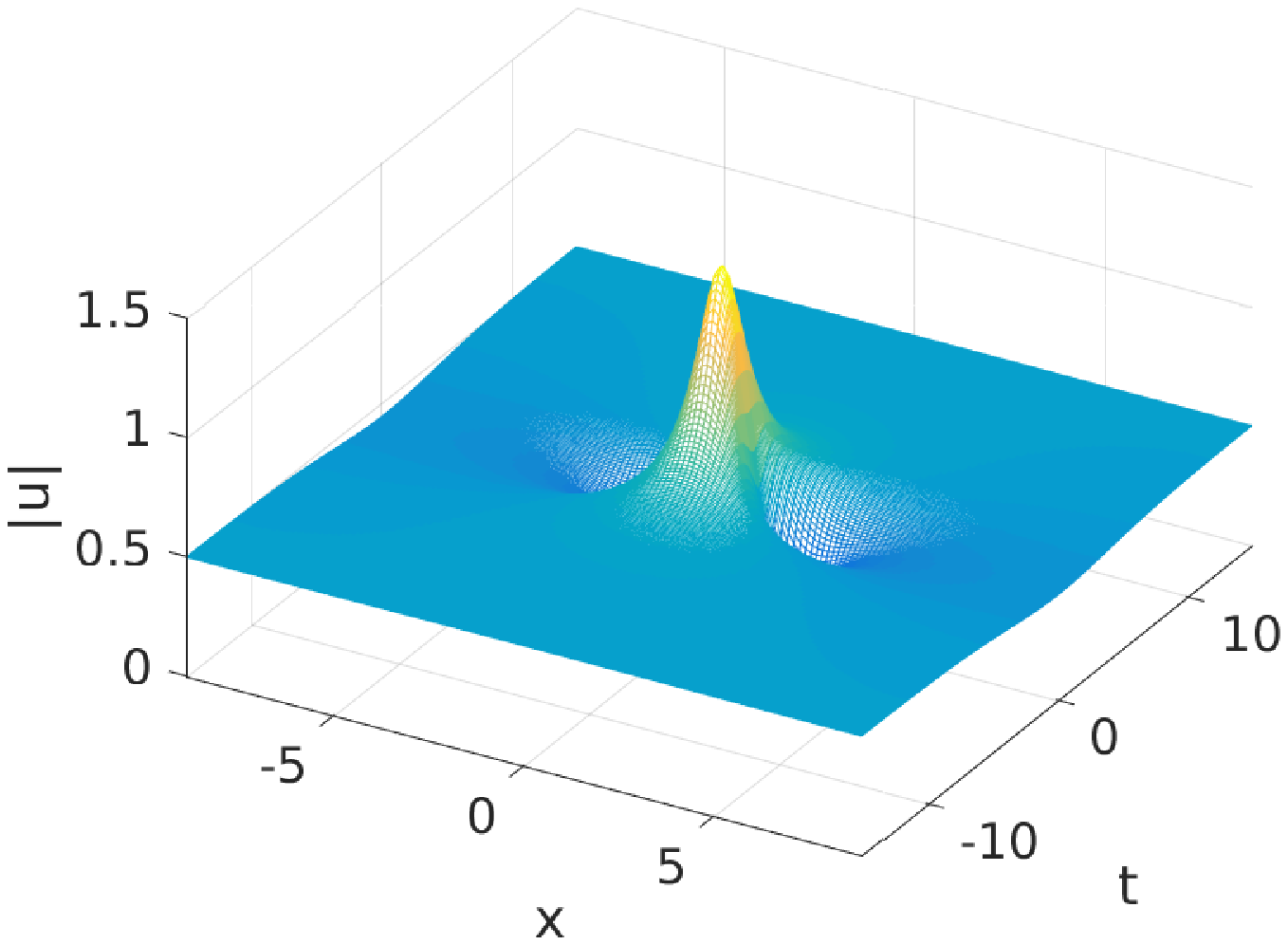}
\includegraphics[width=.33\textwidth]{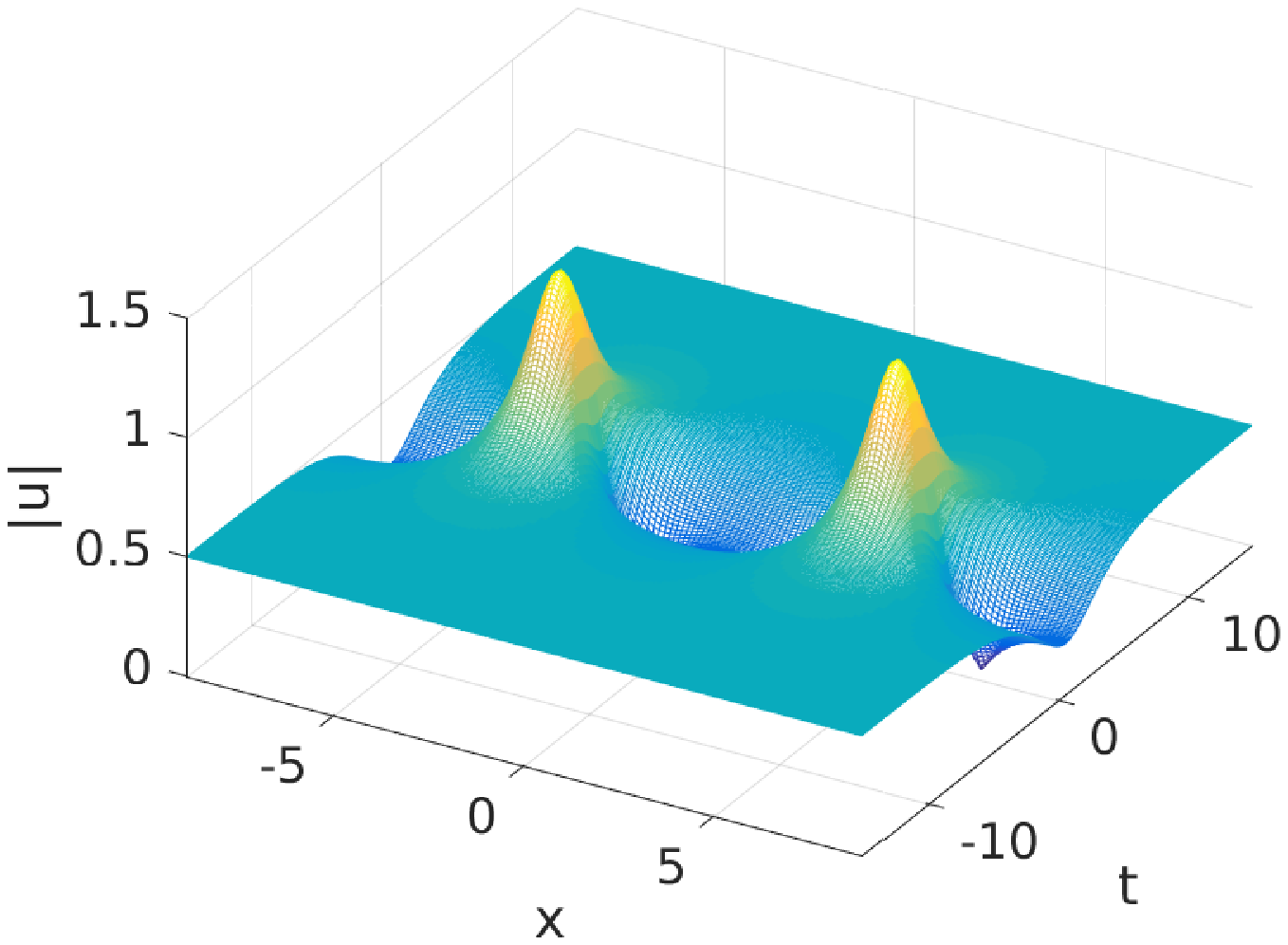}
\includegraphics[width=.33\textwidth]{schober_physD_2C}
  }
  \centerline{\bf A\hspace{2.25in} B\hspace{2.25in} C}
  \caption{SPBs over a Stokes wave with $N = 2$ UMs, $a = .5$,
    $L = 4\sqrt{2} \pi$: Amplitudes of (A-B) single mode SPBs $|U^{(1)}(x,t,\rho)|$ and $|U^{(2)}(x,t,\rho)|$ ($\rho = \beta = 0$) and (C) two-mode SPB
  $|U^{(1,2)}(\rho,\tau)|$ ($\rho = \tau = 0$).}
\label{Figure2}
\end{figure}

Explicit expressions for the spatially periodic breathers (SPBs) can be derived
using the 
B\"acklund-gauge transformation (BT)  for the NLS equation \cite{sz87}.
For an unstable Stokes wave with $N$ UMs, a single BT at the
complex double point
 $\lambda^d_j$, $1 \le j \le N$,  generates the single mode SPB, $U^{(j)}(x,t)$
corresponding to the j-th UM  \cite{aek87}:
\be\label{SPB1}
U^{(j)}(x,t) = a\, e^{2\ri a^2t}\,\left(\frac{1 + 2\, \e^{\sigma_j  t + 2\ri\phi_j+ \gamma}\cos\mu_j x+ A\,  \e^{2(\sigma_j  t + 2\ri\phi_j+ \gamma)}}
{1 + 2\, \e^{\sigma_j  t + \gamma}\cos\mu_j x+ A\,  \e^{2(\sigma_j  t + \gamma)}}\right)
\ee
where $\mu_j = 2\pi j/L$, $\sigma_j = 2i\mu_j\lambda_j$,  $\sin\phi_j= \mu_j/2a$, $A = \sec^2\phi_j$, and $\gamma$ is an arbitrary phase.
$U^{(j)}(x,t)$  is localized in time: as $t \rightarrow \pm\infty$ the SPB
exponentially approaches a phase shift of the Stokes wave at
the rate $\sigma_j$.
Figures~\ref{Figure2}A-B show the amplitudes of the two distinct
single mode SPBs, $U^{(1)}(x,t)$ and $U^{(2)}(x,t)$.

When the Stokes wave has $N \ge 2$ UMs, the BT can be iterated to obtain
multi-mode SPBs.
The family of two-mode SPB with wavenumbers $\mu_i$ and $\mu_j$ is obtained by applying
the BT successively at complex $\lambda_i^d$ and $\lambda_j^d$ and
is given by :
\be
  U^{(i,j)}(x,t;\rho,\tau) =
a\e^{2\ri a^2t} \frac{N(x,t;\rho,\tau)}{D(x,t;\rho,\tau)}.
\label{SPB2}
\ee
The exact formula is provided in  \cite{cs13}. The parameters $\rho$ and $\tau$ determine the time
at which the first and second modes become excited, respectively.
Figure~\ref{Figure2}C shows the amplitude of the ``coalesced'' two mode  SPB
$U^{(1,2)}(x,t;0,0)$  where 
the  two modes are excited simultaneously.

{\bf Note:} The  BT is isospectral:
$\sigma(u_a(t)) = \sigma(U^{(j)}(x,t)) = \sigma(U^{(i,j)}(x,t))$.
The Stokes wave with $N = 2$ UMs and each of the SPBs shown in Figures~\ref{Figure2}A-C share the same Floquet spectrum  (given in Figure~\ref{Figure1}B).

\subsection{Perturbation analysis: Spectrum for damped SPB data}

We are interested in the fate of the complex double points under the NLD-HONLS perturbation as they characterize the SPB initially. 
The linearized initial conditions for the one and two mode SPBs 
are obtained by choosing $t$ and $\gamma$ in formulas \rf{SPB1} and \rf{SPB2}
such that $\eps_s = 4\ri\sin\phi_s\,\e^{\sigma_s t + \gamma}$, $s=i,j$, are small.
Neglecting second order terms we obtain the linearized initial conditions
\bea
U^{(j)}(x,0) &=& a\,\left(1 + \eps_j\, \e^{\ri\phi_j} \cos\mu_j x\right),\nn\\
U^{(i,j)}(x,0) &=& a\,\left(1 + \eps_i\, \e^{\ri\phi_i} \cos\mu_i x 
+ \eps_j\, \e^{\ri\phi_j} \cos\mu_j x\right),\nn
\eea
where $\phi_s =\pi/4$, $s = i,j$.
For small time $h$, we obtain the following first order approximation
\bea
U_{\eps,\beta}^{(j)}(x,h) &=& a\left[1+
\tilde\eps_j\left(\e^{\ri\tilde\phi_j} \cos\mu_j x + r_j \,\e^{\ri\tilde\theta_j}\sin\mu_j x\right)
\right] = a \left[1+\tilde\eps_j f_j\right],\qquad
\tilde\phi_s \neq \tilde\theta_s\nn\\
U_{\eps,\beta}^{(i,j)}(x,h) &=&  a\left[1+\tilde\eps_i f_i + \tilde\eps_j f_j \right]\nn
\eea
where $\tilde\phi_s, \tilde\theta_s,\tilde\eps_s, r_s$ are functions of $h$ and the NLD-HONLS parameters $\eps$ and $\beta$, for $s = i,j$.
Letting $\eps = \tilde\eps_s$ and suppressing their explicit dependence on  $\eps,\beta,h$ the data is of the general form
\be
u = a +  \eps\left[f_i +  Q\,f_j \right] = a + \eps u^{(1)}
\label{hello}
\ee
where $r_s \neq 0$ and $Q$ can be 0 or 1, depending on whether a one or two    mode SPB is under consideration.

In \cite{si21}, via perturbation anlysis we examined the splitting of complex double points for data of the generic form \rf{hello}.
Briefly, at the double points $\lambda_n$ we assume the perturbation expansions
\[\lambda_n = \lambda_n^{(0)} + \epsilon\lambda_n^{(1)} + \epsilon^2\lambda_n^{(2)} + \cdots,\qquad
\mathbf v_n = \mathbf v_n^{(0)} + \epsilon \mathbf v_n^{(1)} + \epsilon^2 \mathbf v_n^{(2)} + \cdots\]
Substituting these expansions into Equation \rf{Lax1} the following 
${\cal O}(\epsilon)$ correction to $\lambda_n$ is obtained :%
\be
\left(\lambda_n^{(1)}\right)^2 = \left\{\ba{ll} \frac{a^2}{4\lambda_n^2}
\left[\sin(\omega_n+\tilde\phi_n)\sin(\omega_n-\tilde\phi_n)\right. &\\
  \qquad+r_n^2\sin(\omega_n+\tilde\theta_n)\sin(\omega_n-\tilde\theta_n)& \quad n = i,j \\
\left.  \qquad+\ri r_n\sin(\tilde\theta_n-\tilde\phi_n)\sin 2\omega_n\right]\\
0 & \quad n \neq i,j\ea\right.
\label{ordereps}
\ee
where $\tan\omega_n = Im\left(\lambda_n^{(0)}\right)/k_n$ and $\tilde\phi_j \neq \tilde\theta_j \pm n\pi$.
At $\mathcal{ O}(\eps)$ there is a correction only to the double points $\lambda_n$, $n = i,j$.
An examination of  $\Delta$ in a neighborhood of $u^{(0)}$ finds the band of continuous spectrum along the imaginary axis splits asymmetrically at $\lambda_n^{(0)}$ into two disjoint bands in the upper half plane.
 As an example, see Figure \ref{Figure3}C.
 The other double points do not experience an $\mathcal{ O}(\eps)$ correction
\cite{si21}.

The ${\cal O}(\epsilon^2)$ correction is of the form
\be
\left(\lambda_n^{(2)} - \beta_n\right)^2 =
\left\{\ba{ll} \alpha_n^+\alpha_n^- & \quad n=2i, 2j, i+j, j-i\\
0 & \quad \mbox{for all other cases}\ea\right.
\label{ordertwo}
\ee

Only the double points $\lambda_n^{(0)}$ with $n=2i, 2j, i+j$, or $j-i$ split at  $\mathcal{ O}(\eps^2)$. All other double points experience  an $\mathcal{ O}(\eps^2)$  translation.
This calculation can be carried to higher order $\mathcal{ O}(\eps^m)$. In the simple example of a damped  single mode SPB, $U_{\epsilon,\gamma}^{(j)}(x,t)$, only
the double points $\lambda_n^{(0)}$ which correspond to a resonant mode $n=mj$ will split at order  $\mathcal{O}(\eps^m)$
whereas the splitting is zero for $\lambda_n^{(0)}$, $n \neq mj$ \cite{si21}.

\subsection{Soliton-like structures}

Under the NLD-HONLS flow, the complex double points present initially  in the
spectrum of an SPB may split, creating additional complex bands of spectrum. As time evolves one or two of the bands may shrink significantly and their lengths become tiny (see e.g. Figure~\ref{Figure4}F,  Figure~\ref{Figure7}C). 
For a qualitative understanding of the waveforms,  we view the  emergence of
the tiny bands of spectrum as corresponding to nonlinear modes which have
developed a one or two ``soliton-like structure''.

Insight into the limiting behavior  of solutions as gaps  in the
spectrum open and close is obtained by considering the family of 3-phase solutions of the  NLS equation 
\cite{aek87}:
 \be 
  u_0(x,t) = \frac{\kappa}{\sqrt{2}}\e^{\ri t}\frac{\cn\left(\sqrt{\frac{1+\kappa}{2}}x, k\right)\,
    \cn(t,\kappa)+
    \ri\sqrt{1+\kappa}\,\dn\left(\sqrt{\frac{1+\kappa}{2}}x,k\right)\,\sn(t,\kappa)}
  {\sqrt{1+\kappa}\,\dn\left(\sqrt{\frac{1+\kappa}{2}}x,k\right) -     \cn\left(\sqrt{\frac{1+\kappa}{2}}x,k\right)\,
    \dn(t,\kappa)},
  \label{3phase}
  \ee
where $0 < \kappa <1$,
and $k = \sqrt{\frac{1-\kappa}{1+\kappa}}$.
  The solutions are
periodic in space  and quasi-periodic in time; the spatial period, $L_x $, and the temporal period of the modulated phase, $L_t $,
    are functions of the complete elliptic integrals of
  the first kind,   ${{\cal K}_x(k)}$ and ${{\cal K}_t(\kappa)}$ respectively.

  The  Floquet spectrum of $u_0(t)$ has two gaps in the spectrum along the imaginary axis as  shown in Figure~\ref{Figure1}C.
As $\kappa \rightarrow 1$ in \rf{3phase}, the gaps close  to complex double points, and $u_0(x,t)$ limits to a scaling of the SPB given in Equation~\rf{SPB1}. On the other
 hand as  $\kappa \rightarrow 0$, $u_0(x,t)$ limits to a scaling of the  NLS soliton
 solution (given in Equation~\rf{one-soliton}) and the bands shrink to  discrete  points. 
%We acknowledge this is a qualitative comparison due to the fact the NLD-HONLS %solution is over a non uniform finite background,  and when appropriate,
%the numerical solution is referred to as having one or  two soliton-like modes.

\subsubsection{Soliton solutions}
Complete information about the one and two soliton solutions, obtained using
the inverse scattering theory, is contained 
in the scattering data \cite{ac91}.
The discrete spectra in the upper half plane, $\lambda_n$, are 
the zeros of the first Jost coefficient.  Given $\lambda_n$, the soliton solutions are as follows:

 {\sl One-soliton:}
Consider the discrete eigenvalue $\lambda = \xi + \ri \eta$. Then the associated one-soliton solution is given by
\be
u(x,t) = 2\ri\eta\,\sech\left\{2\eta\left[x-x_0+4\xi t \right]\right\} \e^{\ri\left[2\xi x+4\left(\xi^2 - \eta^2\right)t + \phi\right]}
\label{one-soliton}
\ee
where $\xi, \eta, x_0, \phi$ are constants physically representing the velocity ($4\xi$), the width and height ($\eta$), the initial center ($x_0$), and the phase ($\phi$) of the envelope, respectively.

{\sl Two-soliton:}
Consider the discrete eigenvalues $\lambda_n = \xi_n + \ri \eta_n$, $n=1,2$. 
Define the quantities
\[\ba{ll}
\phi_n &= \pi/2 - \arctan\left[\frac{(\xi_n - \xi_{\tilde n})^2 + \eta_n^2 - \eta_{\tilde n}^2}{2\eta_{\tilde n} (\xi_n - \xi_{\tilde n})}\right],\quad \tilde n=3-n,\\
\theta_n(t) &= \theta_n^0 - 4\xi_n\eta_n t,\qquad \sigma_n(t) = \sigma_n^0 + 2\left(\xi_n^2 - \eta_n^2\right) t,\\
\Delta(x,t) &= \cosh\left[2(\eta_1 + \eta_2) x - \theta_1(t) - \theta_2(t)\right] \\
&+\frac{4\eta_1\eta_2}{(\xi_1 - \xi_2)^2 + (\eta_1 - \eta_2)^2}\cos\left[2(\xi_1 -\xi_2) x + \sigma_1(t) - \sigma_2(t)\right]\\
&+\frac{(\xi_1 - \xi_2)^2 + (\eta_1 + \eta_2)^2}{(\xi_1 - \xi_2)^2 + (\eta_1 - \eta_2)^2}\cosh\left[2(\eta_1-\eta_2) x - \left(\theta_1(t) - \theta_2(t)\right)\right]
\ea
\]
where $\theta_n(t)$ and $\sigma_n(t)$ give the position and phase of the solitons. Then the associated two-soliton solution, with velocities given in terms of $\xi_n$
, is  given by \cite{pd2007}
\be
\ba{ll} u(x,t) =& -\frac{2}{\Delta}\left(\frac{(\xi_1 - \xi_2)^2 + (\eta_1 + \eta_2)^2}{(\xi_1 - \xi_2)^2 + (\eta_1 - \eta_2)^2}\right)^{1/2}\\
&\times\left[2\eta_1\cosh\left(2\eta_2 x-\theta_2(t)-\mbox{i}\phi_1\right)\mbox{e}^{-2\mbox{i}\xi_1 x - \mbox{i}\sigma_1(t)} +
  2\eta_2\cosh\left(2\eta_1 x - \theta_1(t) -\mbox{i}\phi_2\right)\mbox{e}^{-2\mbox{i}\xi_2 x - \mbox{i}\sigma_2(t)}\right].\ea
\label{two-soliton}
\ee

\section{Numerical Investigations}
In this section we investigate  the effects of nonlinear dissipation and higher order nonlinearities on the routes to stability of the single and multi-mode
SPBs. We
appeal to the Floquet spectral theory of the NLS equation to interpret and
provide a characterization of the perturbed dynamics
in terms of nearby solutions of the NLS equation.
 
The NLD-HONLS equation  is solved numerically 
using a very accurate exponential integrator (ETD4RK).
The ETD4RK integrator combines a Fourier mode decomposition in space with a fourth order exponential Runge-Kutta method in time which uses
Pade approximations of the matrix exponential terms \cite{khaliq2009,liang2015} (see the Appendix for the ETD4RK scheme and its convergence properties).
The number of Fourier modes and the time step used depends on the complexity of the solution. For example, for initial data in the two UM regime,
$L = 4\sqrt{2} \pi$, $N = 256$  Fourier modes are used with time step
$\Delta t = 10^{-3}$.  As a benchmark, with this resolution the
invariants of the HONLS equation, (the mass $ E(t) = \int_0^L|u(x,t)|^2\,dx $, momentum $ P(t) = i\int_0^L\left(u(x,t)^* u_x(x,t) - u(x,t)u_x(x,t)^*\right)dx$, and 
Hamiltonian
\be\label{hamiltonian}
H(t) =  \int_0^L \left\{ | u_x |^2 - | u |^4 - \ri\epsilon\left[
1/4\left( u_x u_{xx}^* - u_x^* u_{xx} \right) 
+ 2 | u |^2\left(u^* u_x - u u_x^*\right) - |u|^2
\left[\mathcal{ H} \left( | u |^2 \right) \right]_x\right]\right\}\,dx
\ee
are preserved with an $\mathcal{O} (10^{-11})$ accuracy for $0 \le t \le 100$.
Furthermore, in our earlier studies of the NLS equation, the ETD4RK scheme was shown to accurately preserve the Floquet spectrum  \cite{HIRS2013}. This is an important feature in our studies as the Floquet spectrum is a significant tool in the  analysis of rogue waves. 

{\sl Setup of the experiments:}
The initial data used in the numerical experiments is generated using  exact SPB solutions of the integrable NLS equation. 
The perturbed SPBs are indicated with subscripts: $U^{(j)}_{\epsilon,\beta}(x,t)$  is the  solution of
the NLD-HONLS  Equation~\rf{dhonls}  for one-mode SPB initial data  $ U^{(j)}(x,T_0) $.
Likewise $U^{(i,j)}_{\epsilon,\beta}(x,t)$
is the solution to (\ref{dhonls}) for two-mode SPB initial data.
The ``$N$ UM regime'' refers to the range in the amplitude $a$
and period $L$ for which the underlying Stokes wave initially has $N$ unstable modes. 

In the current study
we
systematically vary  $T_0 \in [-5,0]$ to allow the NLD-HONLS solutions to be initialized with SPB data at different stages of their development, i.e. from
the early  stage of  the MI when there is strong growth ($T_0 = -5$)
through the nonlinear stage  
when the MI is saturating ($T_0 = 0$).
 For each damped SPB under consideration the complete  set of diagnostics
 is presented for one value of $T_0$.
 Qualitative
 differences in the  the solution as $T_0$ is varied are either 
discussed or summarized graphically.
The perturbation parameters are set at  $\epsilon = 0.05$ and $\beta = 0.2$.
For $U^{(1,2)}_{\epsilon,\beta}(x,t)$, which displays more complex behavior,
$\beta$ is also varied $0.01 \le \beta \le 0.2$.

%{\bf Numerical diagnostics}

{\sl Nonlinear mode decomposition of the NLD-HONLS flow:}
The Floquet spectral decomposition of the NLD-HONLS data is computed
at each time $t$, $0 \le t \le 100$, 
using the numerical procedure developed by
Overman et. al. \cite{omb86}.  After solving 
system \rf{Lax1}, the  discriminant $\Delta$ is constructed.
The zeros of $\Delta \pm 2$ are determined using a root solver 
based on M\"uller's method and then the curves of spectrum are filled in.
The spectrum is calculated with an accuracy of ${\mathcal O}(10^{-6})$.

{\sl Notation used in the spectral plots:} The periodic/antiperiodic spectrum is indicated with
a large ``$\times$'' when $\Delta = -2$ and with a large box when $\Delta = 2$.
The continuous spectrum is indicated with small ``$\times$'' when  $\Delta$ is
negative and with small boxes when $\Delta$ is positive.
Since the  NLS spectrum is symmetric
under complex conjugation, the  
 spectrum  is displayed only in the upper half $\lambda$-plane.

{\sl ``Soliton-like'' Criterion:}
The lengths of the bands of spectrum in the upper half plane associated with the dominant modes are monitored. When  these bands are  tiny we view the  corresponding modes of the $N$-phase solution as having a soliton-like structure. 
 Let  the length
 of a band with endpoints $\lambda_m$ and $\lambda_n$ be given by 
 $\gamma(t;\lambda_m,\lambda_n)$.
 If one or two of the band lengths satisfy
 \be
\gamma(t;\lambda_m,\lambda_n) = |\lambda_m(t) - \lambda_n(t)| < 0.025,
 \label{s_state}
 \ee
 then the 
 spectrum is said to be in a one or two soliton-like configuration.

{\sl Rogue Wave Criterion:}
To determine the occurence of rogue waves we monitor
the {\em strength} function
\[
S(t): = \frac{\max\limits_x |u(x,t)|}{H_s(t)},
\]
where  $H_s(t)$ is the {\em significant wave amplitude}, defined as four times the standard deviation of the surface amplitude. 
A rogue wave is said to occur at $t^*$ if $S(t^*) > 2.2$. In the strength plots the horizontal line indicates the reference strength for a rogue wave.

{\sl Saturation time of the instabilities:}  The spectral analysis is complemented by an examination of the saturation time of the instabilities for the nonlinear damped SPBs.
%$U^{(j)}_{\epsilon,\beta}(x,t)$ and $U^{(j,k)}_{\epsilon,\beta}(x,t)$
This is accomplished by  examining the growth
of small perturbations in
the initial data of the following form,
\be\label{fk}
U^{(j)}_{\epsilon,\beta,\delta}(x,0) = U^{(j)}(x,0) + \delta f_k(x),\qquad\mbox{and}\qquad
U^{(i,j)}_{\epsilon,\beta,\delta}(x,0) = U^{(i,j)}(x,0) + \delta f_k(x)\ee
where
$f_k(x) = \cos \mu_k x + r_k e^{i\phi_k} \sin \mu_k x$, $\mu_k = 2\pi k/L$, $1 \le k \le 2$.

To determine whether $ U^{(j)}_{\epsilon,\beta}(x,t)$ and 
 $U^{(j)}_{\epsilon,\beta,\delta}(x,t)$ remain close as time evolves we monitor 
the evolution of $d(t)$,
    \be
      d(t) = \|u_{\delta}(x,t) - u(x,t)\|_{H^1}
      \label{H1Diff}
 \ee
where
$\|f\|^2_{H^1} = \int_{-L/2}^{L/2} \left(|f_x|^2 + |f|^2\right)\,dx.$
We consider the solution to have stabilized under the NLD-HONLS flow once
$d(t)$ saturates.

From the Floquet perspective, we find the NLD-HONLS 
solutions are stabilized by nonlinear damping once  
 complex double points and complex critical points are eliminated in the
spectral decomposition of the data.

\subsection{The NLD-HONLS SPB in the one UM regime:
$a = 0.5, L = 2\sqrt{2} \pi$}
\begin{figure}[ht!]
  \centerline{
\includegraphics[width=.33\textwidth]{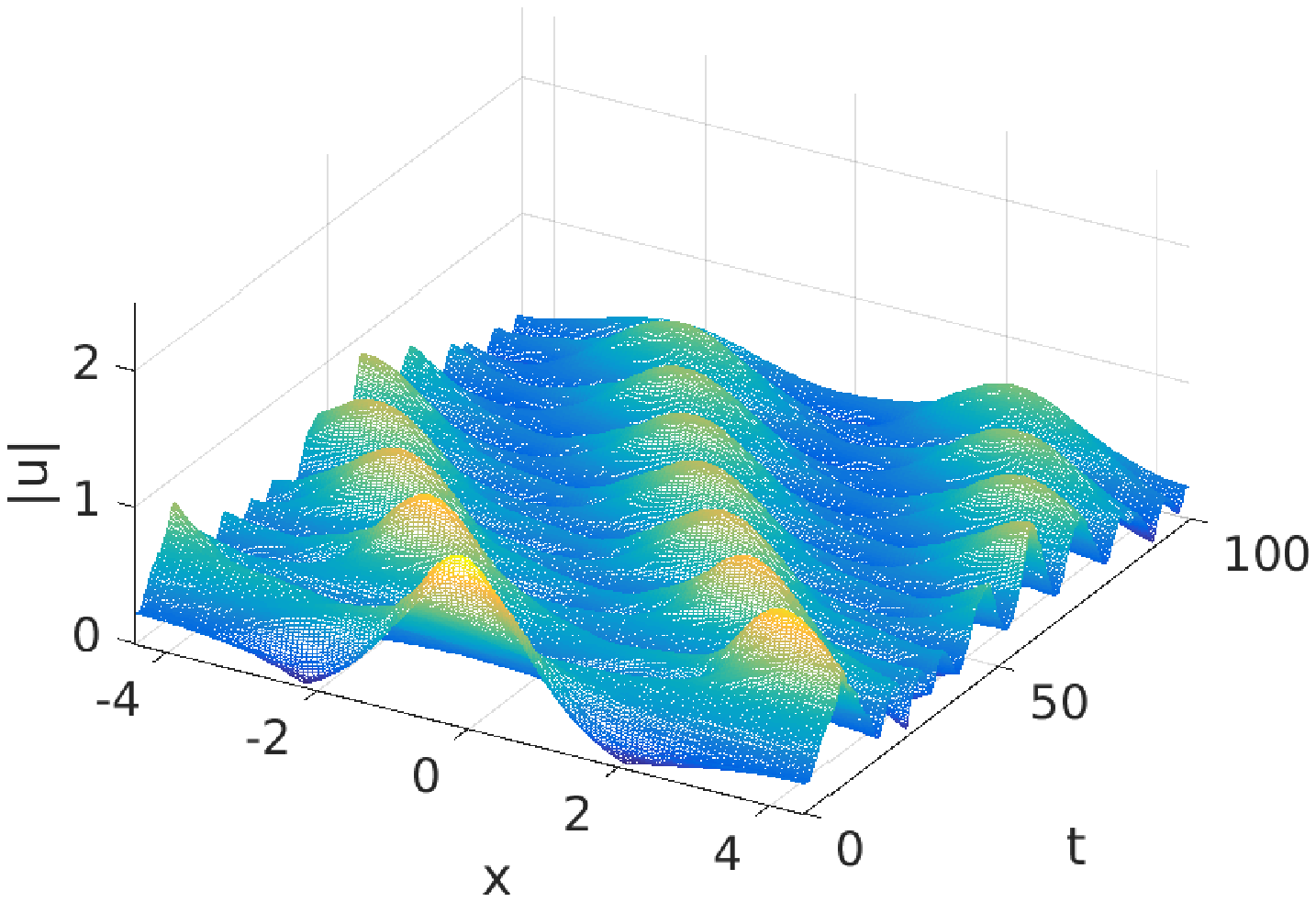}
\includegraphics[width=.33\textwidth]{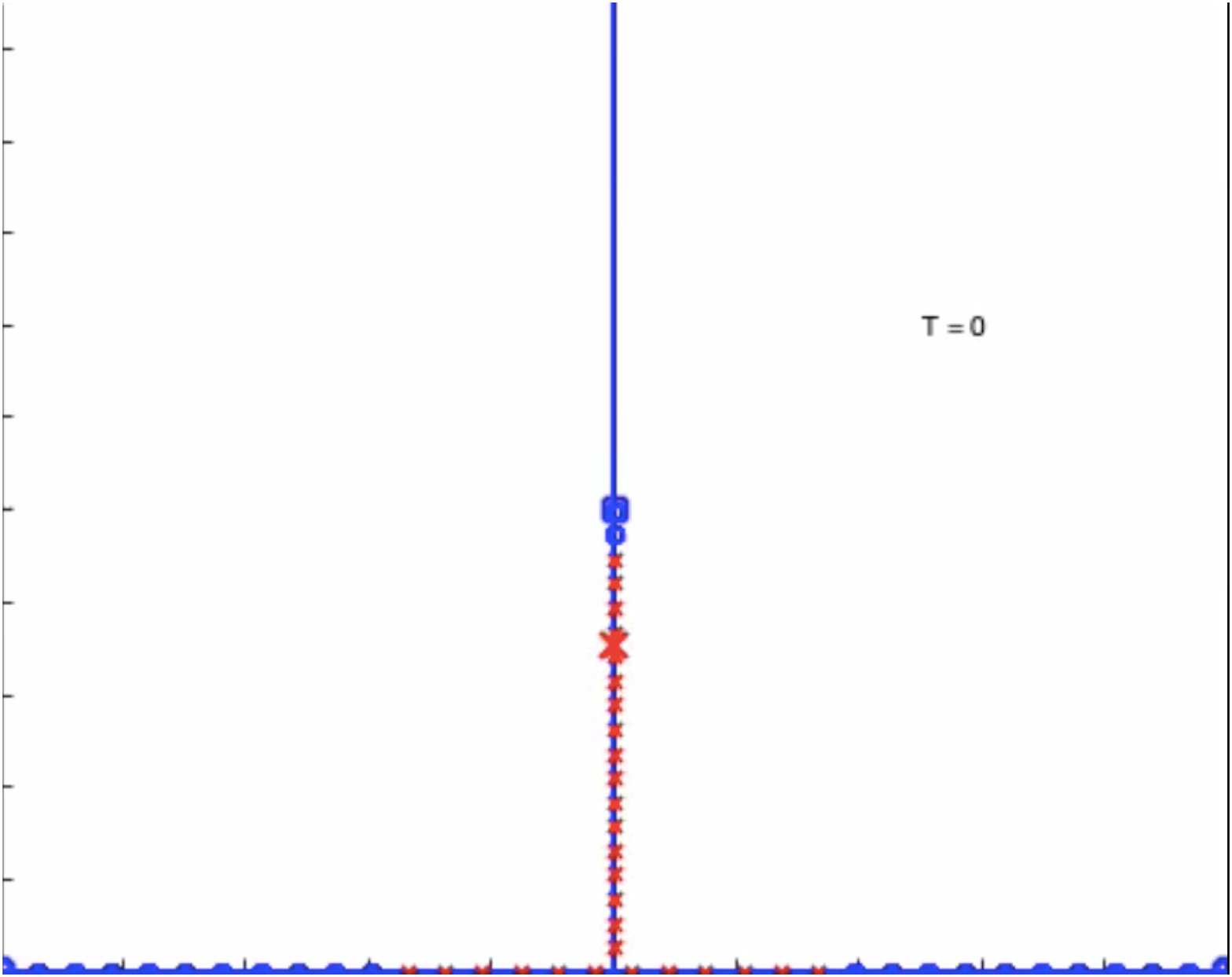}
\includegraphics[width=.33\textwidth]{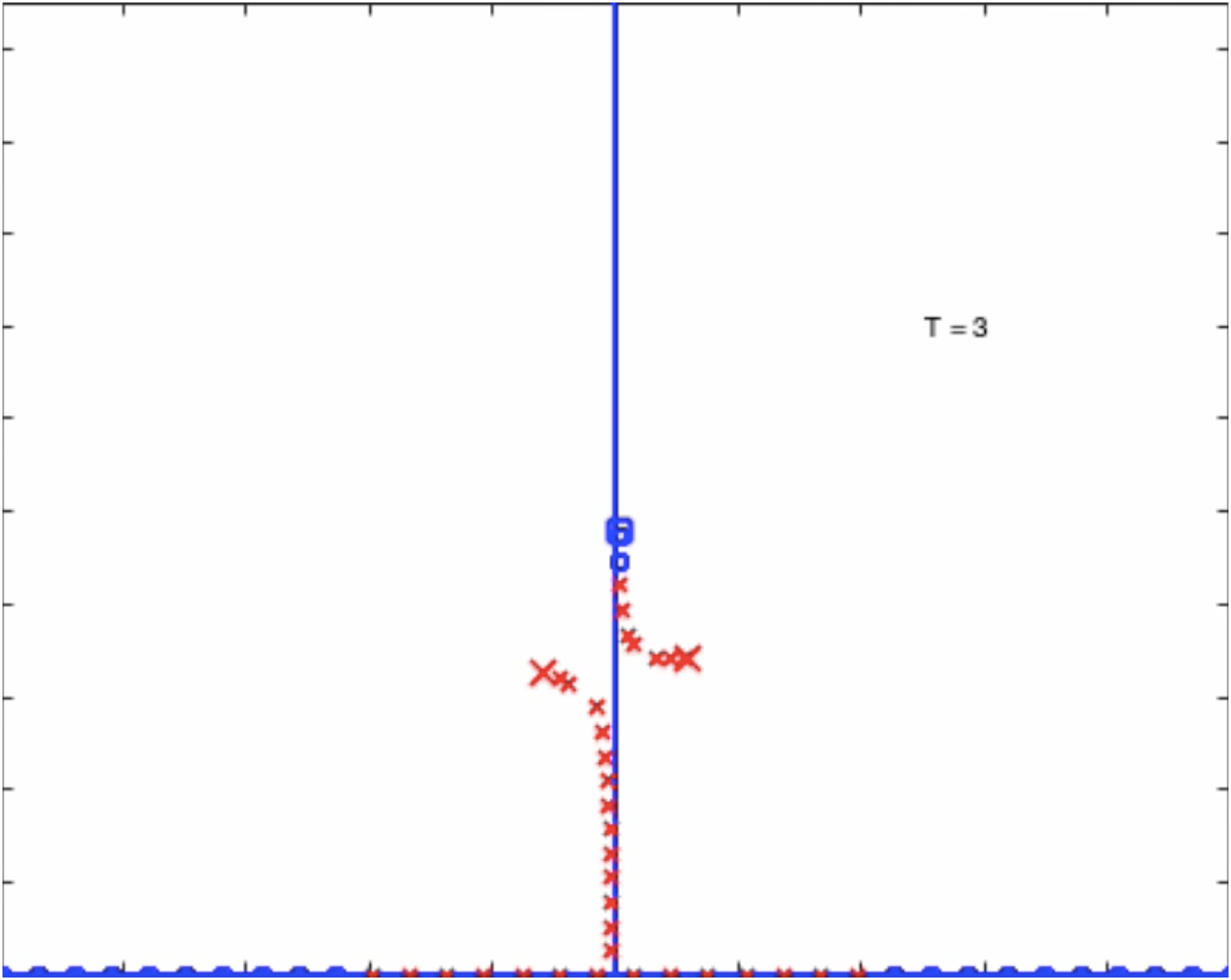}
}
  \centerline{\bf A\hspace{2.25in} B\hspace{2.25in} C}
  \vspace{12pt}
  \centerline{
\includegraphics[width=.33\textwidth]{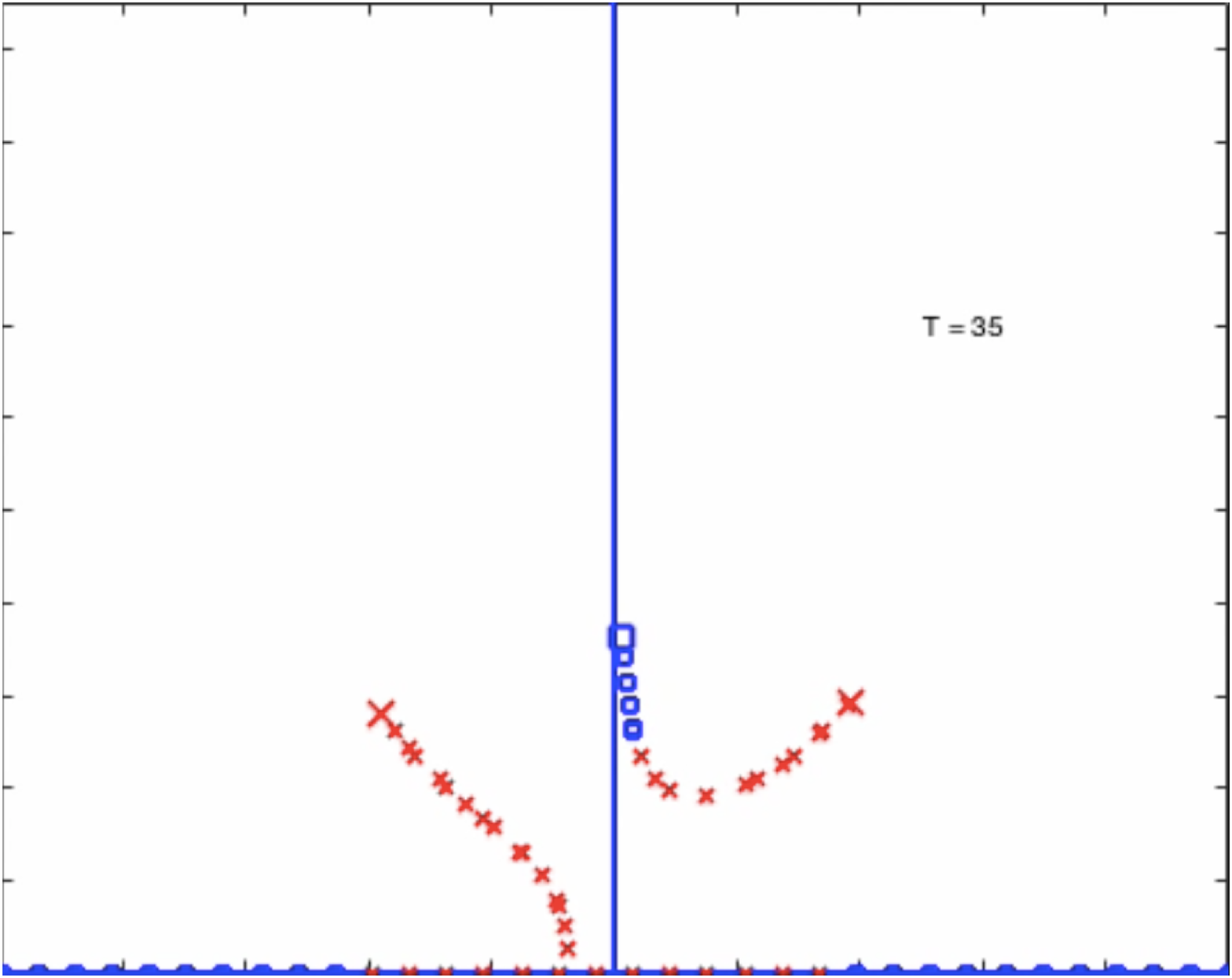}
\includegraphics[width=.33\textwidth]{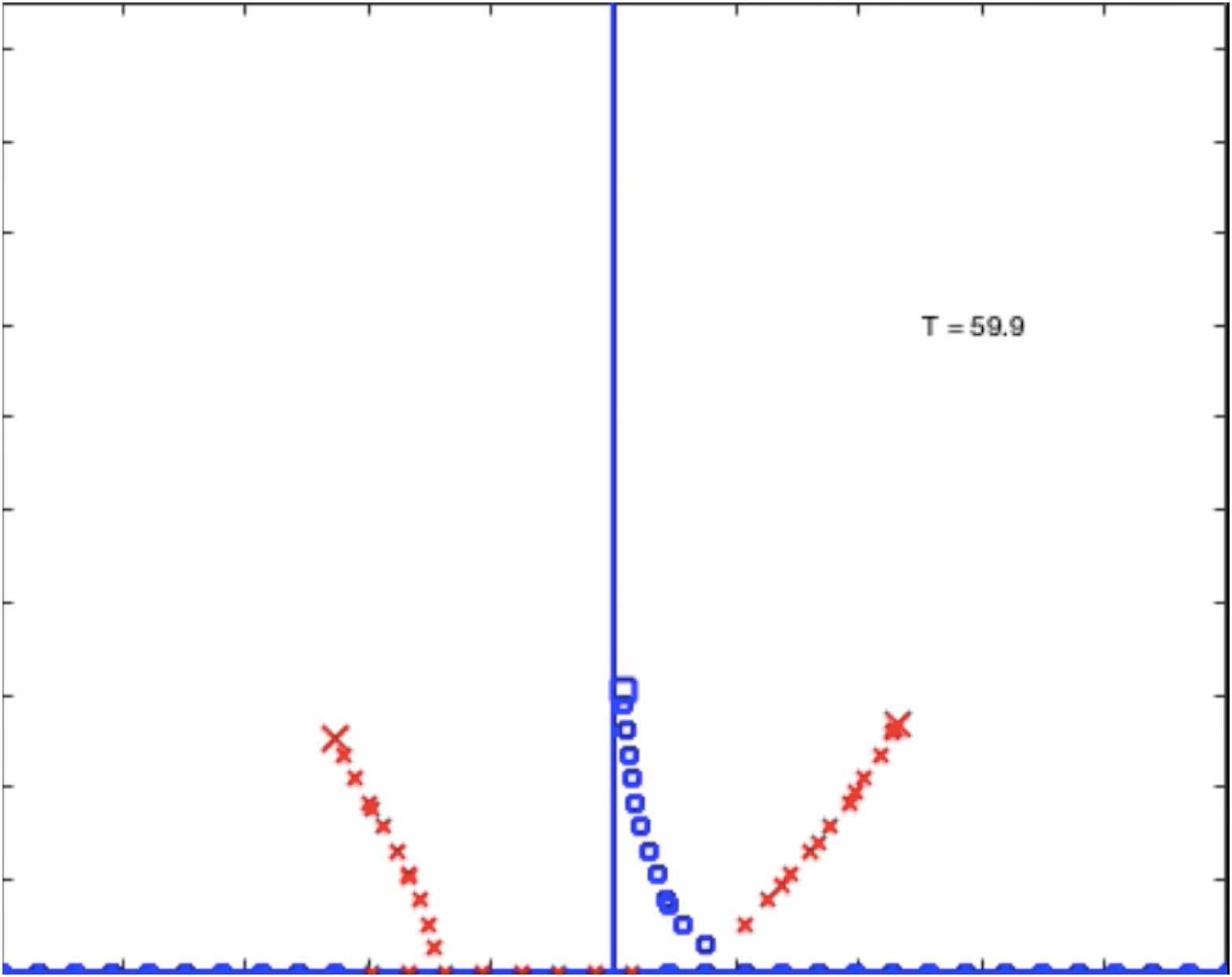}
\includegraphics[width=.33\textwidth]{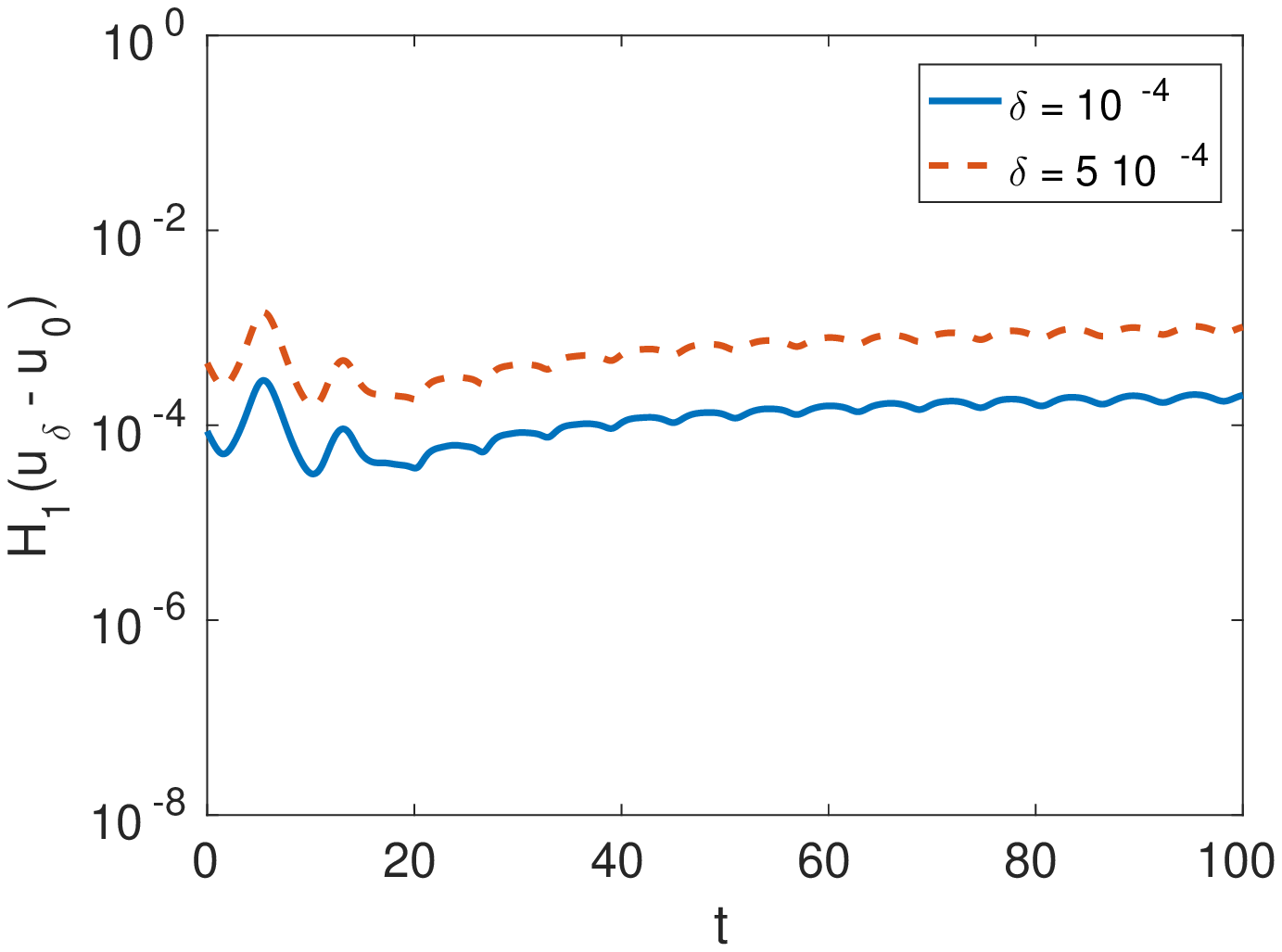}
}
  \centerline{\bf D\hspace{2.25in} E\hspace{2.25in} F}
  \caption{One UM regime: (A) $|U^{(1)}_{\eps,\beta}(x,t)|, T_0 = 0$, for $0\leq t\leq 100$. Spectrum at (B) $t = 0$, (C) $t=3$, (D) $t=35$, (E) $t=59.9$,
and (F) $d(t)$ for $f_1(x)$, $\delta = 10^{-4}, 5\times 10^{-4}$.}
\label{Figure3}
\end{figure}

The NLD-HONLS parameters are $\epsilon = 0.05$ and $\beta = 0.2$.
Figure~\ref{Figure3}A shows the surface
$|U^{(1)}_{\epsilon,\beta} (x,t)|$
for  
initial data generated using  Equation~\rf{SPB1} with $T_0 =  0$ for $ 0 < t < 100$.
%$j = 1$,$  $a = 0.5$  and $L = 2\sqrt{2}\pi$.
This choice of $T_0$ initializes the solution near the peak of the SPB $U^{(1)}(x,t)$.

The  Floquet spectrum evolves 
as follows:
At $t = 0$, the spectrum in the upper half plane consists of a single band of spectrum with
end point at $\lambda_0^s = 0.5\ri$, indicated by a large ``box'',  and
one imaginary double point  at $\lambda_1^d = 0.3535 \ri$,
indicated by a large  ``$\times$'' (Figure~\ref{Figure3}B).
Under the NLD-HONLS flow
$\lambda_1^d$ immediately splits asymmetrically into $\lambda_1^{\pm}$, with the upper band
of  spectrum in the right quadrant and the lower band in the left quadrant
(referred to as  a right state since the  waveform is characterized by a single damped modulated mode
traveling to the right).
The right state is clearly visible at $t = 3$ in
Figure~\ref{Figure3}C.
The numerically observed  evolution of spectrum for $U_{\epsilon,\beta}^{(1)}(x,t)$ for  short time is confirmed by the perturbation analysis. Equation \rf{ordereps}
indicates the complex double point $\lambda_1^d$,
at which the initial SPB was constructed, splits asymmetrically
at $\mathcal{ O}(\eps)$.

As time evolves the spectrum persists in a right state  (Figure~\ref{Figure3}D)
with the upper band increasing in length, although oscillating in length
as it does so.
In  Figure~\ref{Figure3}E the vertex of the loop  eventually touches the origin at $t \approx 59.9$.
Subsequently the  spectrum has three  bands emanating off the real axis,
 with endpoints $\lambda_1^{-}, \lambda_0^{s}, \lambda_1^{+}$ which, as damping
continues,   decrease in length.
Complex double points/critical points do not appear in the spectral evolution for $t >0$, indicating stability.

Figure~\ref{Figure3}F shows the evolution of $d(t)$
for $U_{\epsilon,\beta}^{(1)}(x,t)$ 
  using  $f_1$ and $\delta = 10^{-4}, 5 \times 10^{-4}$ (see Equation \rf{fk}).
There is no significant growth in $d(t)$ and $U_{\epsilon,\beta}^{(1)}(x,t)$ is stable for $t>0$.
The evolution of $d(t)$ is consistent with the prediction of stability from the evolution of the spectrum.
  We do not show a strength plot for $U_{\epsilon,\beta}^{(1)}(x,t)$ as rogue waves do not occur in its evolution.

As $T_0$ is varied in the initial data, the  spectral evolution obtained for $U_{\epsilon,\beta}^{(1)}(x,t)$ is qualitatively similar
(a right state is consistently obtained) to the spectrum shown in
Figure~\ref{Figure3}. 
  Complex critical points or double points do
not occur in the spectral
 evolution for $t >0$.
 As a result, $U_{\epsilon,\beta}^{(1)}(x,t)$
in the one UM regime evolves as a stable damped traveling breather, 
exhibiting regular behavior, and can be characterized as a continuous
deformation of a noneven 3-phase solution.
The  amplitude of the oscillations of $U_{\epsilon,\beta}^{(1)}(x,t)$ decreases and the frequency increases until small fast oscillations about the damped
Stokes wave, visible in Figure~\ref{Figure3}A, appear.

In the one UM regime the main differences in  the NLD-HONLS evolution,
as compared with linear damped HONLS in \cite{si20}, are i) the
lack of complex critical points in the spectral evolution
and ii) the small fast time scale oscillations in the band lengths which
correspond to small higher order oscillations in the temporal frequency.

\subsection{NLD-HONLS SPBs in the two UM regime: $a = 0.5, L = 4\sqrt{2}\pi$} 
We examine the evolution of the two distinct perturbed single mode
SPBs, $U_{\epsilon,\beta}^{(1)}(x,t)$ and $U_{\epsilon,\beta}^{(2)}(x,t)$,  and the perturbed two mode SPB $U_{\epsilon,\beta}^{(1,2)}(x,t)$.
Recall that 
 $\sigma(u_a(t)) = \sigma(U^{(j)}(x,t)) = \sigma(U^{(i,j)}(x,t))$. As a result,
 the Floquet spectrum
 at $t = 0$ is identical for all the numerical experiments in the two
 UM regime. The NLD-HONLS parameters are
 $\epsilon = 0.05$ and  $\beta = 0.2$.

 \subsubsection{Characterization of $U_{\epsilon,\beta}^{(1)}(x,t)$:}
\begin{figure}[ht!]
  \centerline{
\includegraphics[width=.33\textwidth]{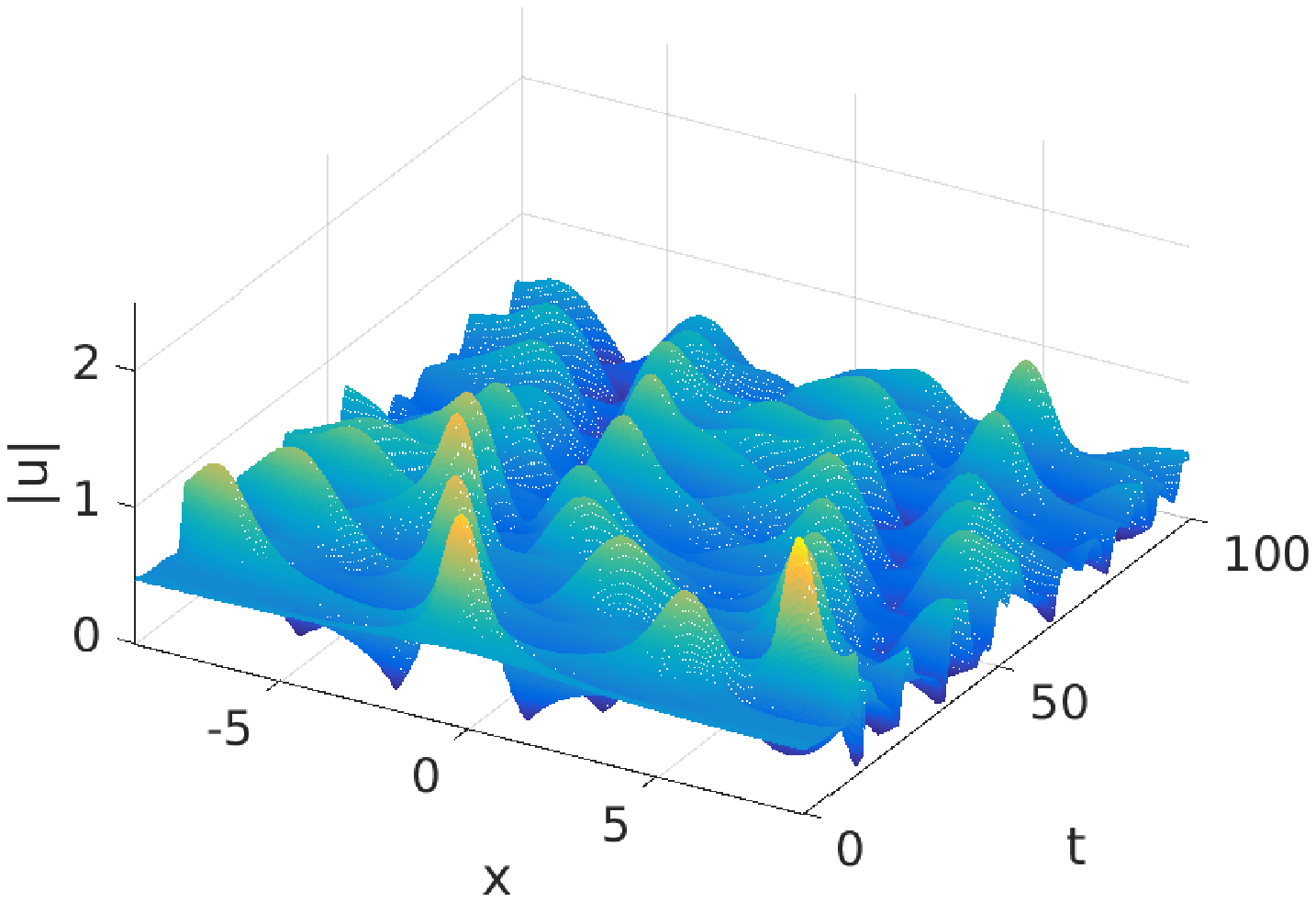}
\includegraphics[width=.33\textwidth]{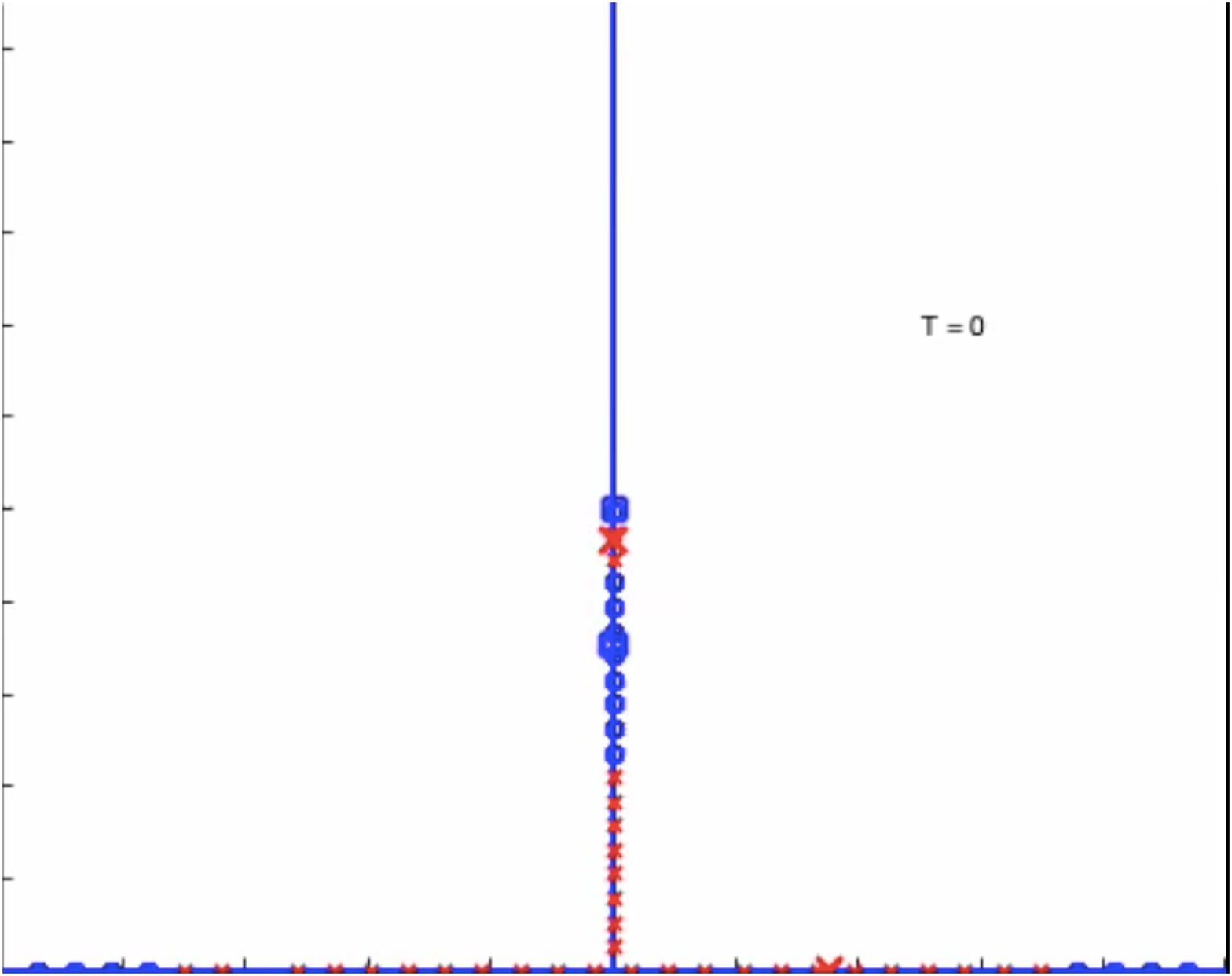}
\includegraphics[width=.33\textwidth]{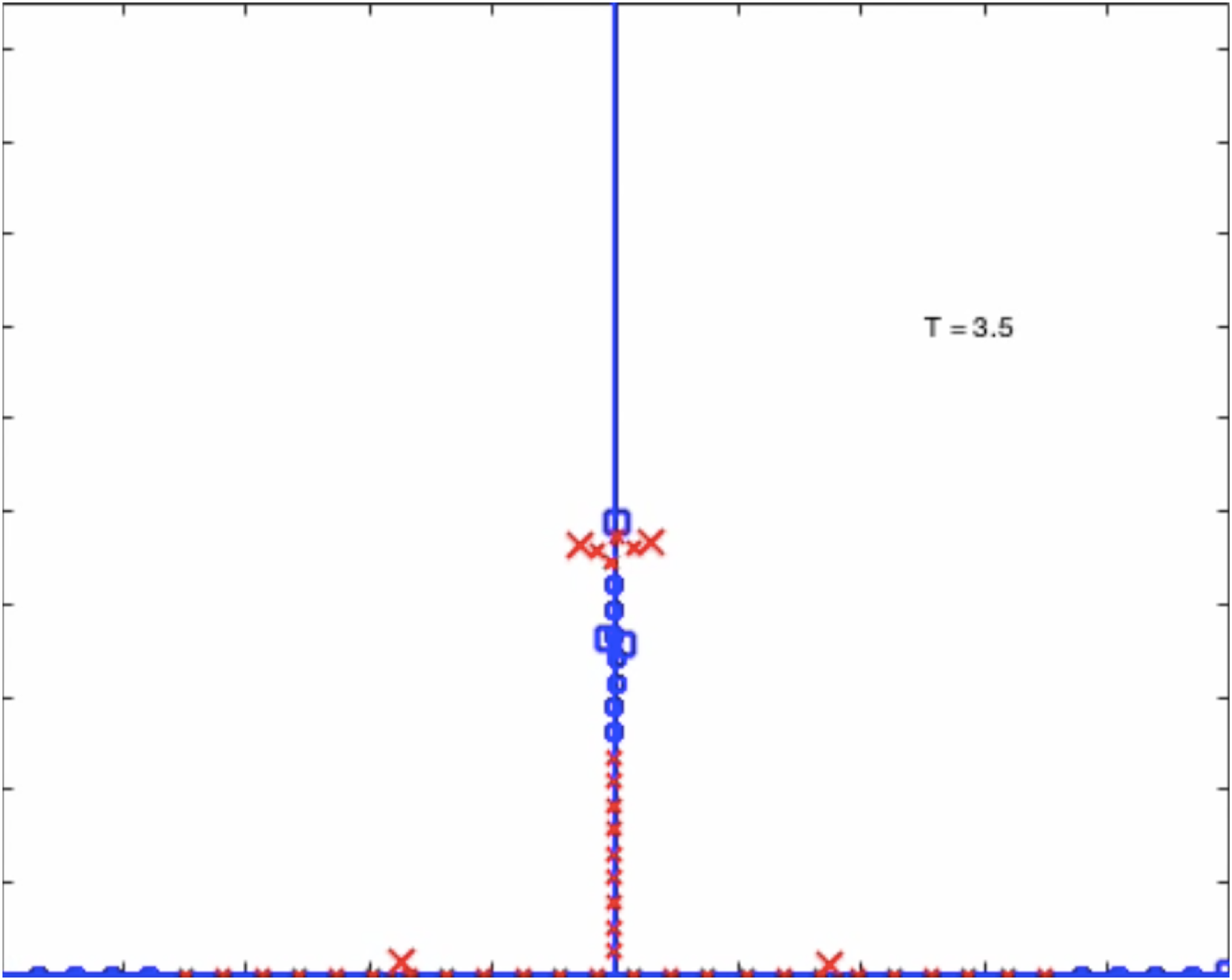}
}
  \centerline{\bf A\hspace{2.25in} B\hspace{2.25in} C}
  \vspace{12pt}
  \centerline{
\includegraphics[width=.33\textwidth]{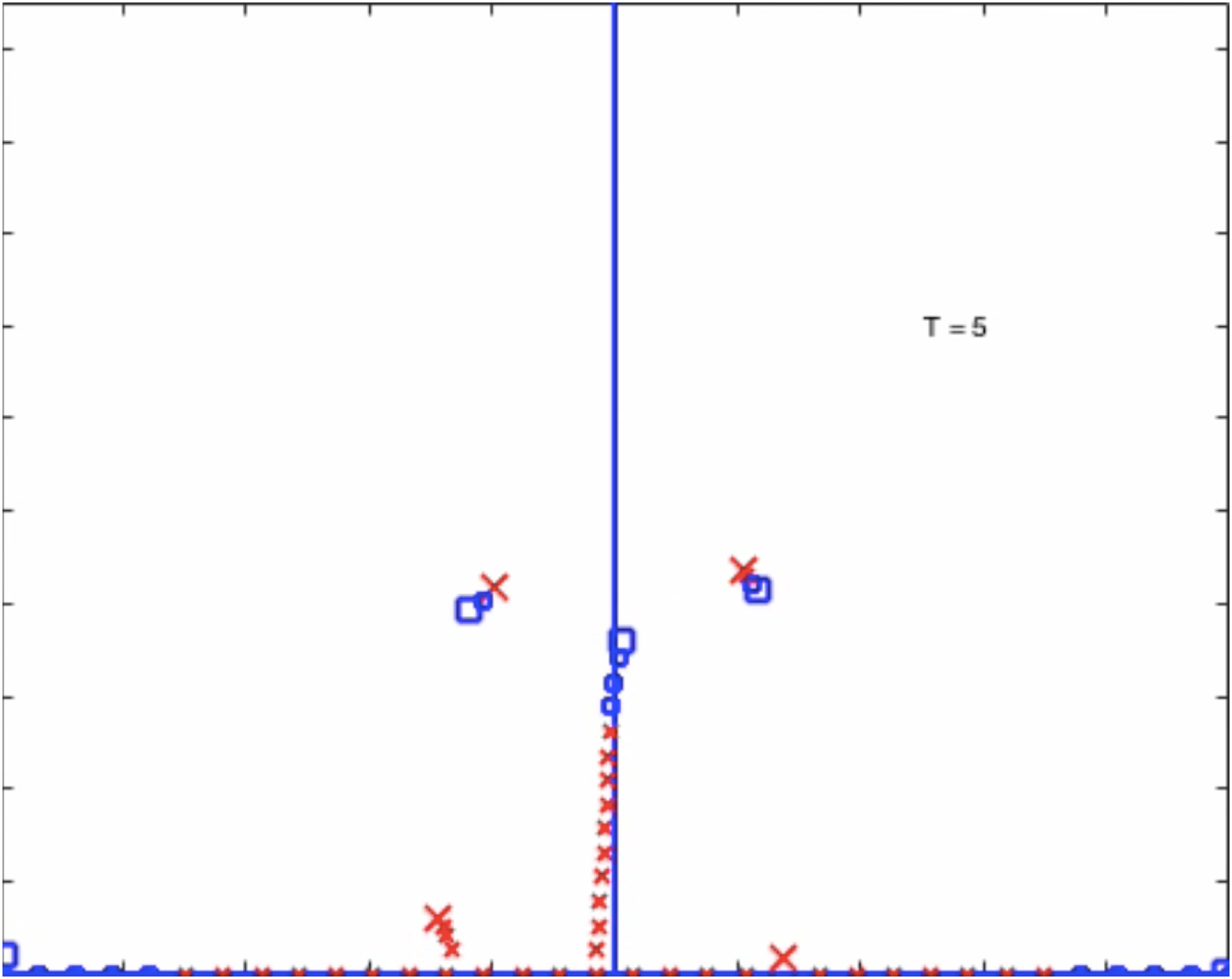}
\includegraphics[width=.33\textwidth]{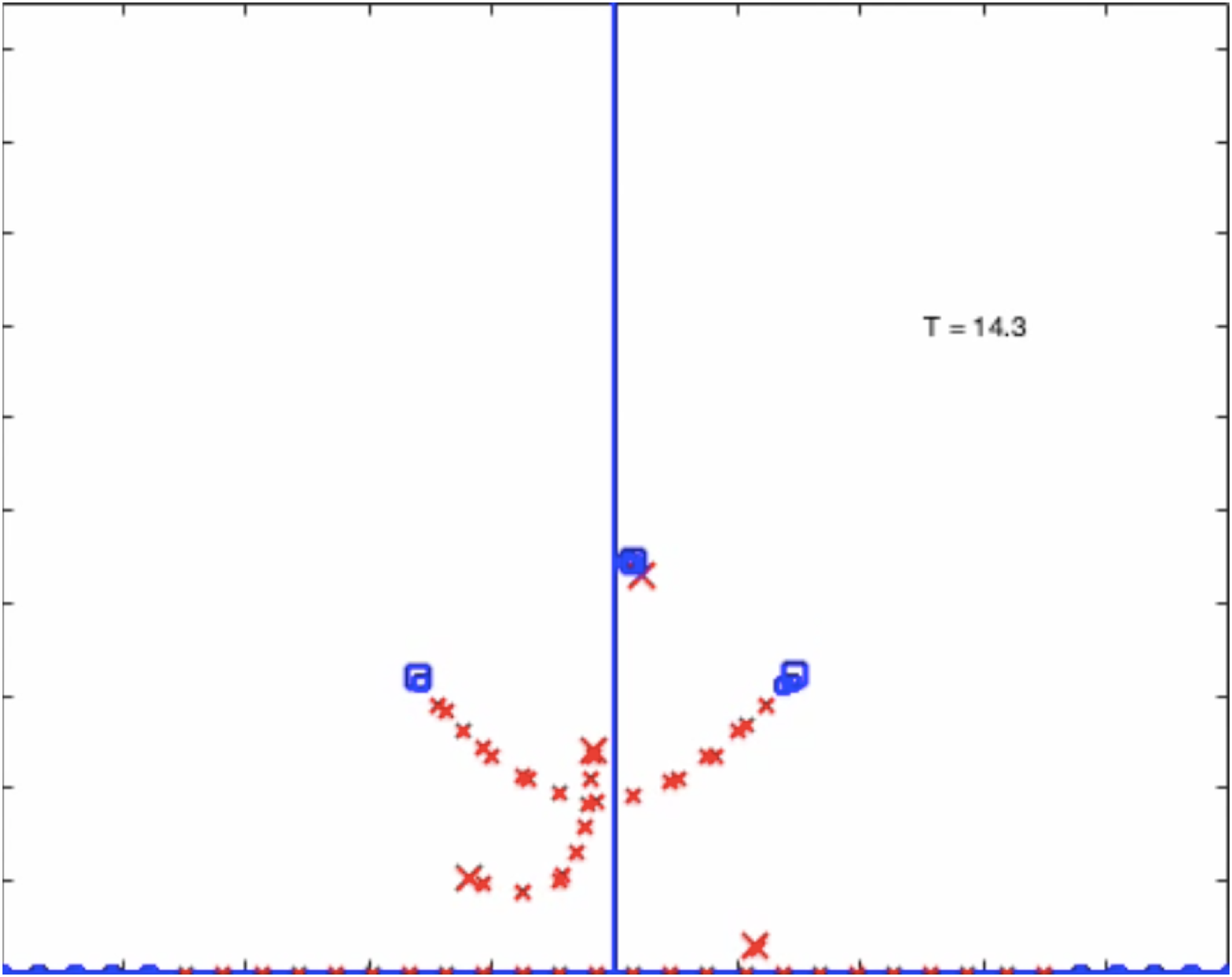}
\includegraphics[width=.33\textwidth]{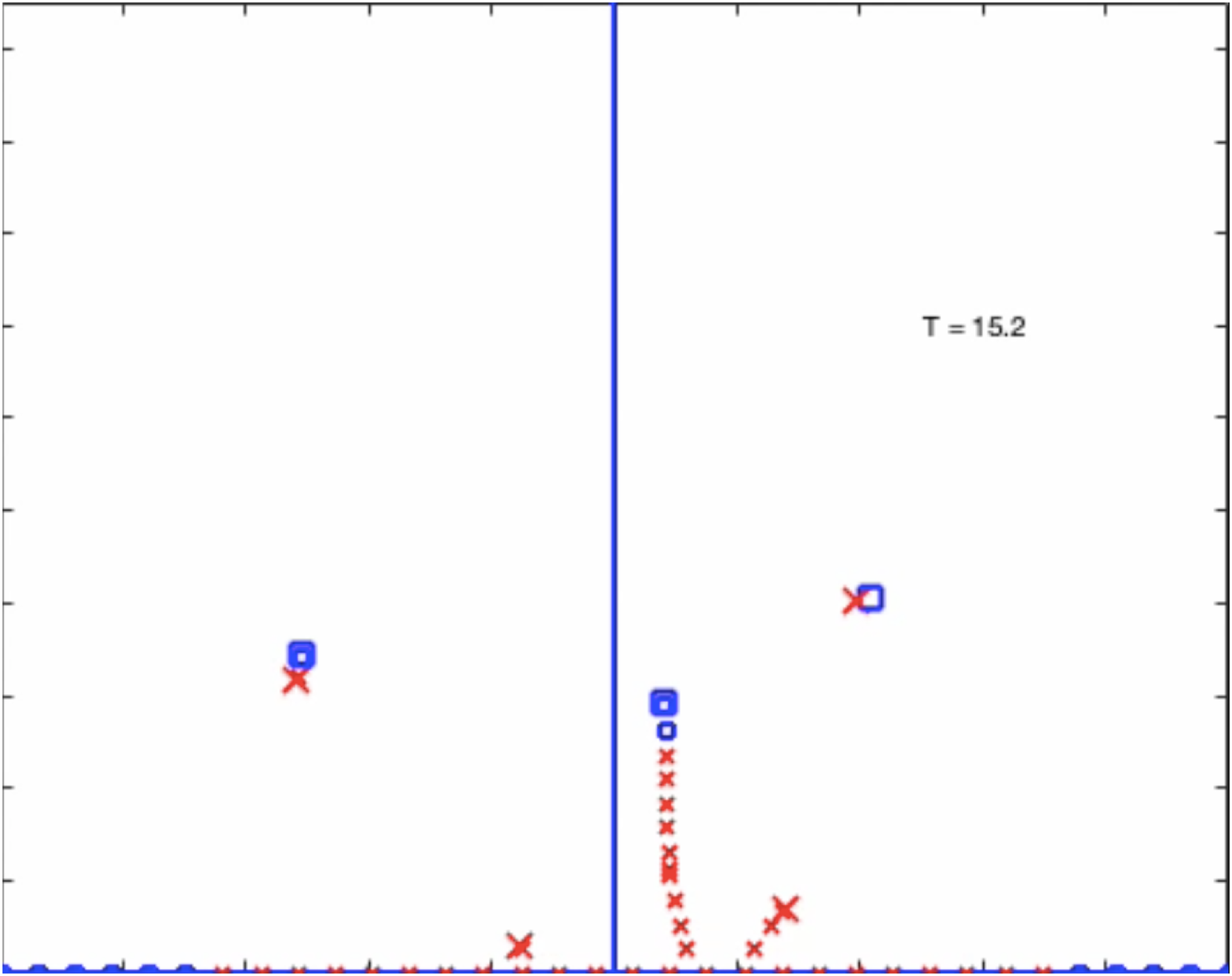}
  }
  \centerline{\bf D\hspace{2.25in} E\hspace{2.25in} F}
  \vspace{12pt}
  \centerline{
\includegraphics[width=.33\textwidth]{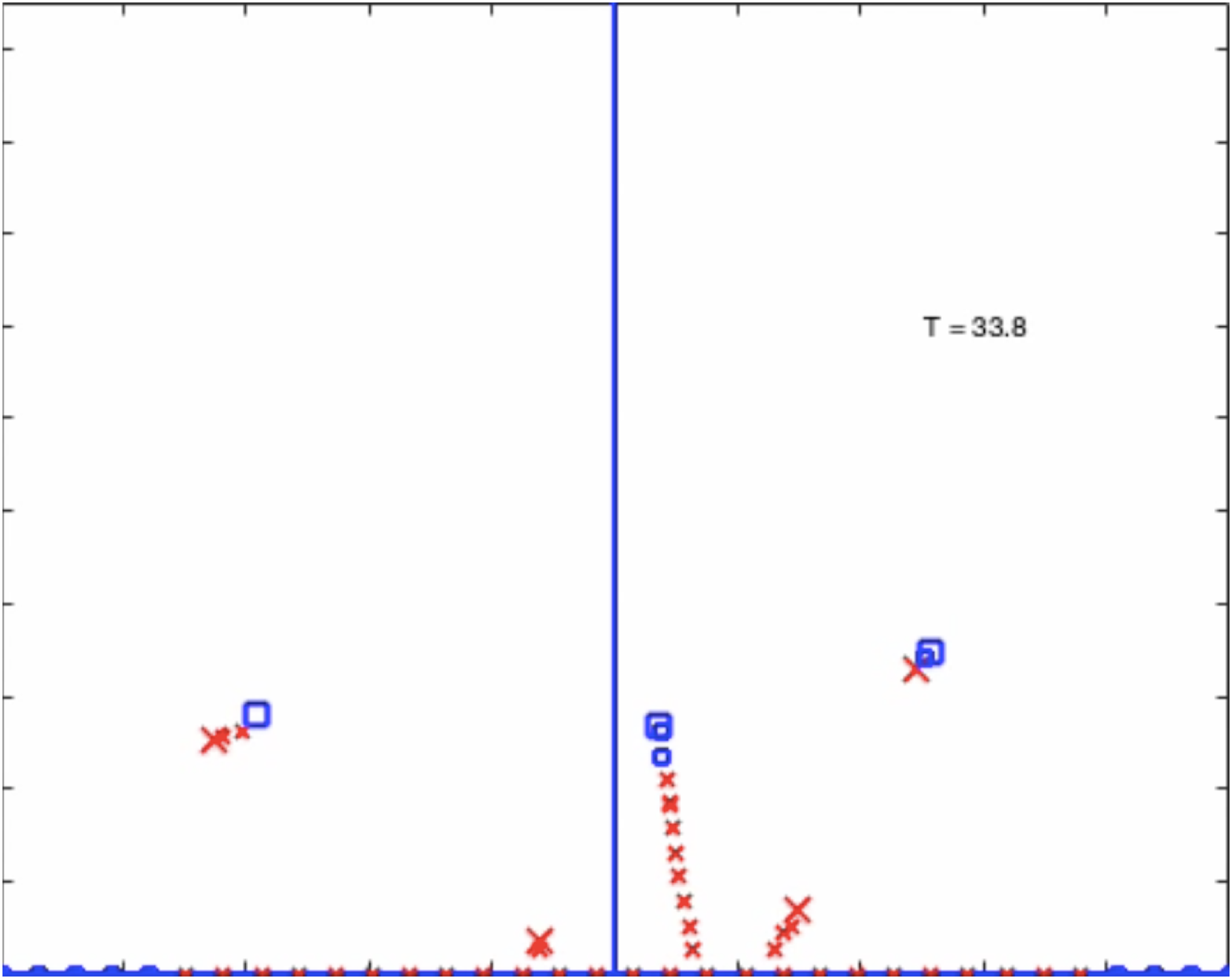}
\includegraphics[width=.33\textwidth]{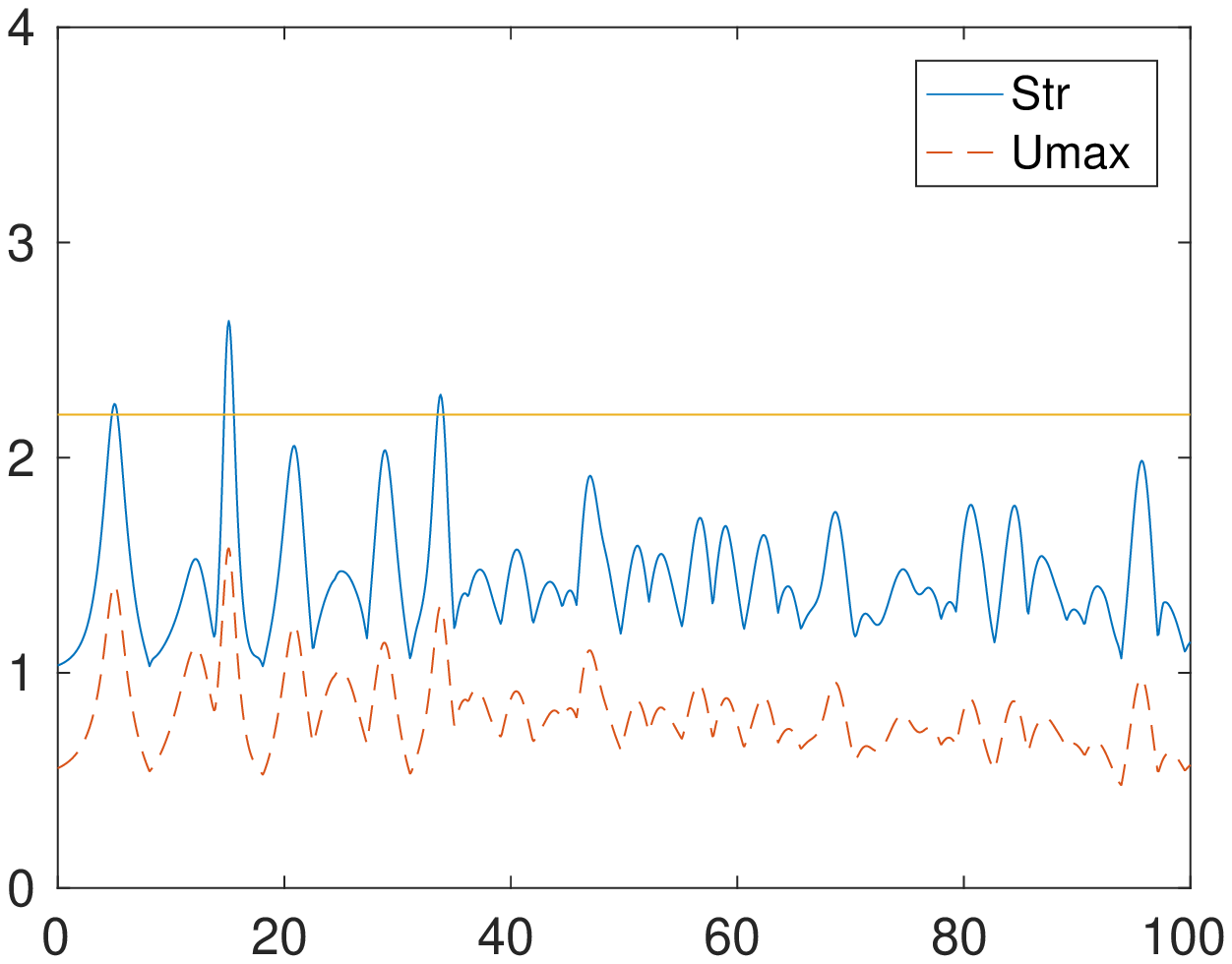}
\includegraphics[width=.33\textwidth]{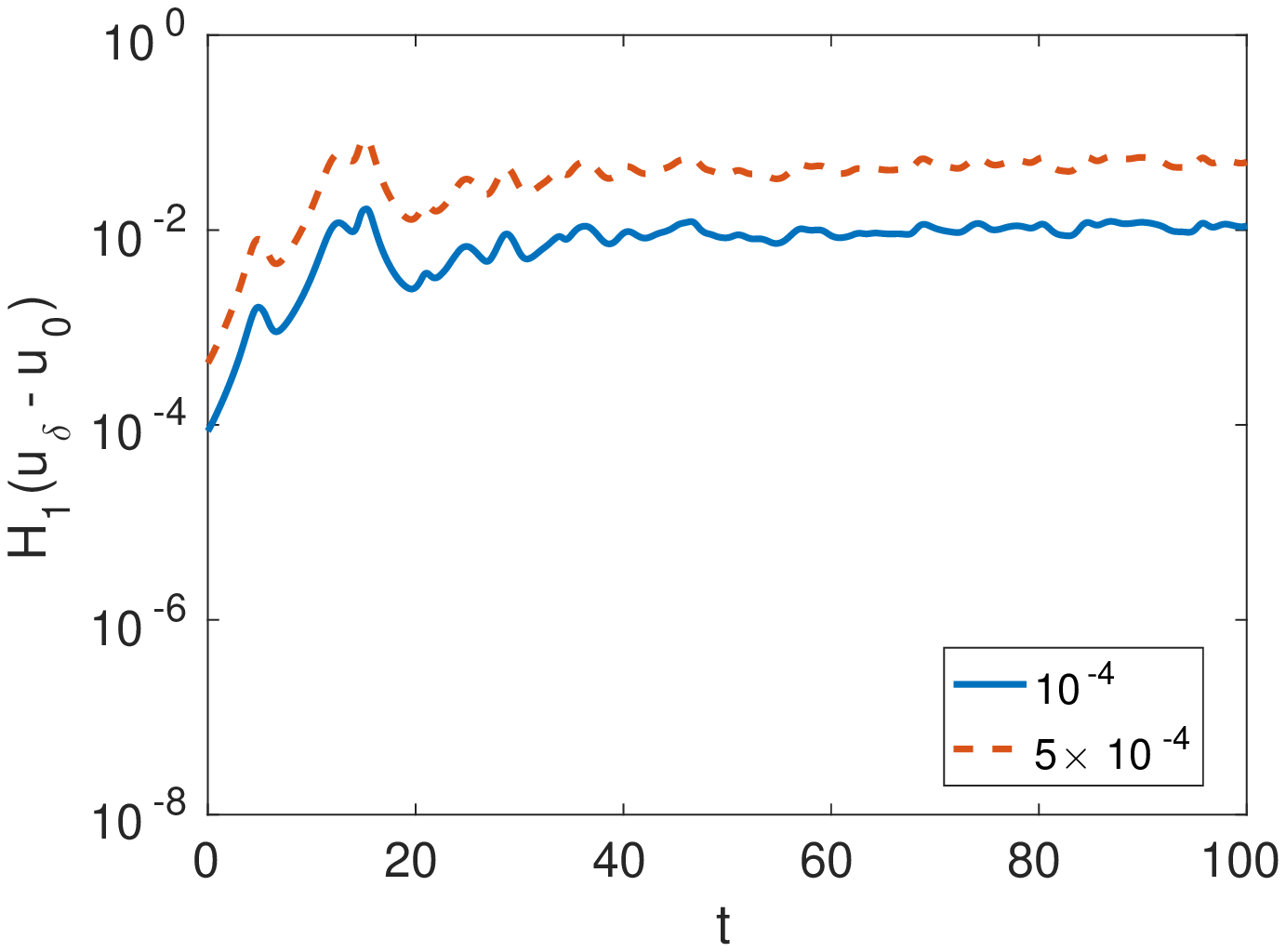}
}
  \centerline{\bf G\hspace{2.25in} H\hspace{2.25in} I}
  \caption{Two UM regime: (A) $|U^{(1)}_{\eps,\beta}(x,t)|, T_0=-5$, for $0\leq t\leq 100$, spectrum at (B) $t=0$, (C) $t=3.5$, (D) $t = 5$, (E) $t=14.3$, (F) $t=15.2$, (G) $t=33.8$, (H) strength, and (I) $d(t)$ for $f_2(x)$, $\delta = 10^{-4}, 5 \times 10^{-4}$. Rogues between: 4.6 - 5.2, 14.6 - 15.5, 33.5 - 34.1. Stability peaks at: 15.2.}
\label{Figure4}
\end{figure}

  Figure~\ref{Figure4}A shows the surface  $|U_{\epsilon,\beta}^{(1)}(x,t)|$
  for $ 0 < t < 100$ for initial data generated using
  Equation~\rf{SPB1} for $T_0 = -5$.
   The Floquet spectrum  at $t = 0$ is given in Figure~\ref{Figure4}B.
The end point of the band of spectrum at $\lambda_0^s = 0.5\ri$ is indicated by a ``box''.
There are two complex double points  at $\lambda_1^d = 0.4677 \ri$ and $\lambda_2^d = 0.3535 \ri$,
indicated by an ``$\times$'' and a ``box'', respectively.
Both complex double points split asymmetrically yielding three bands in the upper half complex plane as seen
in Figure~\ref{Figure4}C
at $t = 3.5$.
The numerically observed  evolution of spectrum for $U_{\epsilon,\beta}^{(1)}(x,t)$ for  short time is confirmed by the perturbation analysis.
Equations \rf{ordereps} - \rf{ordertwo}
indicate both complex double points split asymmetrically: 
$\lambda_1^d$ at which the NLS SPB was constructed 
 splits at ${\cal O}(\epsilon)$ into $\lambda^{\pm}_1$ while 
 $\lambda_2^d$ splits at $\mathcal{ O}(\epsilon^2)$ into  $\lambda_2^{\pm}$.

Subsequently the  two upper bands with end points ($\lambda_0^s,\lambda_1^+$) and
($\lambda_1^-,\lambda_2^+$) decrease in length while they detach and move away from the imaginary axis,
yielding a five-phase solution with one tiny band ($\lambda_0^s,\lambda_1^+$) in the first quadrant and a somewhat larger band ($\lambda_1^-,\lambda_2^+$) in the second quadrant and one band close to the imaginary axis
($\lambda_2^-,\lambda_R$) where $\lambda_ R$ is real, as seen at $t \approx 5$ in  Figure~\ref{Figure4}D
(note: we neglect the
 small bands emanating off the real axis related to the higher modes).
 Applying criteria (\ref{s_state}), we find 
 a soliton-like mode has emerged and the solution is ``close'' to a one soliton-like state.
 
\begin{figure}
  \centerline{
\includegraphics[width=.33\textwidth]{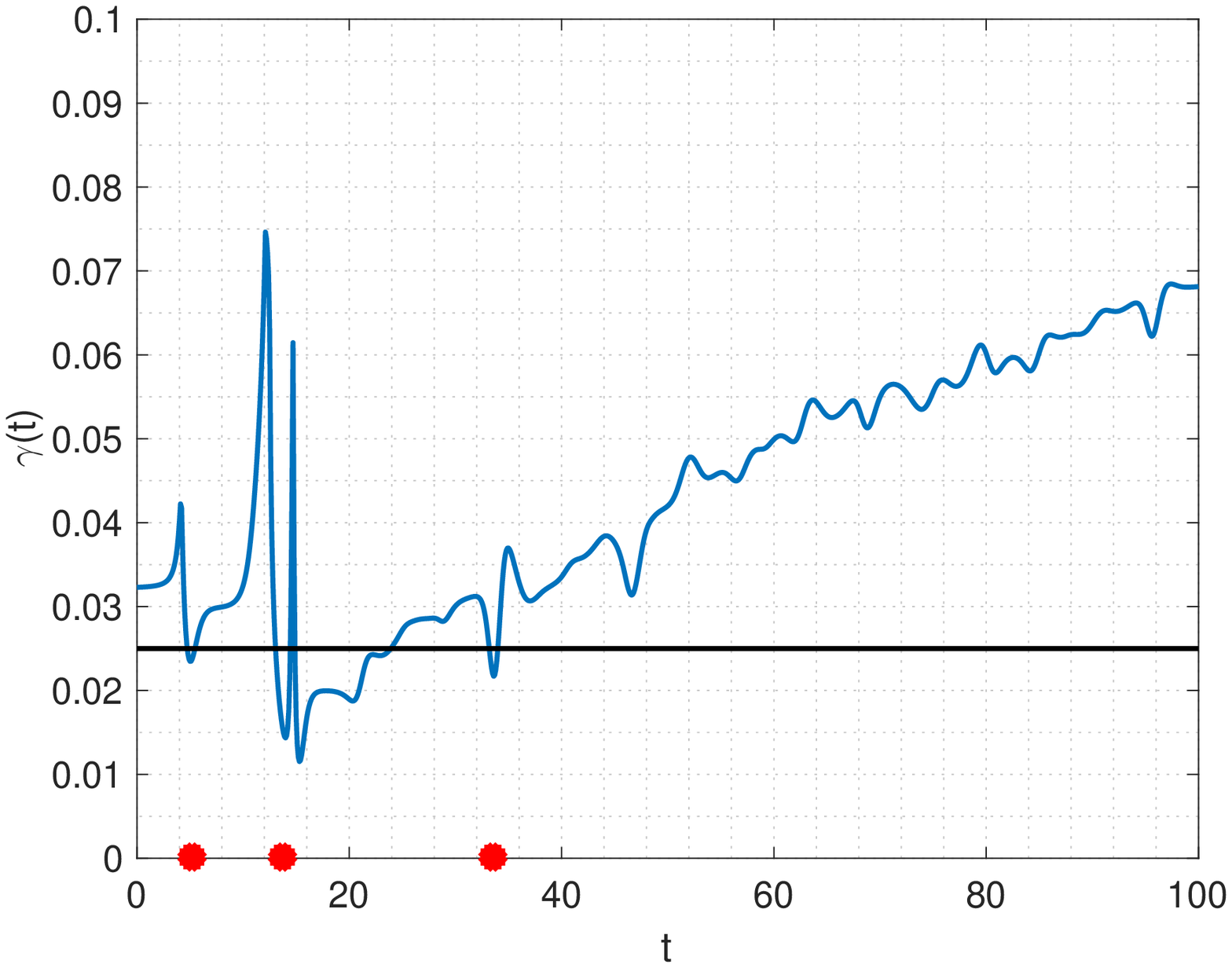}
}  
  \caption{Variation of the length of the band $|\lambda_0^s-\lambda_1^+|$ of spectrum as a function of time. The horizontal line represents the upper bound for a one soliton-like state. The red dots represent the time a rogue wave occurs.}
  \label{Figure5}
\end{figure}

This soliton-like state satisfies the criterion of a  rogue wave.  The rogue wave is visible in
the strength plot, Figure~\ref{Figure4}H, at $t \approx 5$.
The spectrum remains near this state until $t \approx  5.5$.
The lengths of
the bands then increase and a sequence of  bifurcations occur in rapid succession for $11 < t < 14.3$ with complex critical points forming,
 indicating instability of the corresponding nonlinear modes. 
 The last complex critical point at $t = 14.3$ is shown in
 Figure~\ref{Figure4}E.
Subsequently the bands split and their lengths decrease until the band in the first quadrant again satisfies the one soliton-like criteria 
  at $t = 15.2$  in  Figure~\ref{Figure4}F.
A more pronounced rogue wave is observed in the strength plot at this time.
The fluctuation in the band lengths  occur once more, producing
 at  $t = 33.8$ the last soliton-like spectrum (Figure~\ref{Figure4}G) and the last rogue
wave  (Figure~\ref{Figure4}H).
The  spectral band lengths then increase and 
  settle into a typical
  $5$-phase state with no further rogue wave activity.

  The variation in the band length  $\gamma(t) = |\lambda_0^s-\lambda_1^+|$
  for the band in the upper right quadrant are shown in
  Figure \ref{Figure5}. The variation in the band length for the band in the upper left quadrant is not shown as it never satisfies the criteria for a soliton-like state.
  Recalling equation \rf{s_state}, a one soliton-like state  occurs at $t^*$ if  $\gamma(t^*) < 0.025$. In Figure \ref{Figure5} the horizontal line indicates the upper bound on the band length for a soliton-like state. Additionally a red dot on the $t$-axis indicates the time at which a rogue wave occurs providing a correlation between one soliton-like states and rogue waves.

  {\sl \bf Observation:} Soliton-like rogue waves emerge in the NLD-HONLS flow.
  Each time a rogue wave is observed in the strength plot of $U^{(1)}_{\eps,\beta}(x,t)$ (i.e.  $S(t) > 2.2$ at $t = 5$, $t = 15.2$,
  and $t = 33.8$), the corresponding spectrum is in a one soliton-like configuration.

We remark that the  emergence of the one soliton-like structure in the evolution of the nonlinear damped $U_{\epsilon,\beta}^{(1)}(x,t)$
in the two UM regime
was not observed in the linear damped HONLS study \cite{si21}.

The evolution of $d(t)$ 
for $U_{\epsilon,\beta}^{(1)}(x,t)$ is given in Figure~\ref{Figure4}H
with $f_2$ for  $\delta = 10^{-4}, 5\times 10^{-4}$ (see Equation \rf{fk}).
The perturbation $f_2$  is chosen in the direction of the
unstable mode associated with $\lambda_2^d$.
The growth in $d(t)$  saturates by  $t = 15.2$, in agreement with the stabilization time determined by the last critical point in the nonlinear spectral decomposition of the solution at $t = 14.3$.

For $U_{\epsilon,\beta}^{(1)}(x,t)$  in the two-UM regime, both
$\lambda_1^d$ and $\lambda_2^d$ resonate with the perturbation.
The route to stability, for $T_0\in[-5,-1]$ is characterized by the appearance of a one soliton-like structure  which corresponds to rogue wave events.
As $T_0\rightarrow 0$, more complex critical points appear in the spectral decomposition and it takes longer for the damped solution to
stabilize. Fewer rogue waves may occur,  and these weaker rogue waves
are no longer associated with a one soliton-like structure.

\subsubsection{Characterization of $U^{(2)}_{\eps,\beta}(x,t)$:}
\begin{figure}[htp!]
  \centerline{
\includegraphics[width=.33\textwidth]{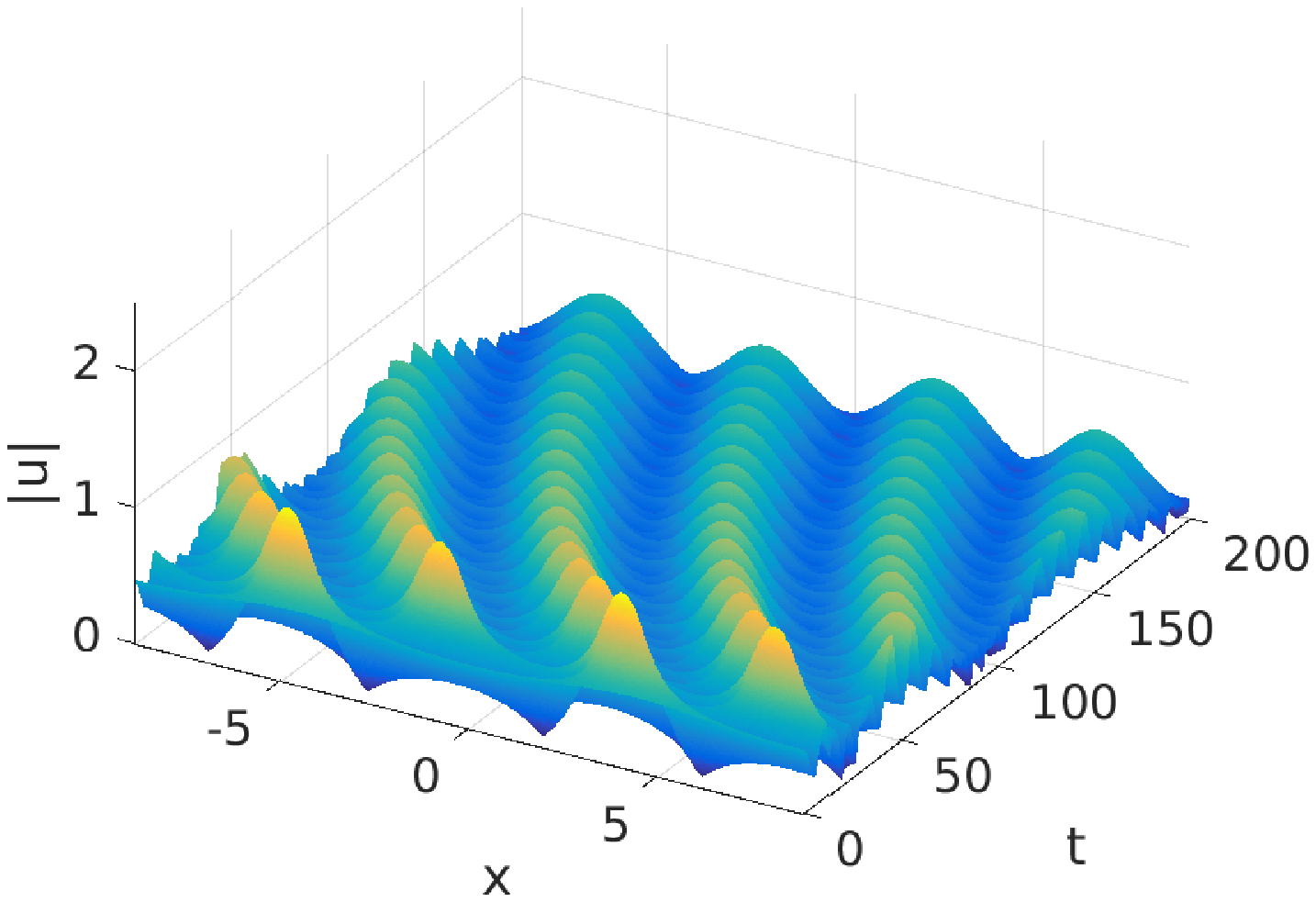}
\includegraphics[width=.33\textwidth]{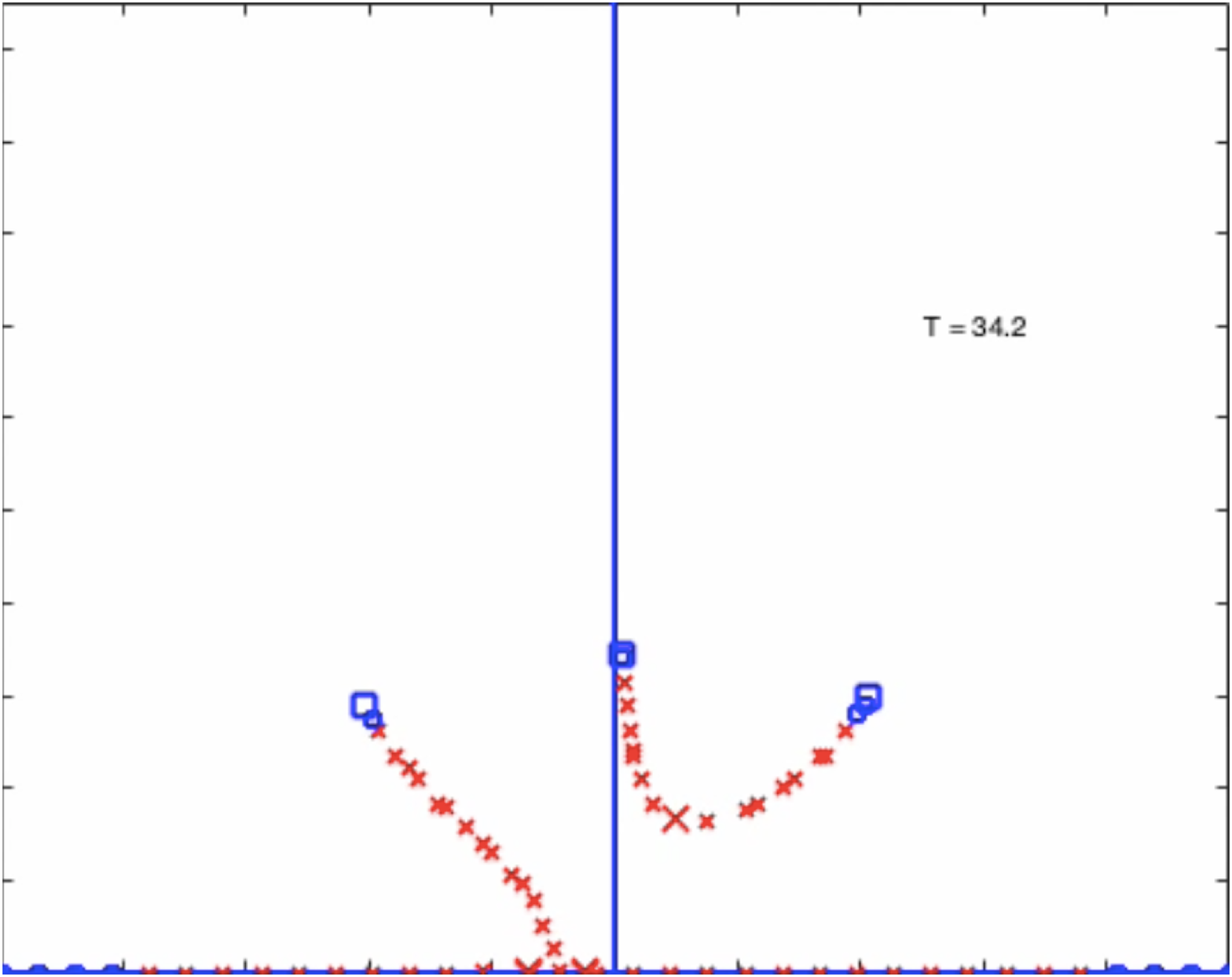}
\includegraphics[width=.33\textwidth]{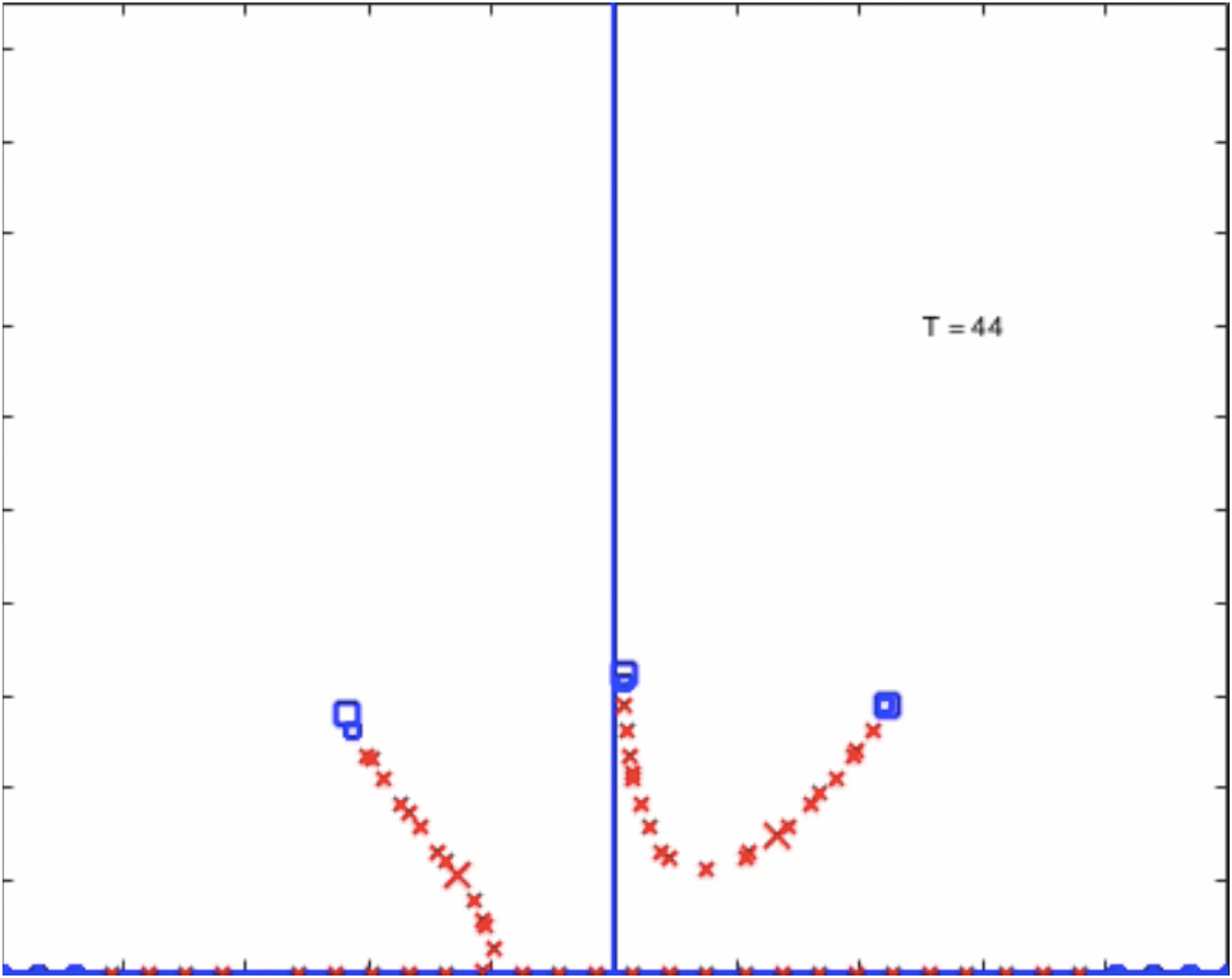}
  }
  \centerline{\bf A\hspace{2.25in} B\hspace{2.25in} C}
  \vspace{12pt}
  \centerline{
\includegraphics[width=.33\textwidth]{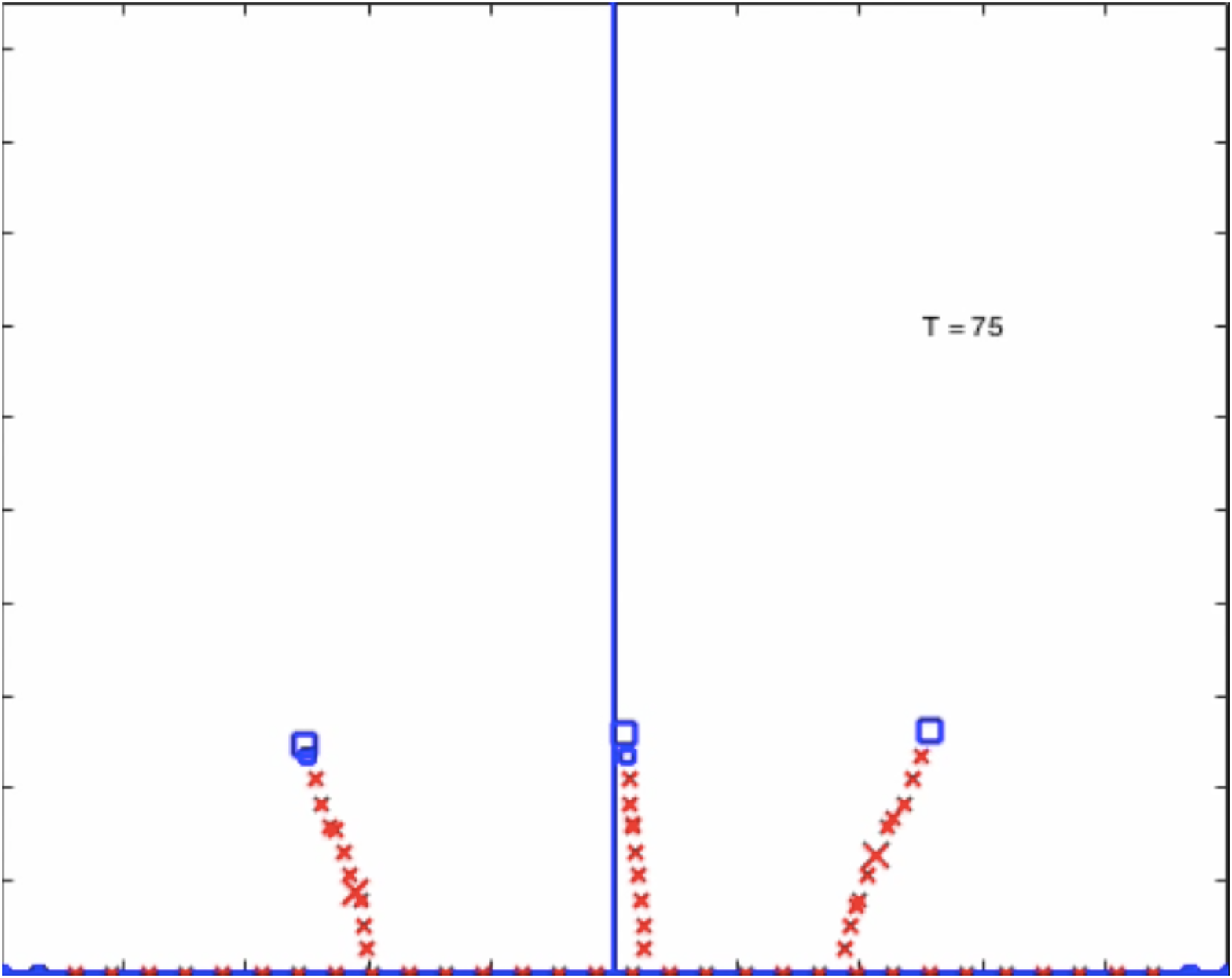}
\includegraphics[width=.33\textwidth]{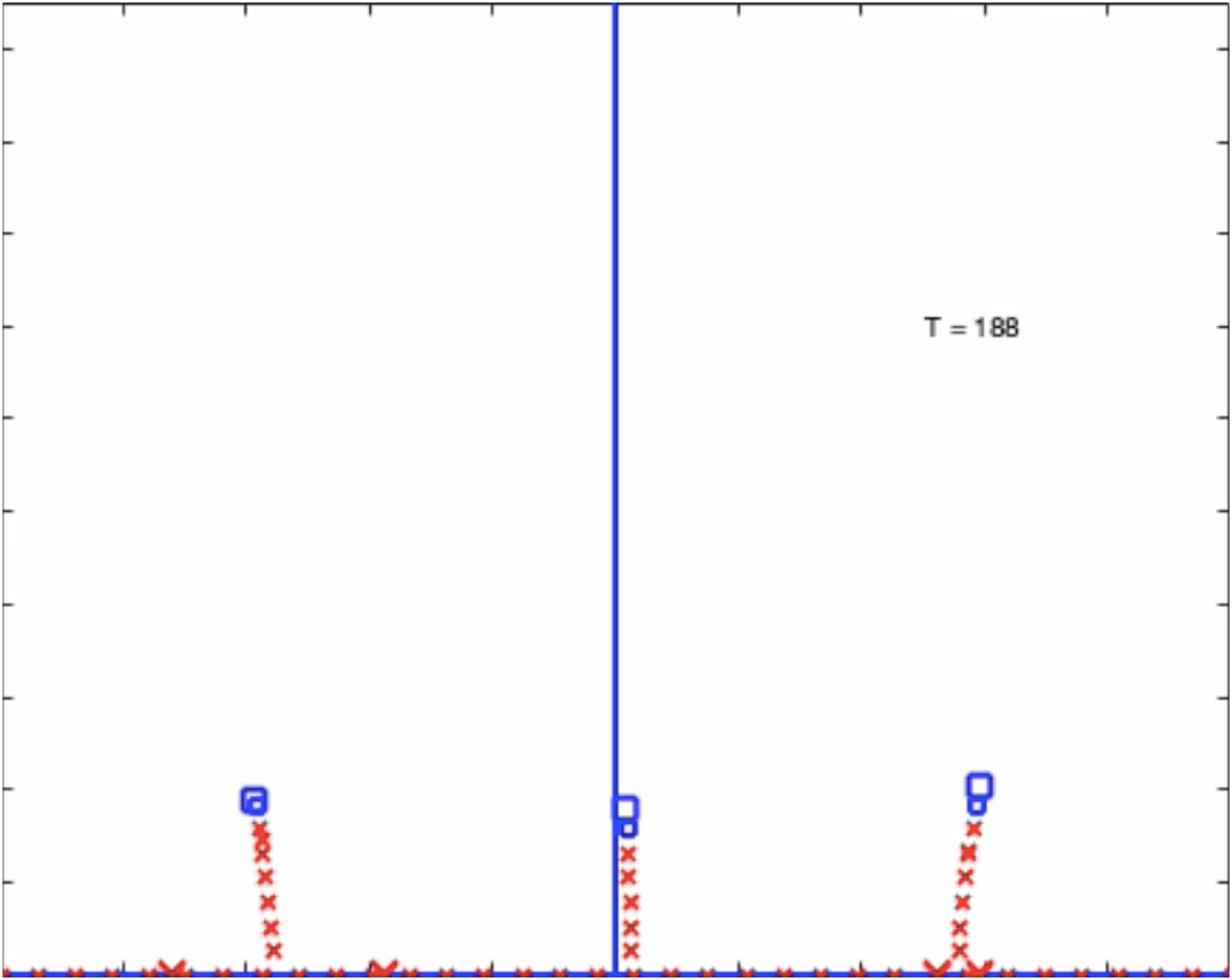}
\includegraphics[width=.33\textwidth]{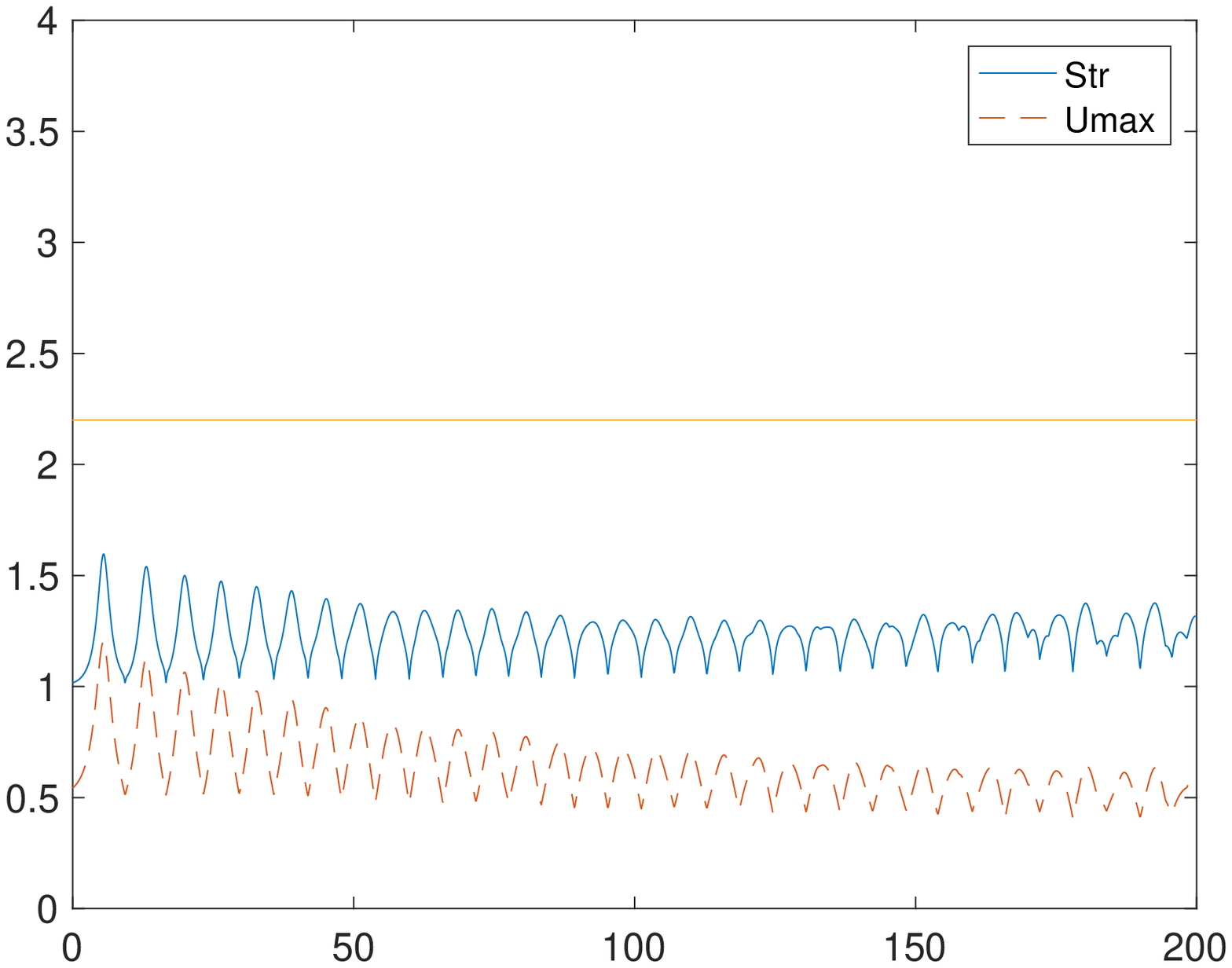}
  }
  \centerline{\bf D\hspace{2.25in} E\hspace{2.25in} F}
  \vspace{12pt}
  \centerline{
\includegraphics[width=.33\textwidth]{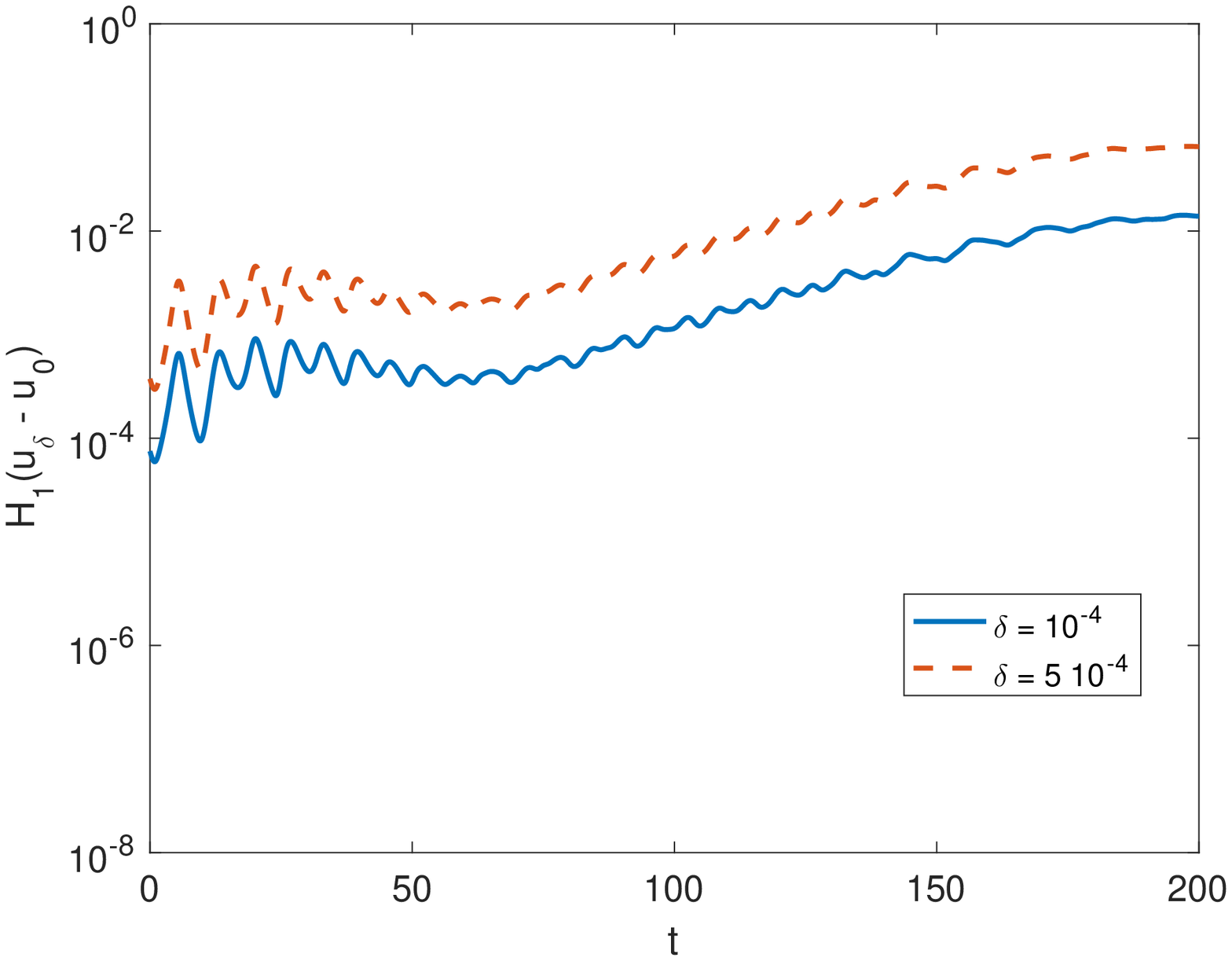}
\includegraphics[width=.33\textwidth]{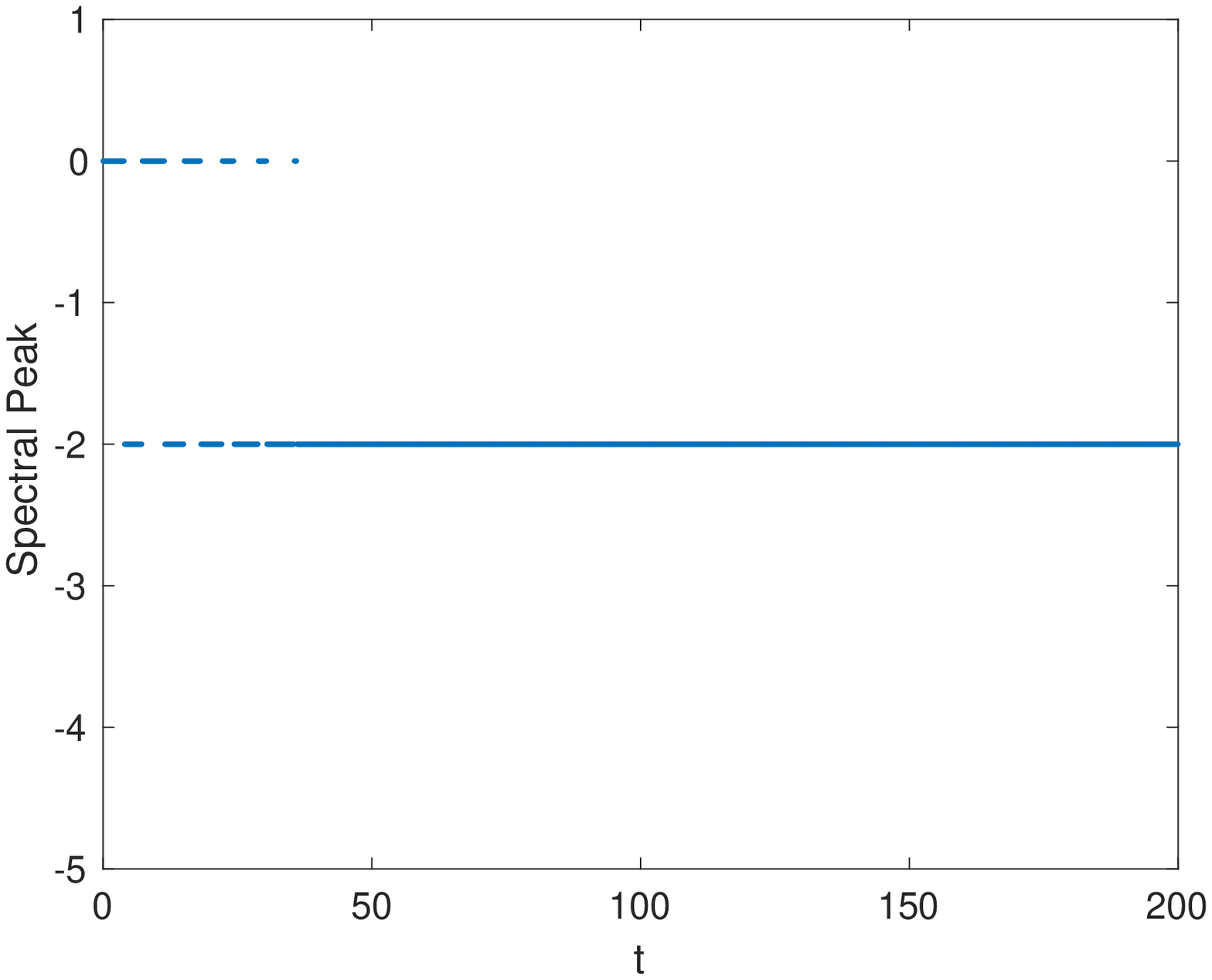}
\includegraphics[width=.33\textwidth]{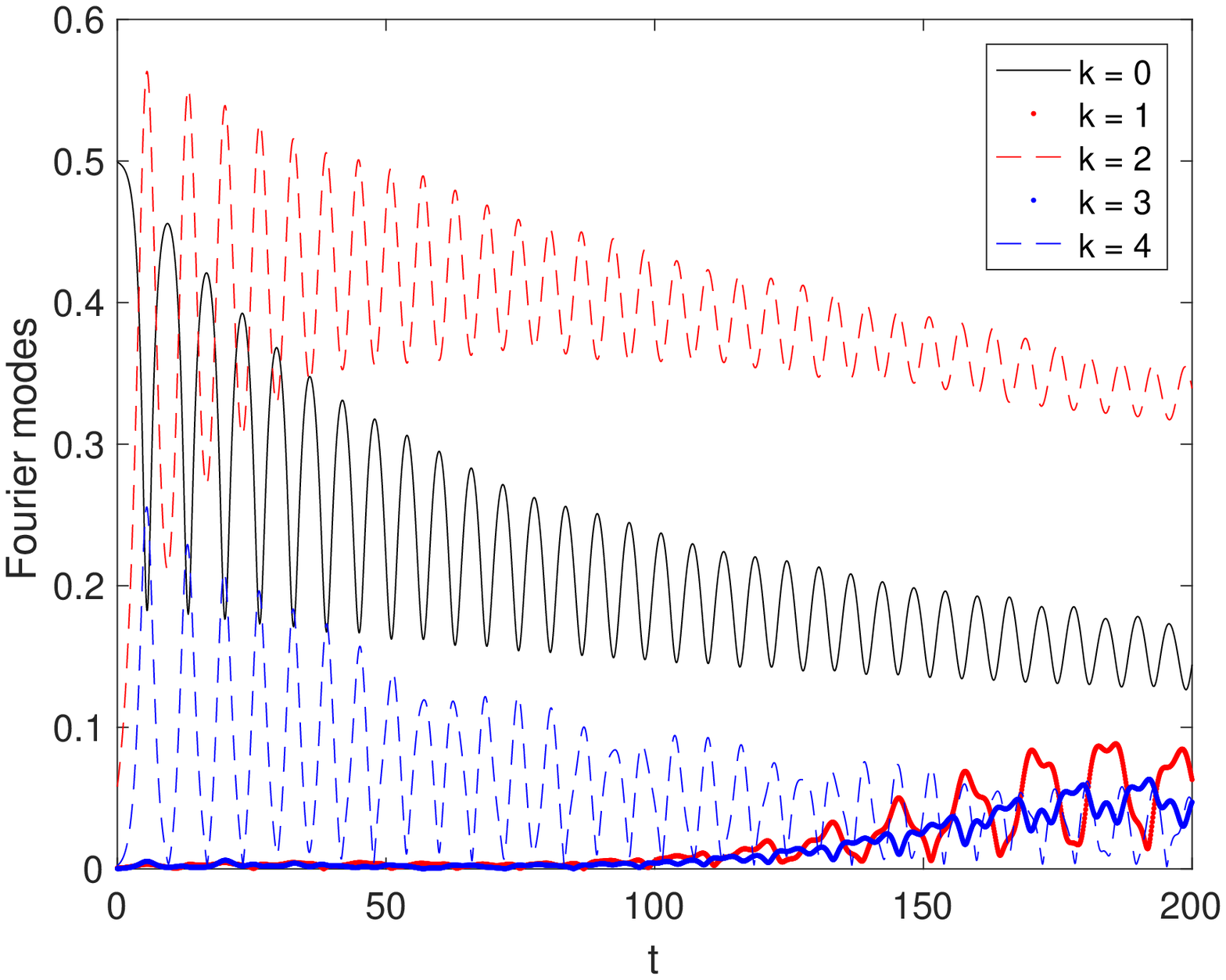}
  }
  \centerline{\bf G\hspace{2.25in} H\hspace{2.25in} I}
  \caption{Two UM regime: (A) $|U^{(2)}_{\eps,\beta}(x,t)|,T_0=-5$, for $0\leq t\leq 100$, spectrum at (B) $t=34.2$, (C) $t=44$, (D) $t=75$, (E) $t=188$,
 (F) strength, (G) $d(t)$ for $f_1(x)$, $\delta = 10^{-4}, 5 \times 10^{-4}$,     (H) spectral peak, (I) Fourier modes.}
  \label{Figure6}
\end{figure}

We now consider $U^{(2)}_{\epsilon,\beta}(x,t)$ with
initial data  generated using  Equation~\rf{SPB1} with $T_0 = -5$.
 $U^{(1)}(x,t)$ and $U^{(2)}(x,t)$ are both single mode SPBs over
the same Stokes wave. Even so,
their respective evolutions under the NLD-HONLS flow
are quite different.
 Notice in Figure~\ref{Figure6}A the surface of $|U_{\epsilon,\beta}^{(2)}(x,t)|$  for $0 \le t \le 100$ is a damped traveling breather,
 exhibiting regular behavior, 
 in contrast to the irregular
 behavior of $|U_{\epsilon,\beta}^{(1)}(x,t)|$ in the two UM regime. Further,
 as shown in the strength plot Figure~\ref{Figure6}F, rogue wave  events
 do not  occur in the evolution of $U^{(2)}_{\eps,\beta}$.

The spectrum of  the initial data
$U_{\epsilon,\beta}^{(2)}(x,0)$  is given  in 
Figure~\ref{Figure4}B and has two complex double points.
Under the NLD-HONLS flow 
$\lambda_2^d$ immediately splits asymmetrically into  $\lambda_2^{\pm}$, forming a right traveling breather with the upper band in the right quadrant and the lower band in the left quadrant. Significantly, the first double point
$\lambda_1^d$  does not split.
Figure~\ref{Figure6}B clearly shows  that $\lambda_1^d$,  indicated by the large ``$\times$''on the upper band, has not split at  $t = 34.2$.

The short time perturbation results,  Equations \rf{ordereps} - \rf{ordertwo}
  show that
only the double point $\lambda_2^d$ associated with the SPB initial data
$U^{(2)}(x,0)$ splits at ${\cal O}(\epsilon)$ under the NLD-HONLS.
The double points $\lambda_{2m}^d$ split
at $\mathcal{ O}(\epsilon^m)$ while  $\lambda_l^d$, $l\neq 2m$ do not split,
 confirming for short time 
the numerically observed evolution of spectrum for $U_{\epsilon,\beta}^{(2)}(x,t)$.
In fact,  $\lambda_1^d$, which corresponds to a non-resonant mode,
is observed to translate along a band of spectrum but does not split
for the duration of
the NLD-HONLS evolution, $0 \le t \le 200$, 
see Figures~\ref{Figure6}B-E.

A distinctive feature of the spectral evolution for $U_{\epsilon,\beta}^{(2)}(x,t)$ is that the solution gains an unstable mode corresponding to
$\lambda_3^d$.
At $t \approx 30$, enough energy has been transferred to the third mode (originally inactive) so that $\lambda_3^d$ starts to move  from the real axis  onto
a band of spectrum in the upper half complex plane. The additional double point is clearly visible in Figure~\ref{Figure6}C at $t = 44$.
The vertex of the upper band of spectrum  touches the real axis at $t = 51.1$ and subsequently, as seen in Figure~\ref{Figure6}D at $t = 75$, there are
three bands of spectrum emanating from the real axis, two of which have complex double points 

As time evolves $\lambda_1^d$ and $\lambda_3^d$
 decrease in magnitude and at $t = 188$ they become real double points,
 Figure~\ref{Figure6}E.
 This agrees with the stability plot, Figure~\ref{Figure6}G, which
 shows   $d(t)$ grows due to the presence of instabilities until
 $t = 188$. 
From the Floquet perspective
 $U_{\epsilon,\beta}^{(2)}(x,t)$ stabilizes slowly, evolving as a regular but
 unstable damped three-phase solution, until the complex double points become real at $t = 188$.

 An interesting aspect of the dynamics is that the growth of the instability associated
 with the complex double point $\lambda_1^{d}$ appears delayed.
 Figure~\ref{Figure6}G shows that the growth in $d(t)$ is more
 significant
 later in the experiment at $t \approx 60$.
 Since $\lambda_1^d$ does not split we would have expected to initially see significant growth in $d(t)$.
 A Fourier decomposition of the data shows that the waveform is undergoing
 frequency  downshifting during the early stage of the experiment.
Defining the 
dominant mode or spectral peak, 
$k_{peak}$, as the wave number $k$ for which $|\hat u_k|$ achieves its maximum,
downshifting occurs when there is a shift down 
in $k_{peak}$ \cite{trul97}.
In Figure~\ref{Figure6}H we find downshifting becomes permanent about $t \approx 40$, with the energy flowing primarily from
the zeroth Fourier mode to the 2nd Fourier mode. 

Figure~\ref{Figure6}I is a plot of the
Fourier modes for NLD-HONLS for the perturbed initial data\newline
$U^{(2)}_{\epsilon,\beta,\delta}(x,0) = U^{(2)}(x,T_0) + \delta f_1(x)$ (see Equation \rf{fk}), $0 < t < 200$.
The dark red and blue curves correspond to the $k= 1$ and $k=3$ mode respectively. The Fourier mode plot  demonstrates that even when the first mode is
excited initially,
frequency downshifting due to nonlinear damping  inhibits the  growth of the first mode  which is then  expressed as a higher order effect.
We find  $d(t)$ grows, i.e. $U_{\epsilon,\beta}^{(2)}(x,t)$
and $U_{\epsilon,\beta,\delta}^{(2)}(x,t)$  grow apart
until $t = 188$. This confirms the stabilization time obtained from the nonlinear spectral analysis and  supports the observation that $\lambda_1$ does not split.

For $U_{\epsilon,\beta}^{(2)}(x,t)$  in the two-UM regime, only  
$\lambda_2^d$  resonates with the perturbation.
Since one of the unstable modes does not resonate with the perturbation,
the dynamics is simplified and organized.
$U_{\epsilon,\beta}^{(2)}(x,t)$ is a damped traveling breather
exhibiting regular behavior and can be characterized as a continuous
deformation of a noneven 3-phase solution.  Rogue waves do not appear in the evolution of $U_{\epsilon,\beta}^{(2)}(x,t)$. This broad characterization does not change as $T_0$ is varied.

\subsubsection{Characterization  of $U^{(1,2)}_{\eps,\beta}(x,t)$:}
The two-mode SPB of highest amplitude is obtained when the modes are excited simultaneously and coalesce, i.e. for $\rho = \tau =0$ in Equation \rf{SPB2}
(see Figure~\ref{Figure2}C).
The coalesced two-mode SPB has come to be viewed as a ``fundamental'' two-mode
rogue wave as it is more robust in the presence of small random variations of initial data \cite{cs14}. In this subsection we examine the evolution of $U_{\epsilon,\beta}^{(1,2)}(x,t)$ for the coalesced SPB initial data as $T_0$ and $\beta$ are varied.

\begin{figure}[ht!]
  \centerline{
\includegraphics[width=.33\textwidth]{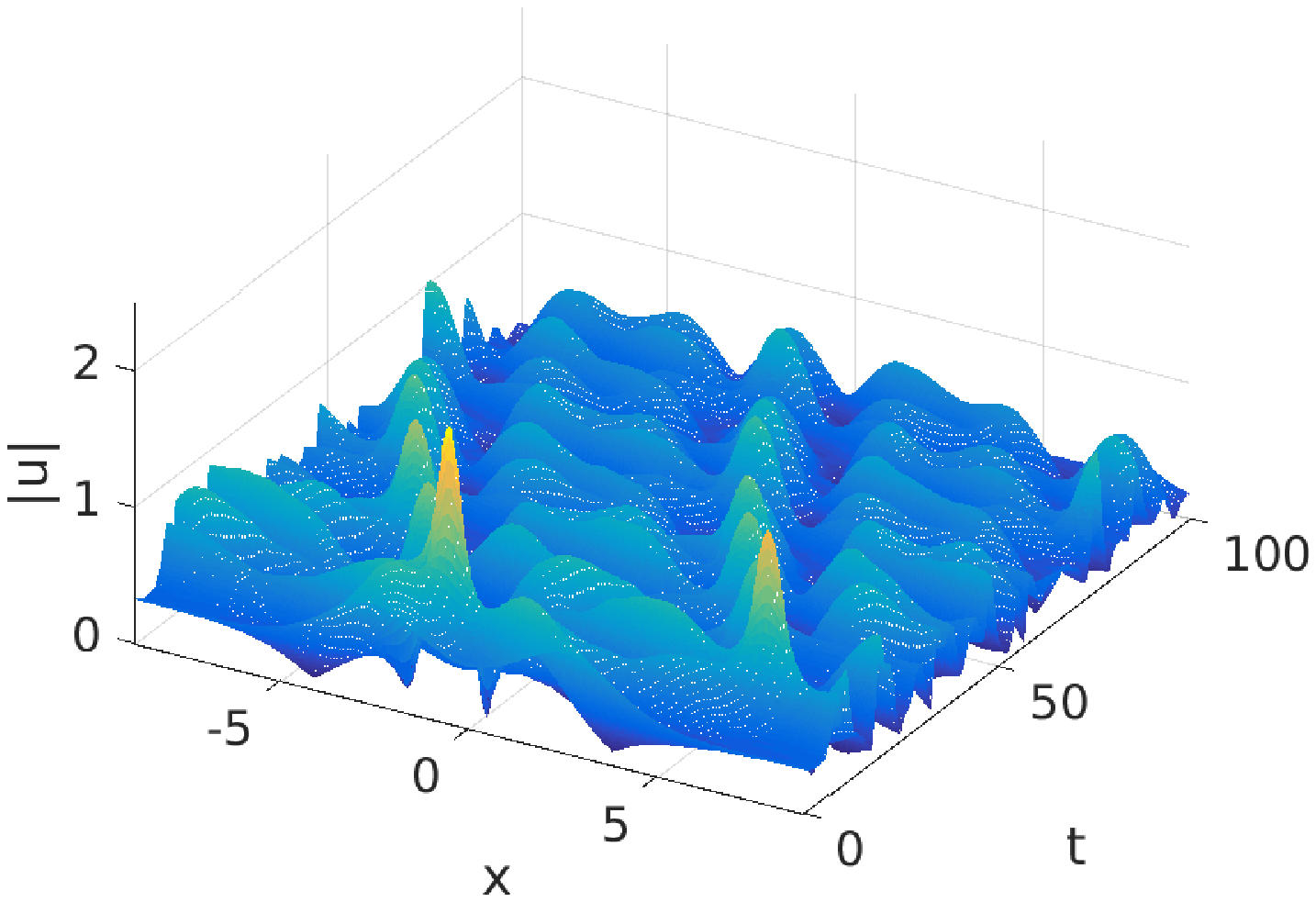}
\includegraphics[width=.33\textwidth]{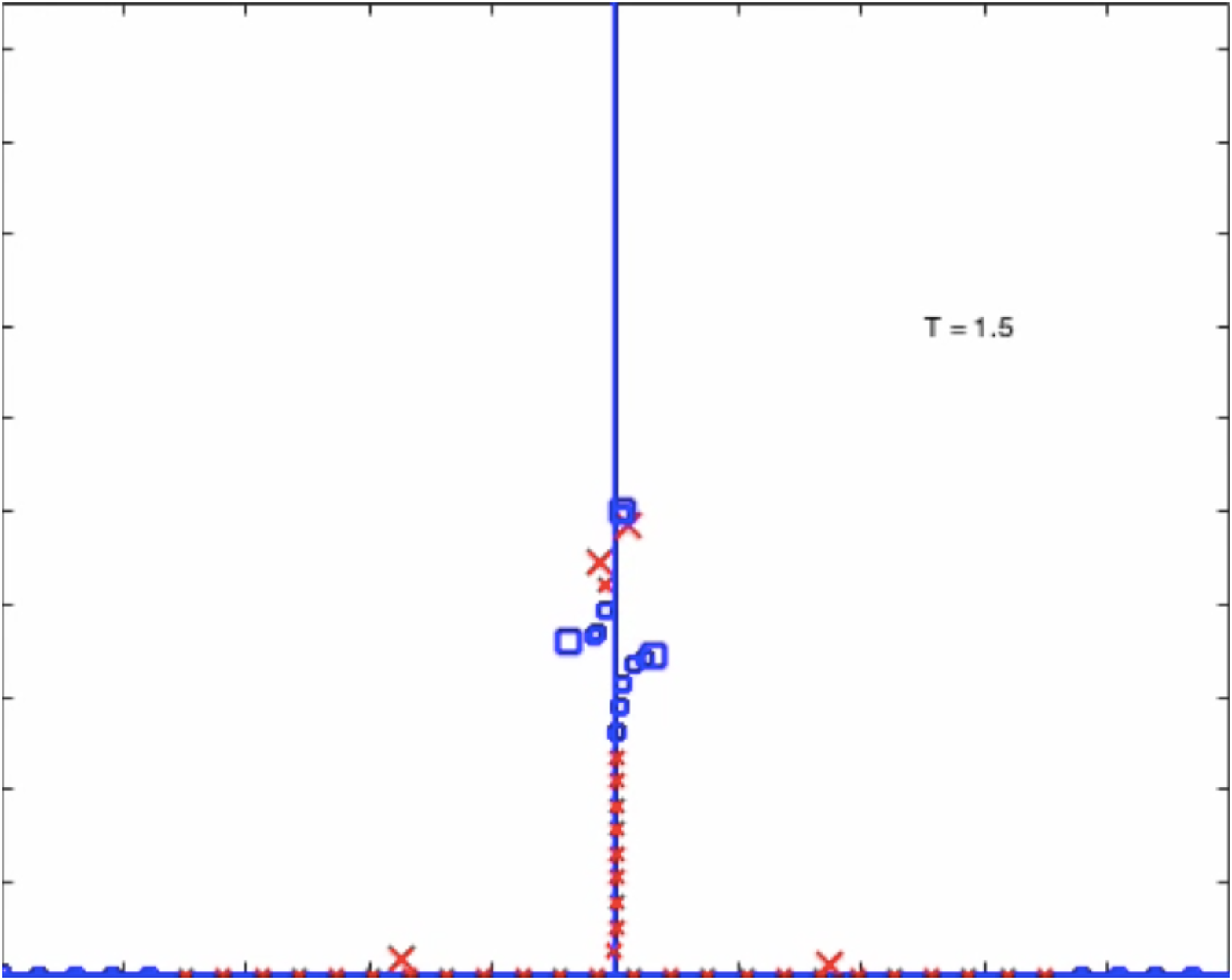}
\includegraphics[width=.33\textwidth]{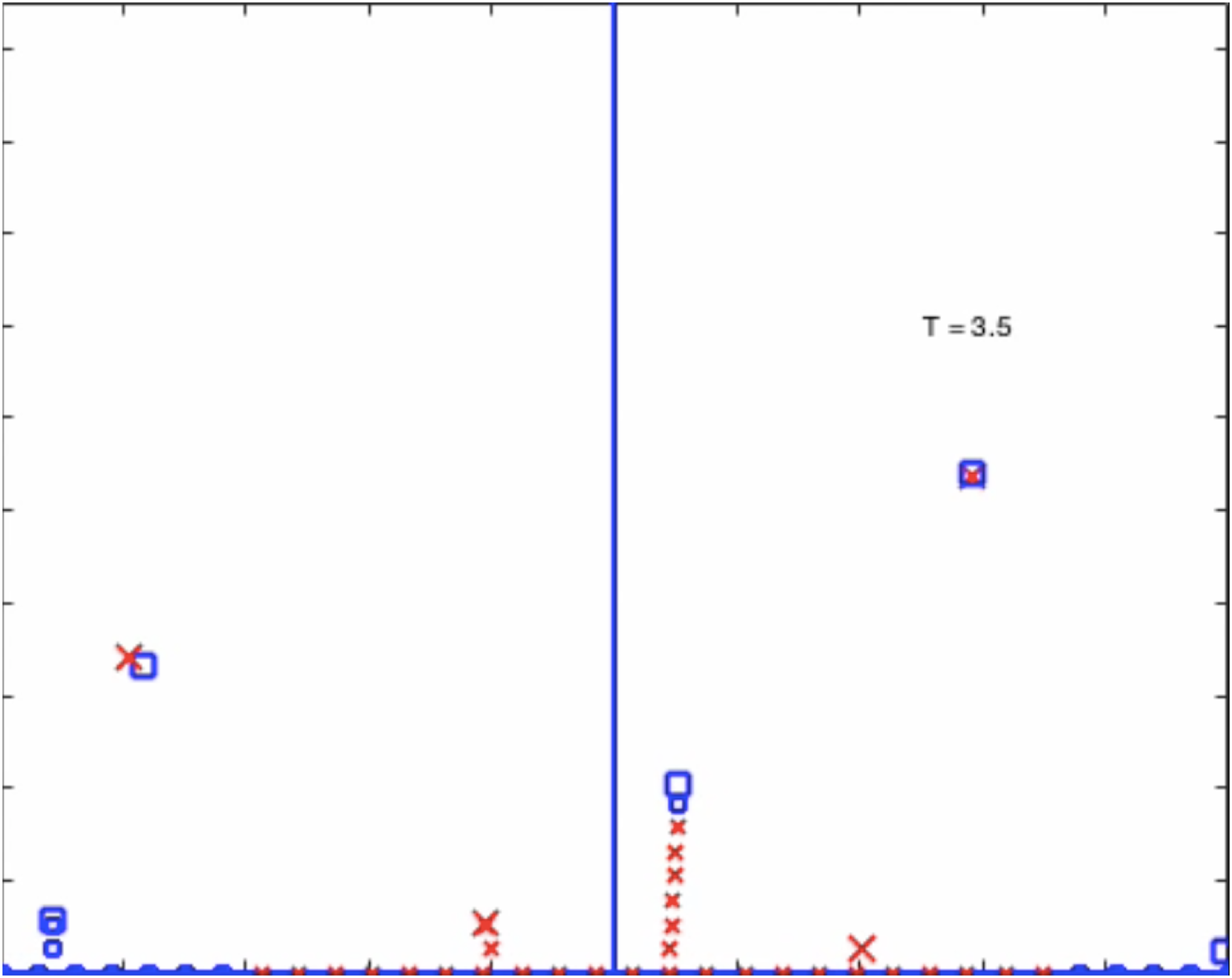}
}
  \centerline{\bf A\hspace{2.25in} B\hspace{2.25in} C}
  \vspace{12pt}
  \centerline{
\includegraphics[width=.33\textwidth]{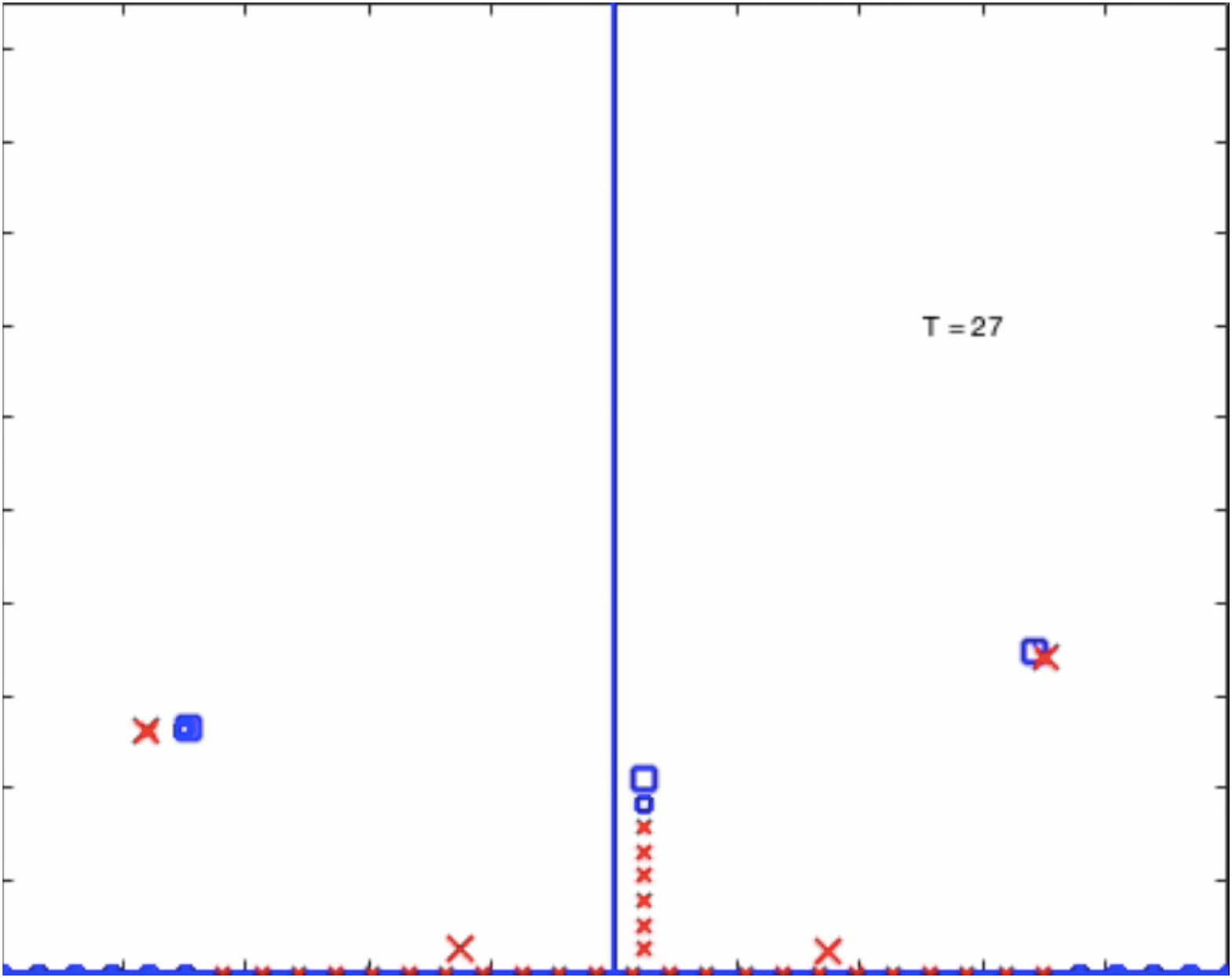}
\includegraphics[width=.33\textwidth]{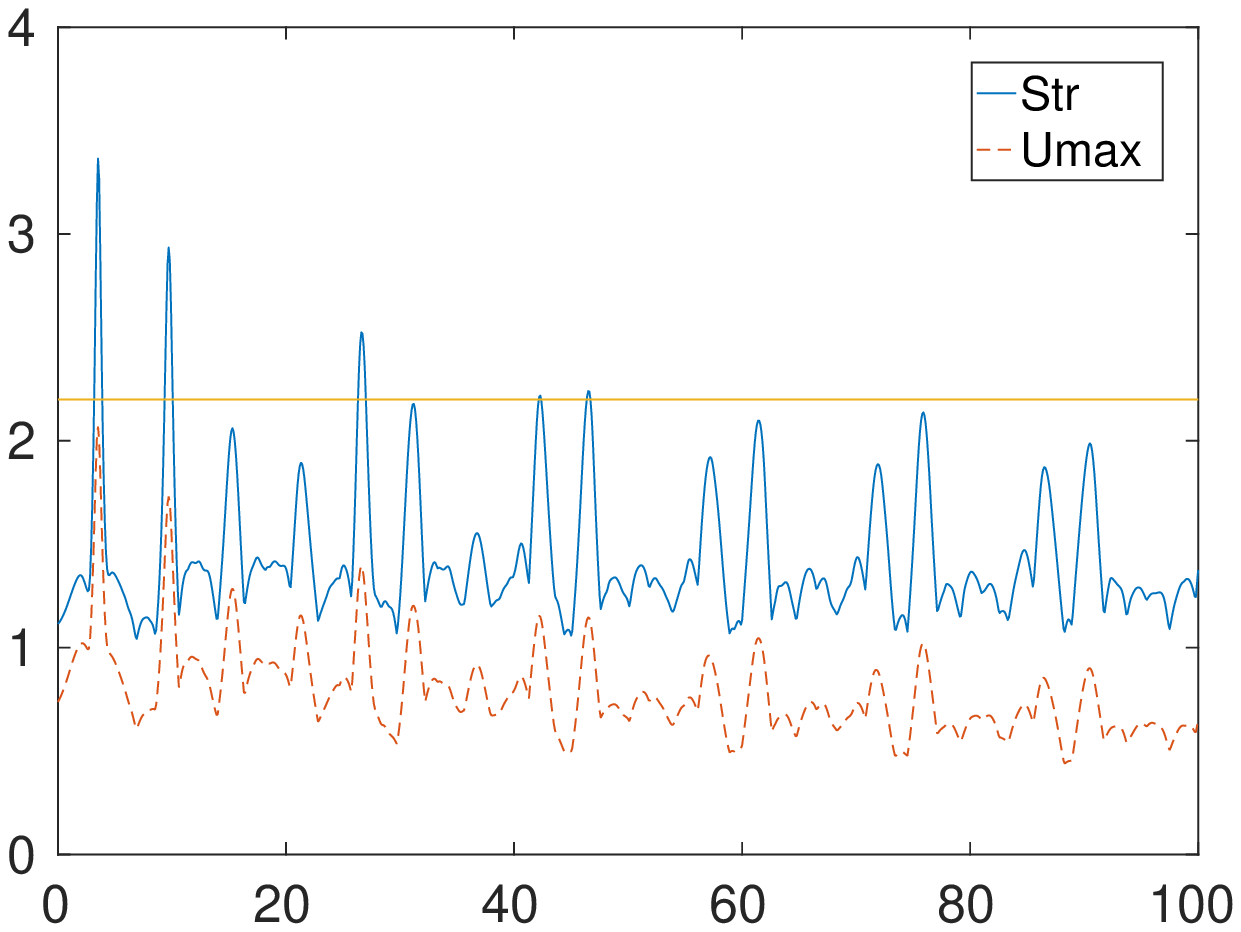}
\includegraphics[width=.33\textwidth]{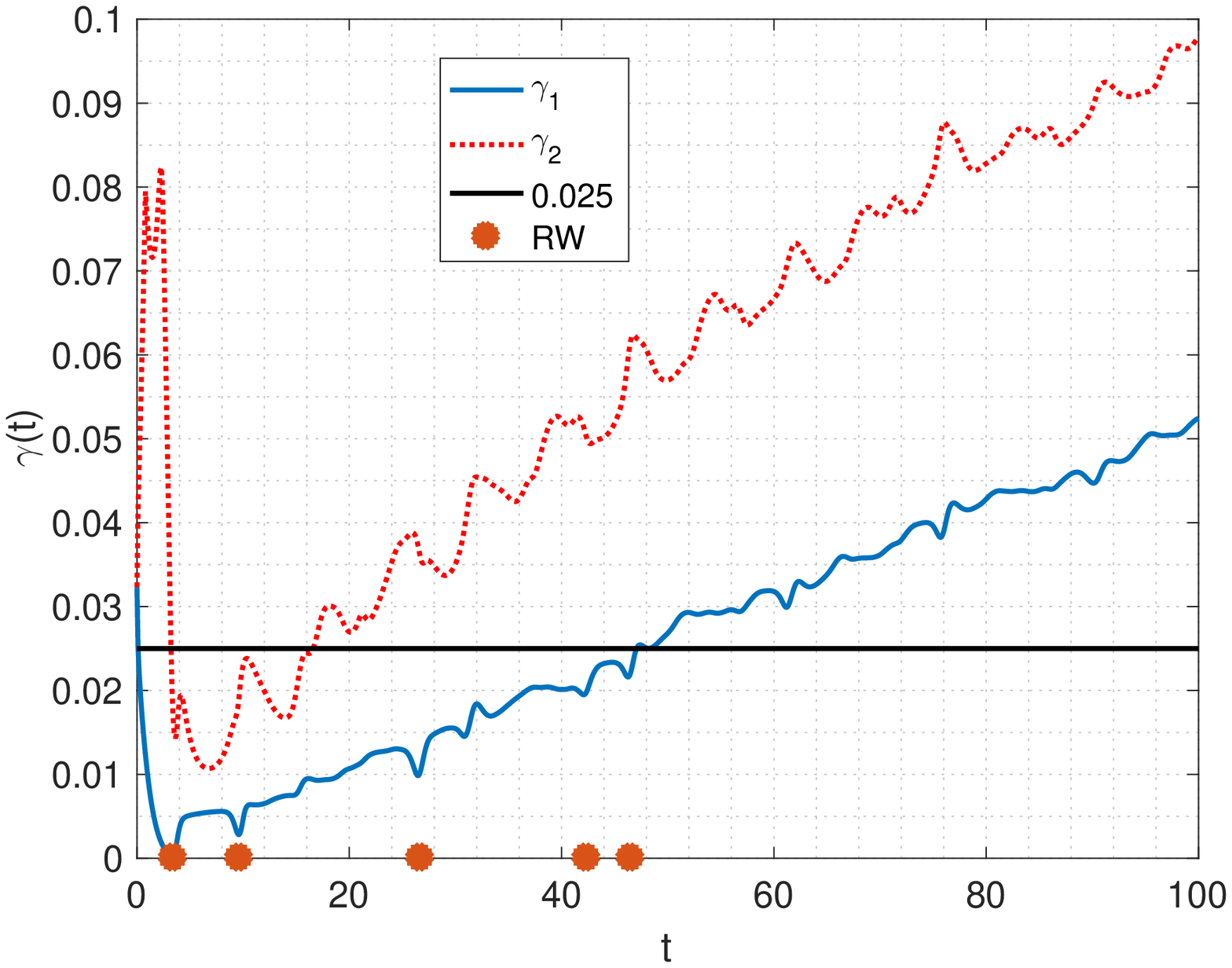}
}
  \centerline{\bf D\hspace{2.25in} E\hspace{2.25in} F}
  \caption{Two UM regime: (A) $|U^{(1,2)}_{\eps,\beta}(x,t)|,T_0=-3.5$, for $0\leq t\leq 100$, spectrum at (B) $t=1.5$, (C) $t=3.5$, (D) $t=27$, (E) strength, (F) Variation of the length of the band $|\lambda_0^s-\lambda_1^+|$ of spectrum as a function of time. The horizontal line represents the upper bound for a one soliton-like state. The red dots represent the time a rogue wave occurs.}
\label{Figure7}
\end{figure}

Figure~\ref{Figure7}A shows the surface
 $|U_{\epsilon,\beta}^{(1,2)}(x,t)|$  for $0 < t < 100$ for 
the coalesced SPB initial data  using $T_0 = -3.5$.
% generated using  Equation~\rf{SPB2} with
% $\rho = \tau = 0$ 
 The Floquet spectrum of the initial data, given in
  Figure~\ref{Figure4}B, contains
 two complex double points $\lambda_1^d$ and  $\lambda_2^d$ which,
under the NLD-HONLS flow,
  immediately split asymmetrically
 into $\lambda^{\pm}_1$ and $\lambda_2^{\pm}$, respectively 
 (Figure~\ref{Figure7}B).
The numerically observed evolution of spectrum for 
$U^{(1,2)}_{\eps,\beta}(x,t)$ for short time is confirmed by the 
perturbation results given in Equation \rf{ordereps} which indicate
 that under the NLD-HONLS flow both complex double points split asymmetrically at ${\cal O}(\epsilon)$.

Subsequently, the two upper bands 
 of spectrum quickly pinch off from the imaginary axis and shrink significantly in length, entering a two soliton-like spectral configuration at $t=3.5$  (Figure~\ref{Figure7}C).
The spectrum is in a two soliton-like state until $t \approx 18$  when
 one of the bands enlarges and the spectrum enters a one soliton state (e.g. see a sample one soliton-like spectrum at $t = 27$ in Figure~\ref{Figure7}D).  The spectral bands continue to enlarge as time evolves and the spectrum settles into a typical spectrum for a $5$-phase solution. 

 The variation in the band lengths  $\gamma_1(t) = |\lambda_0^s-\lambda_1^+|$
 and $\gamma_2(t) = |\lambda_1^- -\lambda_2^+|$
   are shown in
  Figure \ref{Figure7}F. 
 The horizontal line  in Figure \ref{Figure7}F indicates the upper bound (given by equation \rf{s_state}) on the band lengths for a one or two soliton-like state. Additionally a red dot on the $t$-axis indicates the time at which a rogue wave occurs providing a correlation between one or two soliton-like states and rogue waves.

 In this experiment, which used initial data for the coalesced SPB closer to the Stokes wave, the spectrum is in a one or two soliton-like
 configuration  whenever a rogue wave occurs.
The first two 
rogue waves  observed at $t\approx 3.5$
and $t\approx 10$ in the strength plot  (Figure~\ref{Figure7}E) appear when the spectrum is in a two
soliton-like configuration and are considerably larger in amplitude than the
rogue waves at $t \approx 27, 42, 46$ which 
occur when the spectrum is in a one soliton-like configuration.

\begin{figure}[ht!]
  \centerline{
\includegraphics[width=.45\textwidth]{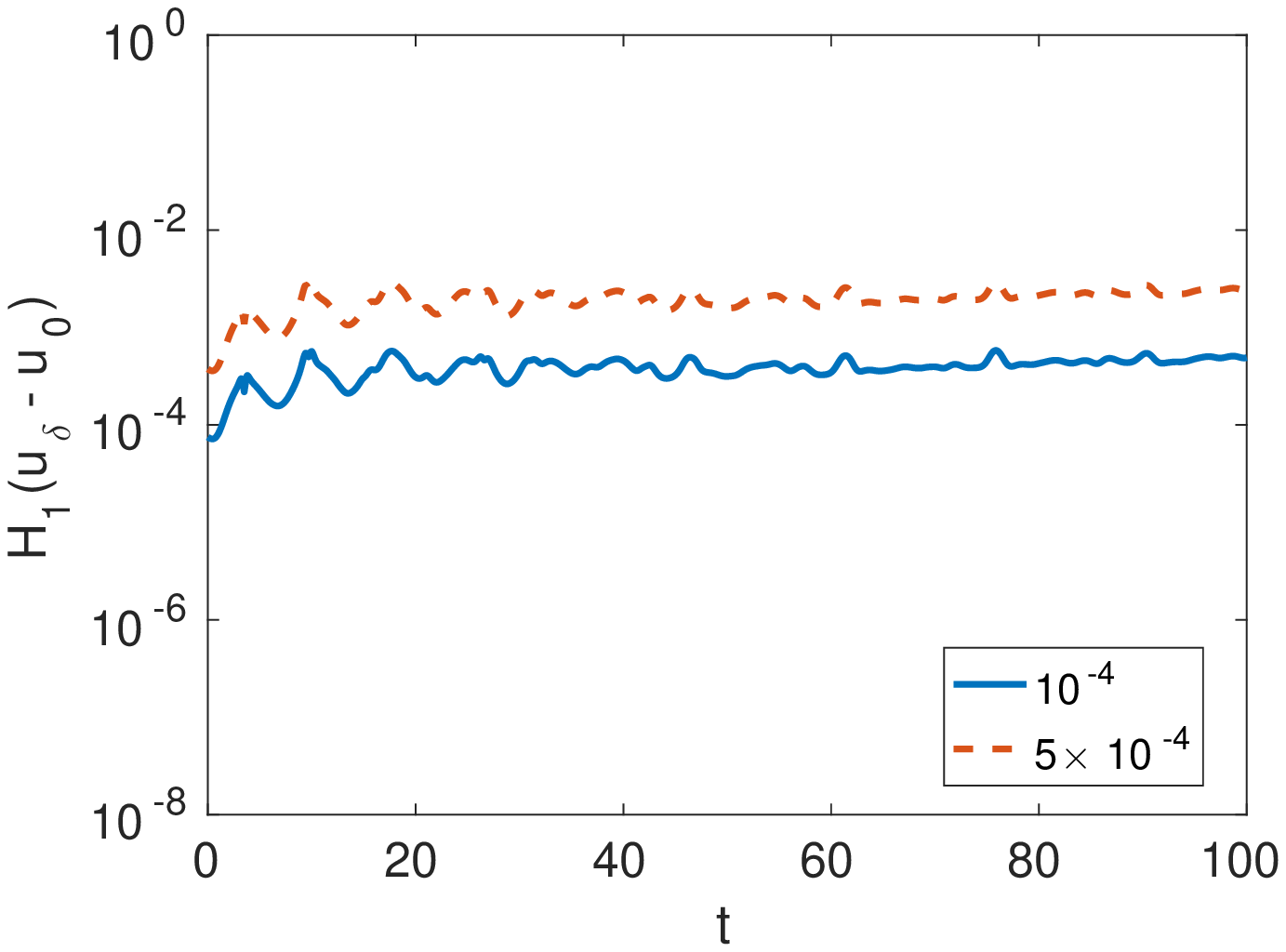}
\includegraphics[width=.45\textwidth]{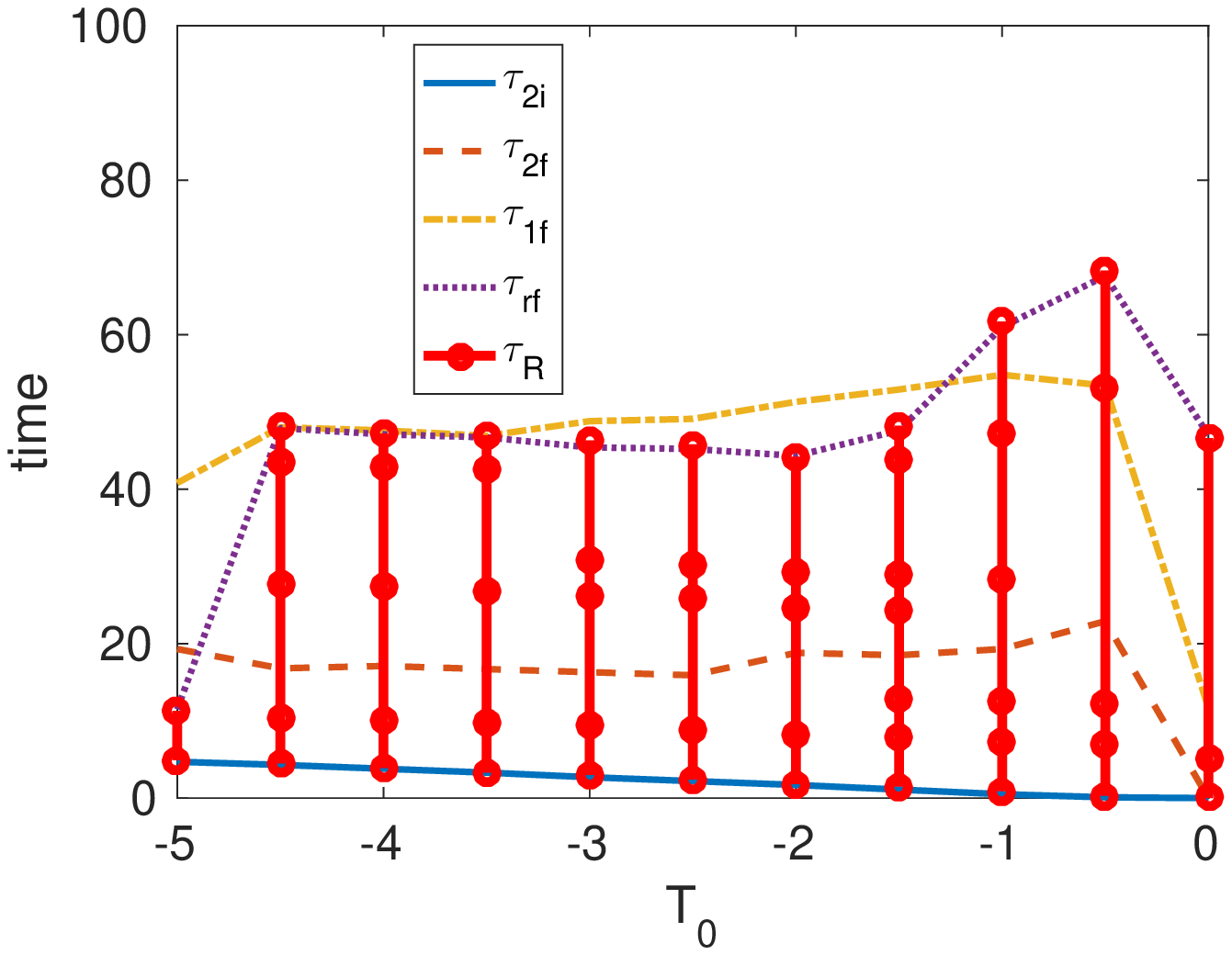}
}
  \centerline{\bf A\hspace{3in} B}
  \caption{Two UM regime: (A) $d(t)$ for $f_1(x)$, $\delta = 10^{-4}, 5 \times 10^{-4}$. (B)  The times $\tau_{2i}$, $\tau_{2f}$, $\tau_{1f}$, $\tau_{R}$, and $\tau_{rf}$, defined by \rf{times}, as $T_0$ varies.}
\label{Figure8}
\end{figure}

Complex critical points do not appear in the spectral evolution for $t >0$.
 In Figure~\ref{Figure8}A there is no significant growth in $d(t)$ and $U_{\eps,\beta}^{(1,2)}$ is  stable for
 $t>0$.  $U_{\eps,\beta}^{(1,2)}$  can be characterized by a continuous deformation of a stable NLS five-phase solution, keeping in mind that the modes determined by the very small bands of spectrum are limiting to a soliton-like structure.

{\sl Dependence of rogue wave activity on $T_0$:} In the $U^{(1)}_{\eps,\beta}(x,t)$ example we found the spectrum transitioned
repeatedly in and out of a one soliton-like  state (see Figure \ref{Figure5}). Transitioning repeatedly between one and two soliton-like states does not occur in $U^{(1,2)}_{\eps,\beta}(x,t)$. Instead, the evolution of the spectrum for $U^{(1,2)}_{\eps,\beta}(x,t)$ has the following pattern for all $T_0\neq 0$:

{\sl Both complex double points split $\rightarrow$  2
 soliton-like $\rightarrow$ 1 soliton-like
 $\rightarrow$ generic $N$-phase spectrum.}

\noindent 
For $T_0\neq 0$ the time at which a two soliton-like spectrum is last observed is the time at which a one soliton-like spectrum is first observed. 
Thus, to examine the dependence of rogue wave activity on $T_0$ for
$U^{(1,2)}_{\eps,\beta}(x,t)$, we determine the following:
  \be
  \ba{rcl}
(a)\;  \tau_{2i} &=& \mbox{the time  at which a two soliton-like spectrum is first observed}\\
(b)\; \tau_{2f} &=& \mbox{the time at which a  two-soliton-like spectrum is last observed}\\
  (c)\; \tau_{1f} &=& \mbox{the time at which a one-soliton-like spectrum is last observed}\\
(d)\; \tau_{R} &=& \mbox{the times at which rogue waves occur}\\
(e)\; \tau_{rf} &=& \mbox{the time of the last rogue wave}
  \ea
  \label{times}
  \ee
  
Figure \ref{Figure8}B provides the final times for :  $\tau_{2f}$  a two soliton-like state, a one soliton-like state $\tau_{1f}$, the last rogue wave $\tau_{rf}$ for 
$U^{(1,2)}_{\eps,\beta}(x,t)$ for  $T_0\in[-5,0)$. $T_0$ is incremented in steps of $0.5$. The red dots  at each sampled $T_0$ provide a time-line for the rogue wave events. Correlating the $\tau$-times  with the timeline for rogue waves we obtain the following 

  {\sl \bf Observation:}   For the coalesced two mode SPB $U^{(1,2)}_{\eps,\beta}(x,t)$, all  rogue waves emerge when  the spectrum is in a one or two soliton-like configuration for all values of
  $T_0 \in[-5,-1.5] $, i.e. for solutions initialized in the early to middle stage of the development of the MI.

\begin{figure}[ht!]
  \centerline{
\includegraphics[width=.33\textwidth]{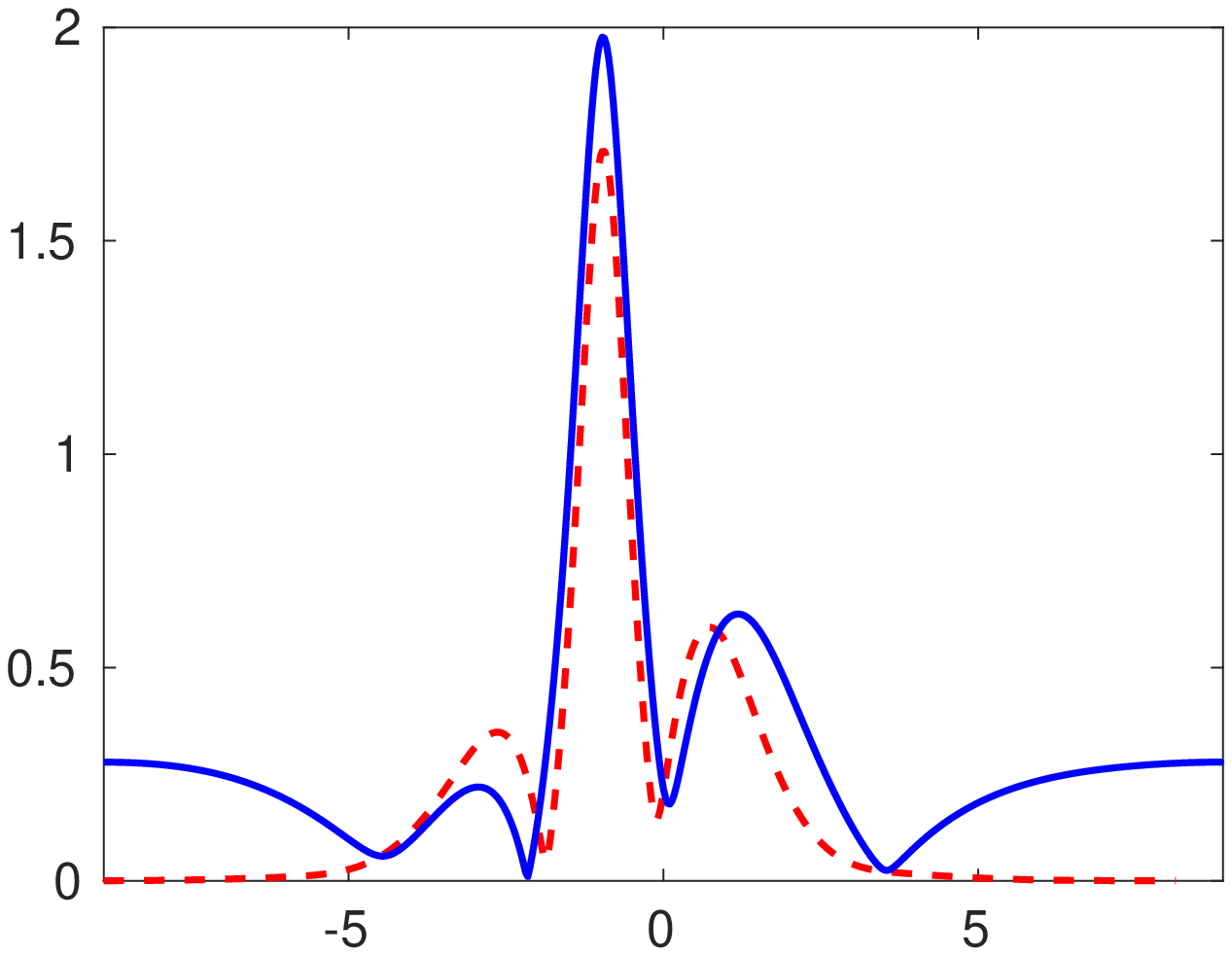}
\includegraphics[width=.33\textwidth]{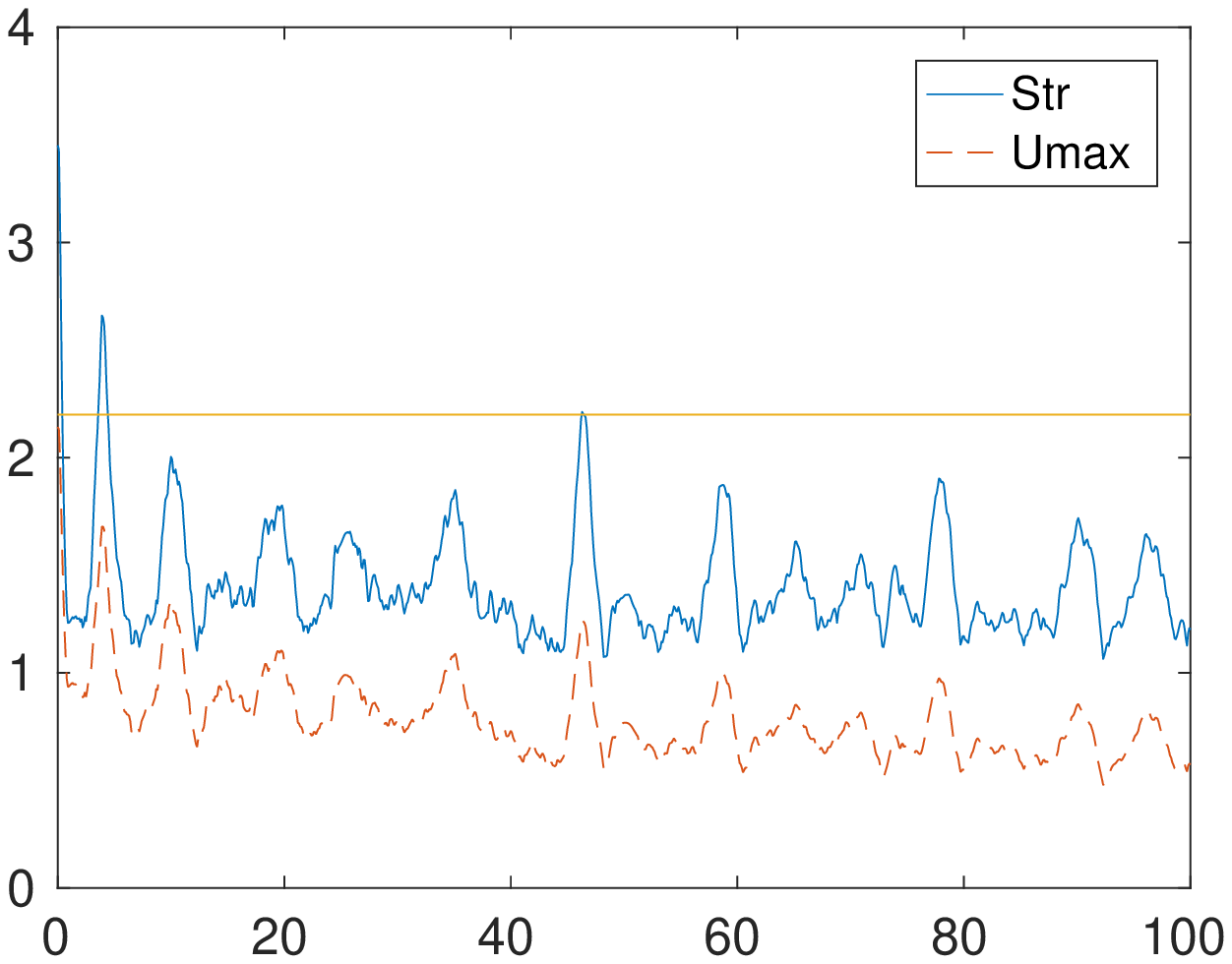}
\includegraphics[width=.33\textwidth]{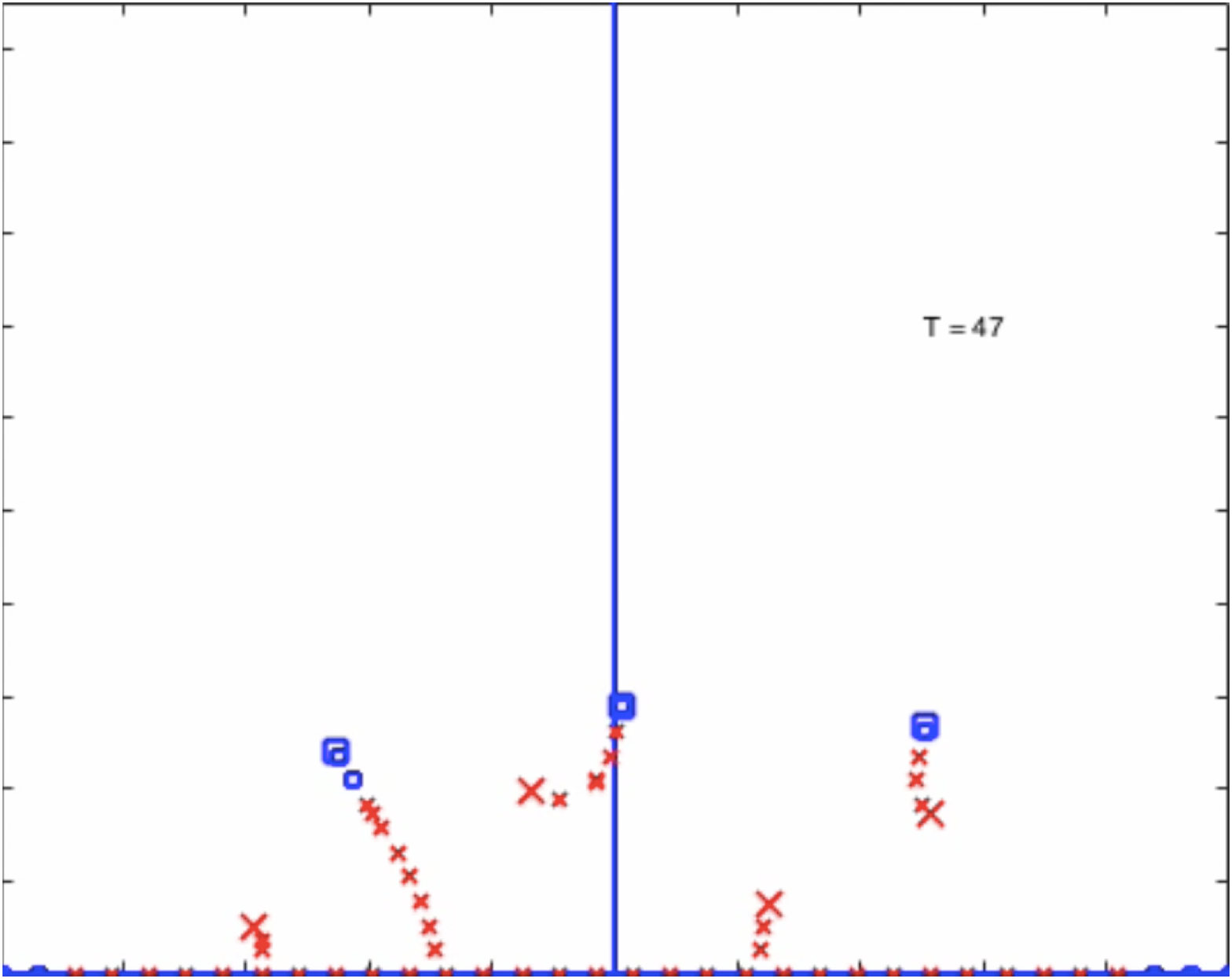}
}
  \centerline{\bf A\hspace{2.15in} B\hspace{2.15in} C}
  \caption{(A) $|U^{(1,2)}_{\eps,\beta}(x,5)|$ vs two-soliton analytical solution for $T_0 = -5$. Two UM regime $|U^{(1,2)}_{\eps,\beta}(x,t)|,T_0=0$: (B) Strength for $0\leq t\leq 100$ and (C) spectrum at $t=47$.}
  \label{Figure9}
\end{figure}

   Figure~\ref{Figure9}A provides the numerical solution $|U^{(1,2)}_{\eps,\beta}(x,5)|$ for $T_0 = -5$ versus the two-soliton analytical solution \rf{two-soliton} which uses the spectral data obtained from the spectral decomposition of the numerical solution.  
In formula \rf{two-soliton}  $\lambda_1$ and $\lambda_2$ are the midpoints of the two small bands that pinched off.
Even though the NLD-HONLS soliton-like solution  is over a non uniform finite background, near it's peak it structurally appears 
in good agreement with the two soliton waveform.

  There are other possible mechanisms for rogue wave development
  when the solutions are initialized as the MI is saturating.
  For $T_0 \in[-1,0]$, the last rogue wave occurs  after the spectrum has left a soliton-like state (for $T_0 =0$ considerably later, see Figure \ref{Figure8}B).
   Figure~\ref{Figure9}B shows the strength plot of
 $U_{\epsilon,\beta}^{(1,2)}(x,t)$ for  initial data
 obtained  using  $T_0 = 0$, i.e. initializing  at the peak of the SPB.
 A rogue wave 
occurs at $t = 47$. Figure~\ref{Figure9}C clearly shows the 
spectrum at $t = 47$ is not in a soliton-like configuration;
the rogue wave occurs due to a superposition of the nonlinear modes.

\section{Appendix}
\subsection{The Exponential Time Differencing Fourth-order Runge-Kutta scheme ETD4RK}
A spectral spatial discretization of the NLS equation (when $\eps = 0$ in Equation \rf{dhonls}) can be written as a system of ordinary differential equations of the form $\hat u_t + L\hat u = F(u)$, where $\hat u$ is the discrete Fourier transform of $u$, $L$ is a diagonal matrix, and $F$ is the discrete Fourier transform of the nonlinear term which is evaluated by transforming to physical space, evaluating the nonlinear terms, and transforming back to Fourier space. Using a fourth-order (2,2)-Pade approximation of $\e^{-z}$  the exponential time differencing fourth-order Runge-Kutta scheme ETD4RK is given by
\be
\hat u_h^{n+1} = R_{2,2}(hL)\hat u_h^n + P_1(hL) F\left(u_h^n\right) + P_2(hL)\left( F\left(a^n\right) +  F\left(b^n\right)\right) + P_3(hL) F\left(c^n\right)
\ee
where
\[\ba{rcl}
R_{2,2}(hL) &=& \left(12I - 6hL + h^2L^2\right)\left(12I + 6hL + h^2L^2\right)^{-1}\\
P_1(hL) &=& h \left(2I-hL\right)\left(12I + 6hL + h^2L^2\right)^{-1}\\
P_2(hL) &=& 4h\left(12I + 6hL + h^2L^2\right)^{-1}\\
P_3(hL) &=& h \left(2I+hL\right)\left(12I + 6hL + h^2L^2\right)^{-1}\\
a^n &=& R_{2,2}(hL/2) \hat u_h^n + P(hL)F\left(u_h^n\right)\\
b^n &=& R_{2,2}(hL/2) \hat u_h^n + P(hL)F\left(a^n\right)\\
c^n &=& R_{2,2}(hL/2) a^n + P(hL)\left(2F\left(b^n\right)-F\left(u_h^n\right)\right)\\
P(tL) &=& 24h\left(48I + 12hL + h^2L^2\right)^{-1}
\ea\]
Using a fixed step size $h$ throughout the integration allows the computation of the diagonal matrices $R_{2,2}$, $P$, $P_1$, $P_2$, and $P_3$ only once at the start of the integration \cite{khaliq2009}.

When applied to the NLS equation ETD4RK has been shown, analytically and numerically, to be highly efficient and accurate \cite{liang2015}.
In simulations of the NLS and higher order NLS equations quantities such as the energy, momentum, and Hamiltonian, are preserved with an ${\cal O}(10^{-11})$ accuracy. Furthermore, for the NLS equation, the ETD4RK scheme was shown to accurately preserve the Floquet spectrum  \cite{HIRS2013}. This is an important feature in our studies as the Floquet spectrum is a significant tool in the  analysis of rogue waves.

\subsection{Three-phase solution}
As an example consider the three-phase solution \rf{3phase} of the NLS given by
\[u(x,t) = \frac{\chi}{\sqrt 2}\e^{\ri t}\frac{A(x)\,\cn\left(t,\chi^2\right) + \ri \sqrt{1+\chi}\, \sn\left(t,\chi^2\right)}
    {\sqrt{1+\chi} - A(x)\,\dn\left(t,\chi^2\right)},\qquad
    A(x) = \frac{\cn\left(\sqrt{\frac{1+\chi}{2}} x, k\right)}{\dn\left(\sqrt{\frac{1+\chi}{2}} x, k\right)},\qquad
    k = \frac{1-\chi}{1+\chi}\]
    with $\chi = 0.98$ and spatial period $L = \frac{4\sqrt 2}{\sqrt{1+\chi}} \int_0^{\pi/2}\left(1+\chi^2\sin^2\theta\right)^{-1/2} d\theta$. The  Hamiltonian $H(t)$ for the NLS equation is given by Equation \rf{hamiltonian} when $\eps = 0$. We apply the ETD4RK scheme to the NLS equation using $N=256$ Fourier modes to simulate the three-phase solution. Figure \rf{Figure10} shows: (A) the surface $|u(x,t)|$, (B) the error $|H(t) - H(0)|$ for $\Delta t = 10^{-3}$, (C) the global error $|H(50) - H(0)|$ as a function of $\Delta t$, and (D) the global $L_2$-error $\|u(x,50)-u_{\Delta t}(x,50)\|_2$ as a function of $\Delta t$. The global error plots show ${\cal O}(\Delta t^4)$ convergence for $\Delta t \ge 10^{-3}$ with minimum errors achieved at $\Delta t \approx 10^{-3}$.

\begin{figure}[ht!]
  \centerline{
\includegraphics[width=.33\textwidth]{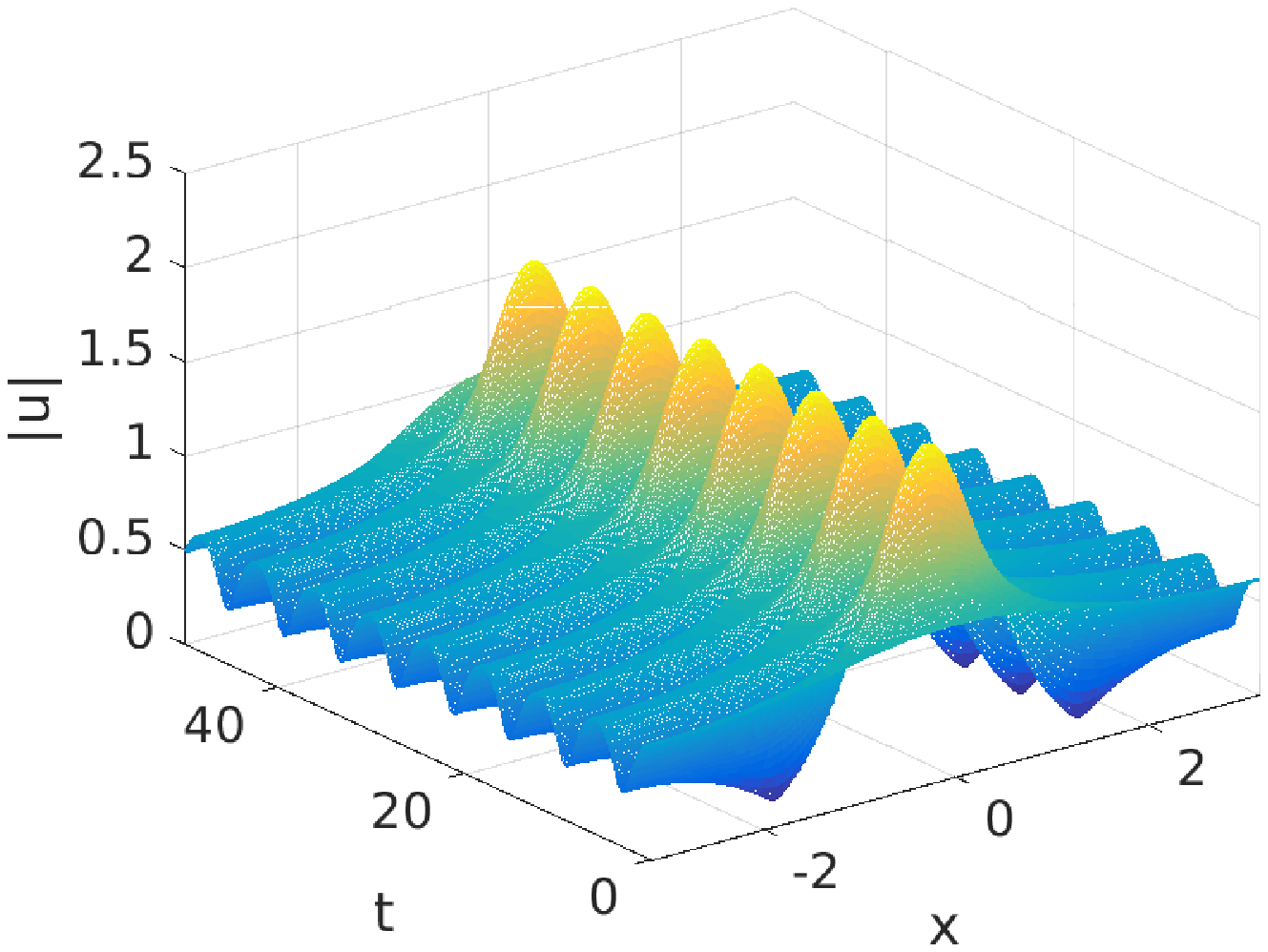}
\includegraphics[width=.33\textwidth]{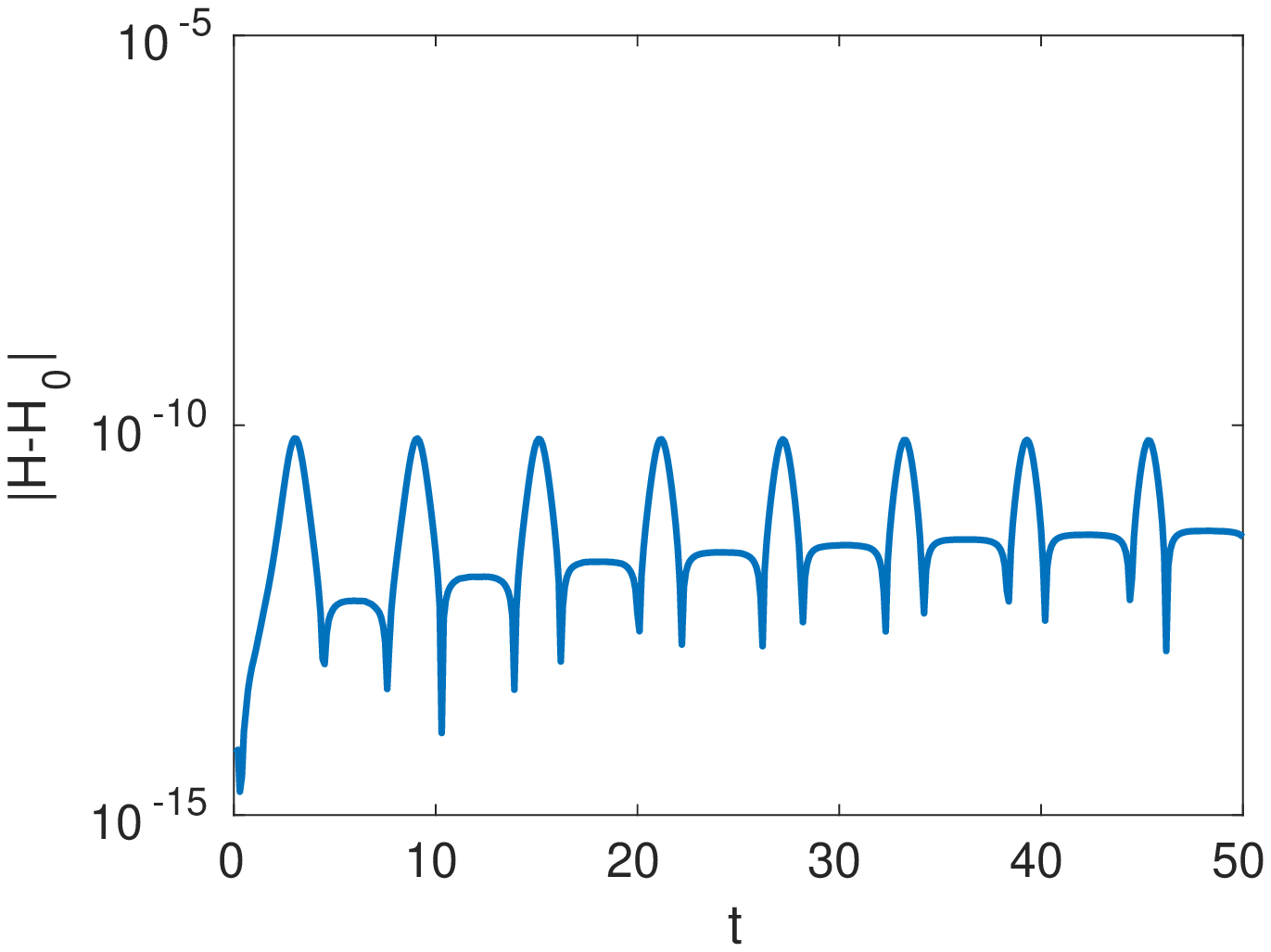}
}
  \centerline{\bf A\hspace{2.35in} B}
  \centerline{
\includegraphics[width=.33\textwidth]{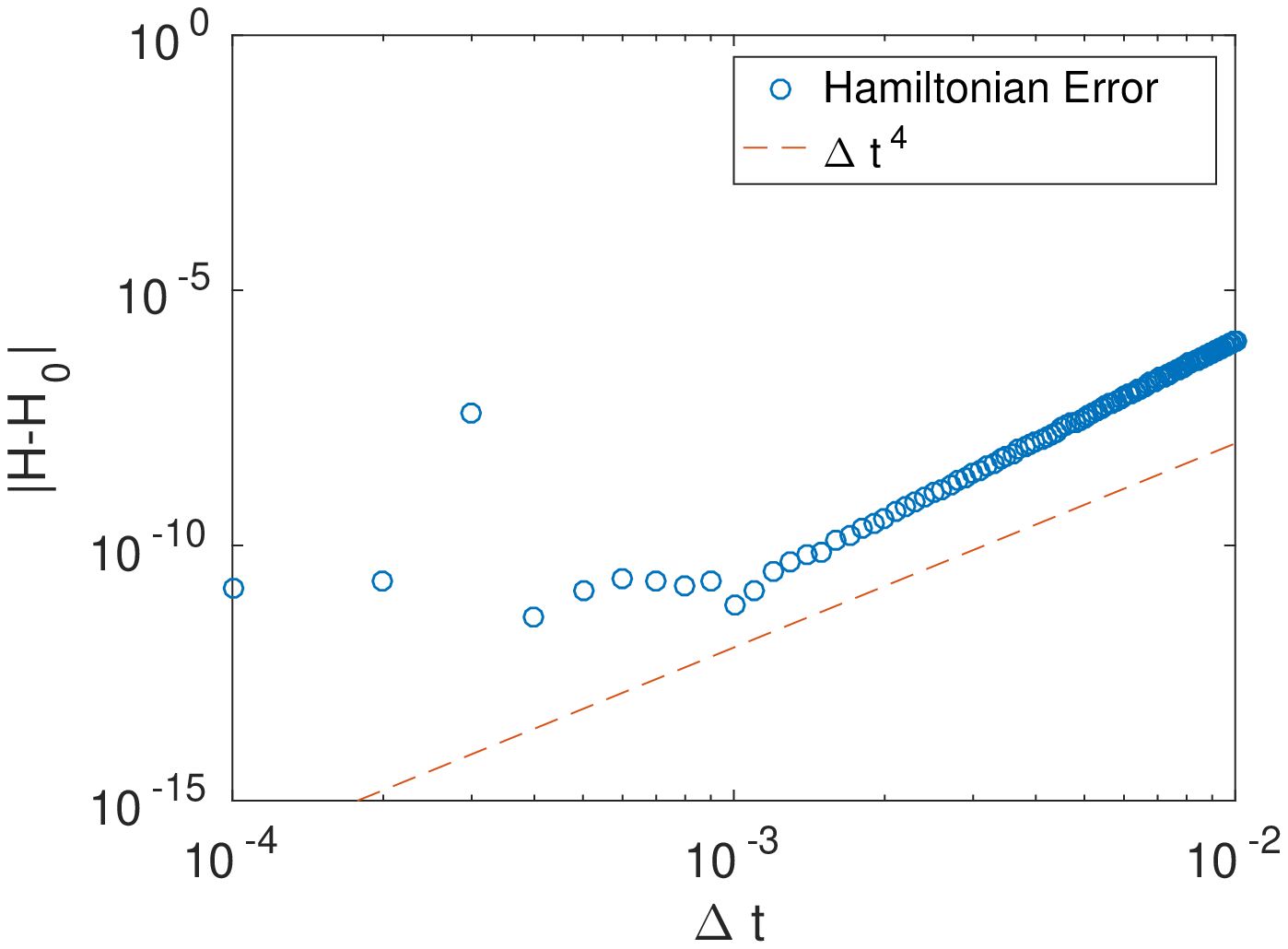}
\includegraphics[width=.33\textwidth]{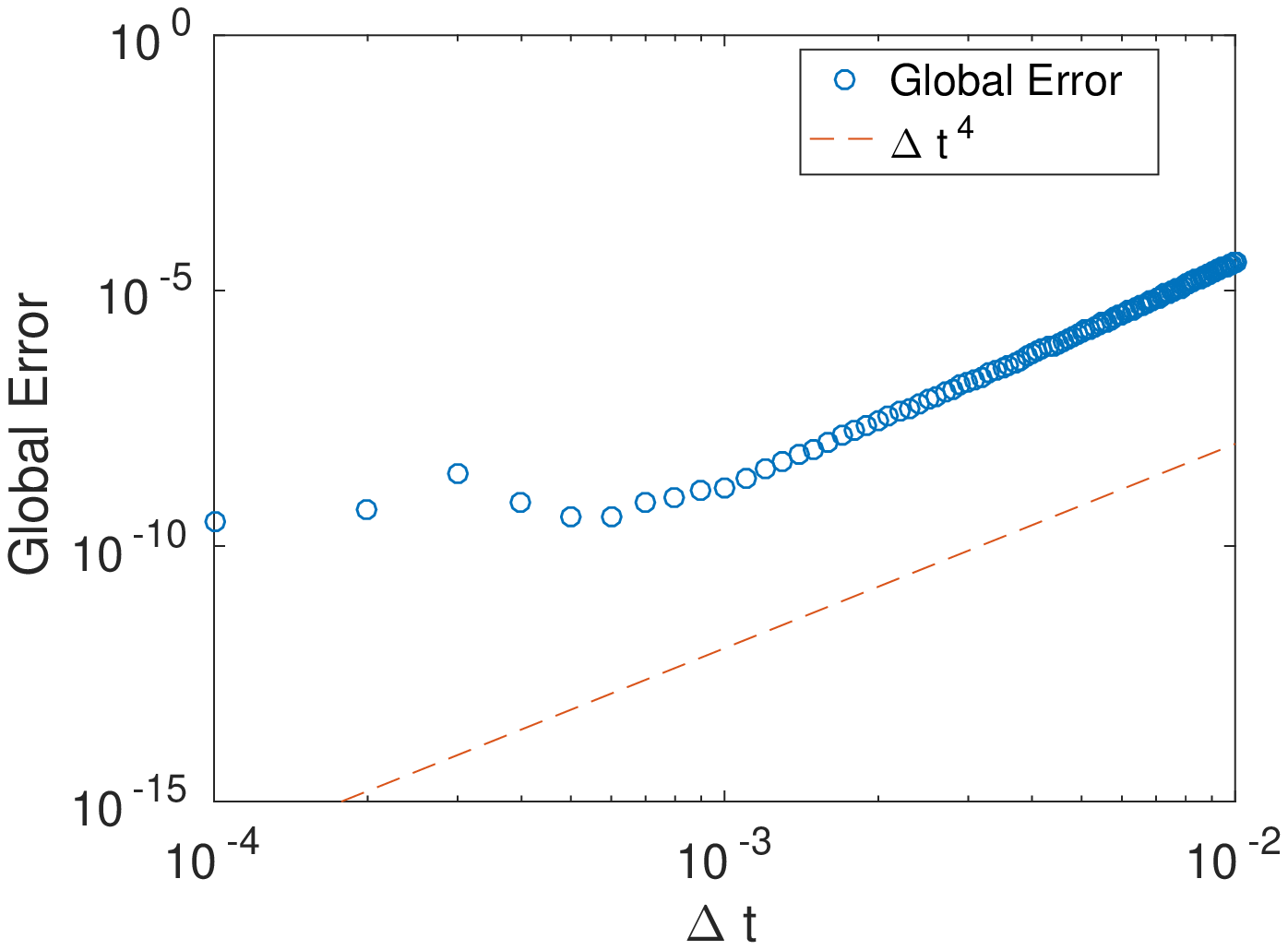}
}
  \centerline{\bf C\hspace{2.35in} D}
  \caption{Three-phase solution: (A) Surface $|u(x,t)|$, (b) error $|H(t)-H(0)|$,  (C) global error $|H(50)-H(0)$ and (D) global error $\|u(x,50)-u_{\Delta t}(x,50)\|_2$.}
\label{Figure10}
\end{figure}

\section*{Funding}
This work was partially supported by Simons Foundation, Grant  \#527565

%\bibliography{master2021}
\hyphenation{Post-Script Sprin-ger}\hyphenation{Post-Script
  Sprin-ger}\hyphenation{Post-Script Sprin-ger}\hyphenation{Post-Script
  Sprin-ger}

\end{document}